# A Software Engineering Perspective on Engineering Machine Learning Systems: State of the Art and Challenges

Görkem Giray

gorkemgiray@gmail.com

0000-0002-7023-9469

Abstract:

*Context:* Advancements in machine learning (ML) lead to a shift from the traditional view of software development, where algorithms are hard-coded by humans, to ML systems materialized through learning from data. Therefore, we need to revisit our ways of developing software systems and consider the particularities required by these new types of systems.

*Objective:* The purpose of this study is to systematically identify, analyze, summarize, and synthesize the current state of software engineering (SE) research for engineering ML systems.

*Method:* I performed a systematic literature review (SLR). I systematically selected a pool of 141 studies from SE venues and then conducted a quantitative and qualitative analysis using the data extracted from these studies.

*Results:* The non-deterministic nature of ML systems complicates all SE aspects of engineering ML systems. Despite increasing interest from 2018 onwards, the results reveal that none of the SE aspects have a mature set of tools and techniques. Testing is by far the most popular area among researchers. Even for testing ML systems, engineers have only some tool prototypes and solution proposals with weak experimental proof. Many of the challenges of ML systems engineering were identified through surveys and interviews. Researchers should conduct experiments and case studies, ideally in industrial environments, to further understand these challenges and propose solutions.

*Conclusion:* The results may benefit (1) practitioners in foreseeing the challenges of ML systems engineering; (2) researchers and academicians in identifying potential research questions; and (3) educators in designing or updating SE courses to cover ML systems engineering.

## TABLE OF CONTENTS

## 1 INTRODUCTION

"Software has been eating the world" (Andreessen, 2011), and with machine learning (ML) capabilities, software has become even more voracious. With the wide availability of digitized data and computational power, ML algorithms, which have been around for many decades, empowered software-intensive systems for providing additional beneficial functionalities. Some of the remarkable tasks that are successfully tackled by ML algorithms include autonomous driving (Chen et al., 2015), social network analysis (Benchettara et al., 2010), natural language processing (Young et al., 2018), image recognition (Zoph et al., 2018), and recommendation (Wei et al., 2017). Compelling examples of ML systems can be seen in various sectors, including finance (Dixon et al., 2020; Heaton et al., 2017), healthcare (Jiang et al., 2017), and manufacturing (Wuest et al., 2016), etc. Despite several promising examples, 47% of AI projects remain prototypes due to the lack of the tools to develop and maintain a production-grade AI system, according to Gartner Research (Gartner, 2020).

ML systems engineering in real-world settings is challenging since it adds additional complexity to engineering "traditional" software. We have separate bodies of knowledge for engineering ML capabilities (Hulten, 2018; Ng, 2018; Rao, 2019) and engineering traditional software (Bourque and Fairley, 2014). On the other hand, ML capabilities are generally served as parts of larger software-intensive systems (besides embedded software in robots or vehicles). Therefore, we need a holistic view of engineering software-intensive systems with ML capabilities (ML systems) in real-world settings. Many researchers from software engineering (SE) (Hesenius et al., 2019; Northrop et al., 2019) and ML (Karpathy, 2017; Menzies, 2019), as well as industry practitioners (Benton, 2020; Booch, 2019; Lin and Kolcz, 2012) have stated the requirement of such a holistic view. In line with this requirement, we can observe the recent events organized to discuss how to engineer ML systems. The Software Engineering for Machine Learning Applications (SEMLA) international symposium (Khomh et al., 2018) was arranged to bring together researchers and practitioners in SE and ML to explore the challenges and implications of engineering ML systems. In 2018, two main topics were addressed intensively in SEMLA (Khomh et al., 2018): (1) How can software development teams incorporate ML related activities into existing software processes? (2) What new roles, artifacts, and activities would be required to develop ML systems? In addition to applying ML techniques to SE tasks (such as defect prediction), researchers started to explore engineering ML systems within the scope of the International Workshop on Realizing Artificial Intelligence Synergies in Software Engineering (RAISE). Another event is the International Workshop on Artificial Intelligence for Requirements Engineering (AIRE) that welcomes submissions addressing requirements engineering and AI, such as (Belani et al., 2019). Christian Kästner at Carnegie Mellon University started to deliver a course called "Software Engineering for AI-Enabled Systems (SE4AI)", which takes an SE perspective on building software systems with a significant ML component (Kästner and Kang, 2020).

Industrial events are being held to discuss the implications of engineering ML systems on the current SE practices. One of such events is QCon.ai, which aims to bring SE and ML practitioners together to exchange experiences and thoughts on all aspects of SE for ML[1]. In such circumstances, speakers started to address the SE viewpoint of ML systems. For instance, Kishau Rogers suggested a roadmap for adopting enterprise architecture to AI capabilities[2] at O'Reilly Software Architecture Conference. Sriam Srnivasan from IBM presented SE practices for data science and ML lifecycle in DataWorks Summit 2018[3]. Peter Norvig from Google delivered talks on the intersection of SE and ML[4]. Recently, Ivica Crnkovic addressed the new challenges in architecting and managing the lifecycle for AI-based systems at the European Conference on Software Architecture (ECSA) 2020[5].

This study aims to present the state of the art on software engineering for ML systems. The distinction between SE and AI/ML has been blurred by as many disciplines and research areas. Researchers and practitioners from both fields need to understand the other side's concerns and have a holistic view of engineering ML systems. This paper can serve as a starting point to obtain such a holistic view and a repository of papers to explore this topic.

The goal of this paper is to summarize the state-of-the-art and identify challenges when engineering ML systems. Section 2 sets out the context and the vocabulary to be used in this paper. Section 3 presents the related work. Section 4 details the research objectives and method. Section 5 involves the results of detailed analysis and synthesis. Section 6 explains the findings by discussing potential research directions and threats to validity. Section 7 shall conclude the paper.

## 2 BACKGROUND

This section introduces the foundational concepts and describes the vocabulary by defining the essential concepts used in this paper.

---

[1] "The Applied AI Software Conference for Developers", https://qcon.ai/

[2] "Enterprise architecture for artificial intelligence", https://conferences.oreilly.com/software-architecture/sa-eu-2018/public/schedule/detail/71390

[3] "Software engineering practices for the data science and machine learning lifecycle" at "DataWorks Summit" in 2018, https://www.youtube.com/watch?v=IA3UXH7XBXw

[4] For instance, "SE with/and/for/by ML", https://www.youtube.com/watch?v=MP-wAc-XNUU

[5] "AI engineering – new challenges in system and software architecting and managing lifecycle for AI-based systems", at ECSA 2020, https://ecsa2020.disim.univaq.it/details/ecsa-2020-keynotes/1/AI-engineering-new-challenges-in-system-and-software-architecting-and-managing-lif

## 2.1 ARTIFICIAL INTELLIGENCE, MACHINE LEARNING, AND DEEP LEARNING

Artificial intelligence (AI) is the name of the field involving the efforts for building intelligent agents (Russel and Norvig, 2013). Intelligent agents perceive their environment and try to achieve their goals by acting autonomously (Russel and Norvig, 2013). Machine learning (ML) is a subfield of AI, which tries to acquire knowledge by extracting patterns from raw data (Ashmore et al., 2021; Goodfellow et al., 2016) and solve some problems using this knowledge. Deep Learning (DL) is a subfield of ML that focuses on creating large neural network models capable of making accurate data-driven decisions (Kelleher, 2019). DL has emerged from research in AI and ML and is particularly suited to contexts where large datasets are available and the data is complex (Kelleher, 2019).

Based on the dataset representation and the approach to defining the candidate models and final model (or function), three different categories of ML are identified: supervised, unsupervised, and reinforcement learning (Kelleher, 2019). In supervised learning, an ML model (or a function mapping inputs to outputs) is constructed using a training dataset with labels. Classification and regression problems are typical examples of supervised learning. In unsupervised learning, a function to describe a hidden structure is inferred from unlabeled data. Common problems for unsupervised learning are clustering and association rule learning. In reinforcement learning, an agent learns from a series of reinforcements, i.e., rewards and punishments (Russel and Norvig, 2013). Reinforcement learning finds lots of uses in video games.

## 2.2 TRADITIONAL SOFTWARE AND MACHINE LEARNING SYSTEMS

Engineering traditional software (or conventional software (Druffel and Little, 1990)) is about the implementation of programs (arithmetic & logic operations, a sequence of if-then-else rules, etc.) explicitly by engineers in the form of source code, which can be decomposed into units (e.g., classes, methods, functions, etc.) (Ma et al., 2018b). As the left side of Figure 1 shows, the input and hand-designed program are provided to the computer, and an output is generated. On the contrary, in ML systems, ML algorithms search through a large space of candidate programs, driven by training experience, to find a program that optimizes the performance metric (Jordan and Mitchell, 2015) (i.e., fulfills the requirements). As the right side of Figure 1 shows, the computer extracts patterns using input and output data; in other words, learn a function that maps from inputs to outputs.

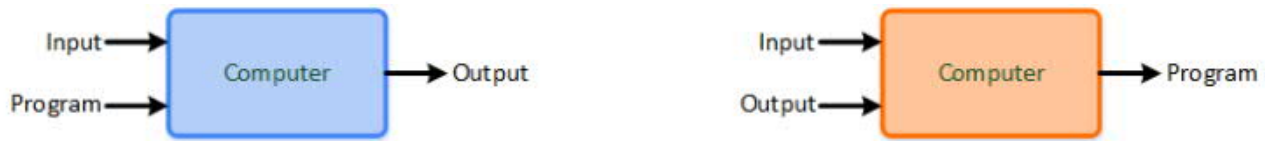

Figure 1. Traditional software development (on the left) vs. ML system development (on the right) (Domingos, 2015)

One of the success factors for ML system development is the identification of a subset of input data (the process of feature selection) that is informative for a given task. Traditional ML algorithms entail a manual feature selection process, which may require domain expertise, statistical analysis of input data, and experiments for building models with different feature sets (Kelleher, 2019). As the left side of Figure 2 shows, engineers provide hand-designed features to computers to have a function (or program) that maps inputs to outputs. Since the design of features often involves a significant amount of human effort in traditional ML, DL takes a different approach to feature selection by automatically learning features that are most useful for a given task from the input data (Kelleher, 2019). As the right side of Figure 2 shows, features are learned by computer using DL algorithms. DL systems development is generally possible when large enough datasets are available and is particularly useful for the tasks in complex high-dimensional domains, such as face recognition and machine translation (Kelleher, 2019).

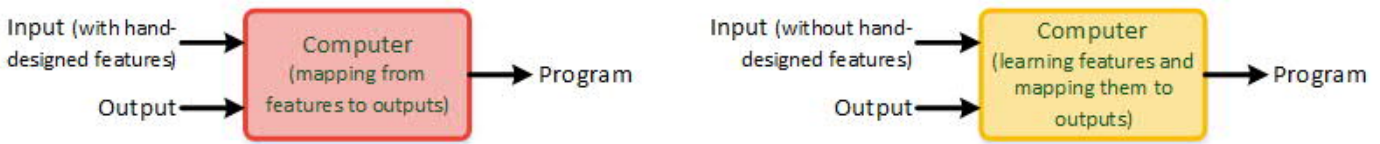

Figure 2. ML system development (on the left) vs. DL system development (on the right) – adapted from (Domingos, 2015) and (Goodfellow et al., 2016)

Practitioners and academicians do not use a standard term to name systems that incorporate AI/ML/DL capabilities. Software that applies AI techniques can be referred to "AI software" (Nemecek and Bemley, 1993) or "AI-based software" (Druffel and Little, 1990). In Cyber-Physical systems, a component whose behavior is driven by an ML/DL model obtained via training and updated through a learning process is called a "learning-enabled component" (Hartsell et al., 2019). Recently, in parallel with the success of ML, such systems are referred to as ML systems (Kuwajima et al., 2020)/applications (Kriens and Verbelen, 2019)/solutions (Bosch et al., 2020). In this paper, I use the term "ML system" as either a software framework, tool, library, or component that provides ML (including DL) functionalities or software systems that include ML components (Wan et al., 2019).

## 2.3 SE FOR ML AND ML FOR SE

Researchers have been addressing the interplay between SE and ML (as well as AI more broadly and DL more narrowly) for many years (Partridge, 1988; Rech and Althoff, 2004; Zhang and Tsai, 2003). On the other hand, some researchers point out the rift between SE and ML communities. One of the reasons for this rift may be these communities' focus: the ML community focuses on algorithms and their performance, whereas the SE community focuses on implementing and deploying software-intensive systems (Khomh et al., 2018). Bringing together the knowledge and experience of these two communities uncovers two areas of synergy:

1. *SE for ML* refers to addressing various SE tasks for engineering ML systems, i.e., designing, developing, and maintaining ML-enabled software systems. Researchers are trying to identify the different aspects of engineering ML systems compared to traditional software and develop new techniques and tools to cope with these differences.
2. *ML for SE* refers to applying or adapting AI technologies to address various SE tasks (Xie, 2018), such as software fault prediction (Malhotra, 2015), code smell detection (Azeem et al., 2019), reusability metrics prediction, and cost estimation (Panigrahi et al., 2019), etc. Researchers utilize ML models obtained from SE data (source code, requirement specifications, test cases, etc.) to engineer software more efficiently and effectively.

This paper focuses on SE for ML by systematically reviewing the SE literature on engineering ML systems.

## 3 RELATED WORK

According to my literature search, whose summary is presented in Table 1, researchers started to allocate effort in identifying SE aspects of engineering ML systems. Masuda et al. (2018) aimed to discover the techniques for evaluating and improving the quality of ML systems. Ashmore et al. (2021) and Liu et al. (2019) focused on ML systems' safety and security aspects, respectively. Sherin and Iqbal (2019) and Felderer et al. (2019) conducted a systematic mapping study (SMS) to identify, analyze, and classify the literature on ML testing. Zhang et al. (2020c) worked a more comprehensive SLR on various aspects of testing ML systems, including testing properties (correctness, robustness), testing components (data, ML framework), and testing workflow (generation and evaluation). Braiek and Khomh (2020) reported challenges, existing solutions, and gaps in testing ML systems using a smaller set of primary studies compared to Zhang et al., (2020). Riccio et al. (2020) focused on the existing solutions and open challenges for functional testing of ML systems. Watanabe et al. (2019) published an SLR involving seven papers to understand the current practices for developing ML systems. Washizaki et al. (2019b) collected good and bad SE design patterns for ML systems from the academic and grey literature.

To the best of my knowledge, Kumeno (2019) published the first SLR addressing the challenges of engineering ML systems. Unlike Kumeno (2019), I was able to identify more papers published in SE venues (141 vs. 47) and provided a taxonomy of challenges and proposed solutions classified under SE's main knowledge areas. Lwakatare et al. (2020) have recently published an SLR to identify the challenges of developing and maintaining large-scale ML-based systems in industrial settings. The primary study pool of (Lwakatare et al., 2020) includes papers mainly from journals and conferences on ML/DL, data engineering, big data, knowledge discovery, and data mining. Around five primary studies of Lwakatare et al. (2020) were published in SE venues. In another recent study, Nascimento et al. (2020) investigated SE practices applied and challenges faced in the development of AI/ML systems by analyzing 55 papers.

I compare this study with the existing works in three aspects:

1. *Time frame:* This review on the state-of-the-art is the most contemporary as it is the most up-to-date review. While none of the existing reviews addressing SE aspects (Kumeno, 2019; Lwakatare et al., 2020; Nascimento et al., 2020) cover the papers published in 2020, this study covers the papers published until the end of 2020 in SE venues (75 papers in total).
2. *Comprehensiveness:* The number of papers analyzed in this study is higher than the existing reviews addressing SE aspects. The included number of papers in Kumeno (2019), Lwakatare et al. (2020), Nascimento et al. (2020), and this study are 115, 72, 55, and 141, respectively.

3. *Analysis:* The challenges and the proposed solutions are mapped in this study. The potential SE-related research directions are provided in Section 6.1.

Table 1. Summary of related work

| Ref | Time Frame | Objective | Research Method & Source of Information | Knowledge Area Focus |
|-----|-----------|-----------|----------------------------------------|----------------------|
| Masuda et al., 2018 | 1987 – 2017 | discover techniques to evaluate and improve the software quality of ML systems | SLR using DB search on a limited scope of venues<br><br>78 papers from 16 academic conferences & 5 academic magazines on AI & SE | Quality |
| Liu et al., 2019 | 2001 – 2018 | pinpoint challenges and future opportunities for developing secure DL systems from a software quality assurance perspective | SLR using DB search on the conferences listed on the Computer Science Rankings within the scope of AI & ML, SE, and security<br><br>223 papers | Quality (Security) |
| Sherin and Iqbal, 2019 | 2007 – Jan 2019 | identify, analyze and classify the existing literature on testing ML systems | SMS using DB search, snowballing, and manual search<br><br>37 papers | Testing |
| Kumeno, 2019 | 2003 – Sep 2019 | clarify the SE challenges for ML applications | SLR using DB search and snowballing<br><br>115 papers (47 SE papers + 68 ML papers) | SE + ML |
| Watanabe et al., 2019 | 2010 – Mar 2019 | understand current practices and to help developers standardize ML system development processes | SLR using DB search<br><br>7 papers | SE Process |
| Washizaki et al., 2019b | 2008 – Aug 2019 | collect good/bad SE design patterns for ML systems | MLR using DB search and snowballing<br><br>38 papers/documents (academic & grey literature) | Design |
| Zhang et al., 2020c | 2007 – Jun 2020 | analyze and report data on the research distribution, datasets, and trends that characterize the ML testing literature; identify challenges, open problems, and promising research directions for ML testing | SLR using DB search, snowballing, and contacting researchers working on ML testing<br><br>138 papers | Testing |
| Braiek and Khomh, 2020 | 2007 – 2019 | identify challenges, report existing solutions, and identify gaps in testing ML systems | SMS using DB search and backward snowballing<br><br>37 papers | Testing |
| Riccio et al., 2020 | 2004 – Feb 2019 | identify the existing solutions and open challenges for functional testing of ML systems | SMS using DB search and snowballing<br><br>70 papers | Testing |
| Lwakatare et al., 2020 | 1998 – Feb 2019 | survey the literature related to the development and maintenance of large scale ML-based systems in industrial settings to provide a synthesis of the challenges that practitioners face | SLR using snowballing and DB search<br><br>72 papers from SE, ML, and data science venues | SE |
| Nascimento et al., 2020 | 1990 – 2019 | investigated SE practices applied and challenges faced in the development of AI/ML systems | SLR using DB search and snowballing<br><br>55 papers | SE |
| This study | 2007 – Dec 2020 | identify, analyze, summarize, and synthesize the current state of SE research for engineering ML systems | SMS using DB search, snowballing, and manual search<br><br>141 papers | SE |

In addition to the academic literature, practitioners and researchers in the industry have also begun to address the SE aspects of developing and maintaining ML systems. Lorica and Loukides (2018) stated that "Machine learning is poised to change the nature of software development in fundamental ways, perhaps for the first time since the invention of FORTRAN and LISP." Heck identified "engineering ML systems" as a new discipline, which needs further work on methods, tools, frameworks, and tutorials (Heck, 2019). Sato et al. (2019) argued that the process for developing, deploying, and continuously improving ML systems is more complex than traditional software, such as a web service or a mobile application. As these examples show, the industry is calling for action to resolve the challenges of engineering ML systems and to propose new techniques to cope with the additional complexity of ML systems.

## 4 RESEARCH OBJECTIVES AND METHOD

This section describes the research objectives and the method used in this study. An SLR approach was adopted to synthesize the knowledge of engineering ML systems from an SE perspective. The research method was based on established guidelines (Kitchenham and Charters, 2007; Wohlin, 2014), some previous good examples (Durelli et al., 2019; Mohanani et al., 2018), and my previous experience in conducting SLR (Garousi et al., 2019; Giray and Tüzün, 2018; Tarhan and Giray, 2017). Table 2 summarizes the SLR protocol used in this study (the format was taken from Motta et al. (2018)). The details are described in the following subsections.

Table 2. Protocol summary

| Research questions | RQ1. What research methods were used? RQ2. What application scenarios and datasets were used for in experiments and case studies? RQ3. Which challenges and solutions for engineering ML systems were raised by SE researchers? | |
|---|---|---|
| Search string | Population | "software engineering" |
| | Intervention | "machine learning" OR "deep learning" |
| Search strategy | - DB search: ACM, Google Scholar, IEEE, ScienceDirect, Springer, Wiley <br> - Backward and forward snowballing using Google Scholar <br> - Manual search: <br>    o International Workshop on Artificial Intelligence for Requirements Engineering (AIRE): 2014 – 2020 <br>    o International Workshop on Realizing Artificial Intelligence Synergies in Software Engineering (RAISE): 2012 – 2020 <br>    o IEEE/ACM International Conference on Software Engineering: 2017 – 2020 <br>    o ACM Joint Meeting on European Software Engineering Conference and Symposium on the Foundations of Software Engineering: 2017 – 2020 | |
| Inclusion and Exclusion criteria | Inclusion: <br> - The paper is written in English. <br> - The paper is published in a scholarly SE journal or conference/workshop/symposium proceedings. <br> - The paper explicitly involves at least one SE challenge of engineering ML systems. | |
| | Exclusion: <br> - The paper is published in a scholarly AI/ML/Data Science journal or conference/workshop/symposium proceedings. <br> - The paper focuses only on an ML-specific aspect (such as proposing a new ML algorithm, hyperparameter optimization, etc.). <br> - The paper is an editorial, issue introduction or secondary study (literature review, SMS, SLR). | |
| Study type | Primary studies | |

## 4.1 GOAL AND REVIEW QUESTIONS

The scope and goal of this study were formulated using the Goal-Question-Metric approach (Basili et al., 1994) as follows.

*Analyze the state-of-the-art in engineering machine learning systems*

*for the purpose of exploration and analysis*

*with respect to the reported challenges; proposed solutions; the intensity of the research in the area; the research methods*

*from the point of view of software engineering researchers*

*in the context of software engineering.*

As Kitchenham et al. (2015) pointed out, research questions (RQs) must embody secondary studies' goals. Accordingly, the purpose of this study can be broken down into the following three main RQs.

RQ1. What research methods were used?

RQ2. What application scenarios and datasets were used for in experiments and case studies?

RQ3. Which challenges and solutions for engineering ML systems were raised by SE researchers?

## 4.2 PRIMARY STUDY SELECTION

Figure 3 displays the primary study selection process used in this study.

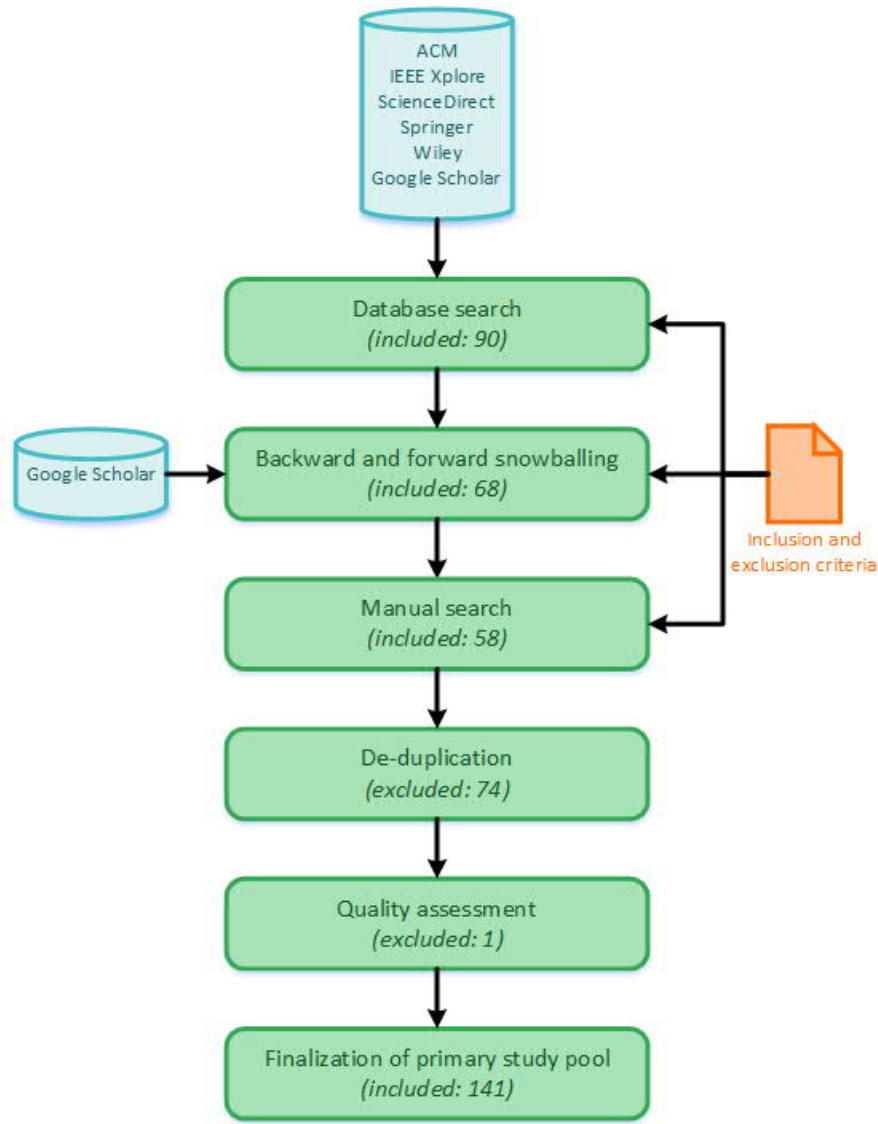

Figure 3. The primary study selection process

### 4.2.1 Inclusion and exclusion criteria

To identify the relevant primary studies, I defined the inclusion and exclusion criteria listed in Table 3. To present a SE perspective and keep the size of the final primary study pool manageable, I only considered papers published in the SE journals and conference/workshop/symposium proceedings, as others have done (Ferreira et al., 2019). I searched for the papers that involve at least one SE challenge of engineering ML systems. This assessment was based on two criteria: (1) The paper should be about an ML system or component; (2) The paper should mention at least one challenge on requirements engineering, design, software development and tools, testing and quality, maintenance and configuration management, software engineering process and management, or organizational aspects. Some examples related to testing are as follows: (1) "*Existing testing methodologies always fail to include rare inputs in the testing dataset and exhibit low neuron coverage.*" (Guo et al., 2018); (2) "*Deep neural networks lack an explicit control-flow structure, making it impossible to apply to them traditional software testing criteria such as code coverage.*" (Sekhon and Fleming, 2019); (3) "*Unlike software bugs, model bugs cannot be easily fixed by directly modifying models. Existing solutions work by providing additional training inputs. However, they have limited effectiveness due to the lack of understanding of model misbehaviors and hence the incapability of selecting proper inputs. Inspired by software debugging, we propose a novel model debugging technique that works by first conducting model state differential analysis to identify the internal features of the model that are responsible for model bugs and then performing training input selection that is similar to program input selection in regression testing.*" (Ma et al., 2018c). A challenge reported by Nguyen-Duc et al. (2020) in requirements engineering area is "*However, it is common that business opportunities involving with AI systems are not validated and there is lack of business-driven metrics that guide the development of AI systems.*" An example of a design-related obstacle is "*When adopted by a software system, AI influences and significantly changes its architecture due to its complexity, as well as due to a need to adjust the existing system to use AI.*" (Jahić and Roitsch, 2020). Reimann and Kniesel-Wünsche (2020) addressed a challenge in development: "*Development of machine learning (ML) applications is hard. Producing successful applications requires, among others, being deeply familiar with a variety of complex and quickly evolving application programming interfaces (APIs).*" Wu et al. (2020) mentioned an issue related to maintenance: "*Since DL software adopts the data-driven development paradigm, it is still not clear whether and to what extent the clone analysis techniques of traditional software could be adapted to DL software. In this paper, we initiate the first step on the clone analysis of DL software at three different levels, i.e., source code level, model structural level, and input/output (I/O)-semantic level, which would be a key in DL software management, maintenance and evolution.*" An instance of process-related difficulty is "*The increasing reliance on applications with machine learning (ML) components calls for mature engineering techniques that ensure these are built in a robust and future-proof manner.*" (Serban et al., 2020).

Since I limited my search space to SE venues, I did not face with papers addressing the internal details of ML (such as ML/DL algorithms, learning theory, optimization, or probabilistic inference) without a context of engineering a software-intensive system.

I did not include secondary studies (literature review, SMS, SLR) as well since they usually do not present any new finding. They analyze a set of primary studies and aggregate the results from these in order to provide stronger forms of evidence about a particular phenomenon (Kitchenham et al., 2015).

Table 3. Inclusion and exclusion criteria

| Inclusion criteria | Exclusion criteria |
| --- | --- |
| The paper is written in English. | |
| The paper is published in a scholarly SE journal or conference/workshop/symposium proceedings. | The paper is published in a scholarly AI/ML/Data Science journal or conference/workshop/symposium proceedings. |
| The paper explicitly involves at least one SE challenge of engineering ML systems. Based on the classification in Wan et al. (2019), which was derived from SWEBOK (Bourque and Fairley, 2014), an SE challenge can be in one of the following knowledge areas: Requirements Engineering, Design, Software Development and Tools, Testing and Quality, Maintenance and Configuration Management, Software Engineering Process and Management, Organizational Aspects | The paper focuses only on an ML-specific aspect, such as ML/DL algorithms, learning theory, optimization, probabilistic inference, etc. |
| | The paper is an editorial, issue introduction, or secondary study (literature review, SMS, SLR). |

### 4.2.2 Database search

I started by applying the database (DB) search method to identify relevant primary studies. I used five widely used online databases, i.e., ACM, IEEE Xplore, ScienceDirect, Springer, and Wiley. Besides, I used Google Scholar to enrich the pool of candidate primary studies. I used two search strings to query online databases: Query 1: "software engineering" AND "machine learning"; Query 2: "software engineering" AND "deep learning." I used general terms for the search to have high recall and relatively lower precision. Although this required more effort for screening, I obtained a broader initial set of papers and substantially decreased the possibility of missing some relevant studies.

I searched each of the six online databases using the defined search strings in January 2020 and March 2021. In January 2020, I applied a filter for publication date and included the studies that were published until the end of 2019. In March 2021, I searched the databases for the papers published only in year 2020. Table 4 shows the number of results obtained via two search strings.

Table 4. DB search results

|  | Search in January 2020 | | Search in March 2021 | |
| --- | --- | --- | --- | --- |
| Database | Query 1 | Query 2 | Query 1 | Query 2 |
| ACM | 6,889 | 782 | 896 | 415 |
| IEEE Xplore | 1,817 | 485 | 489 | 417 |
| ScienceDirect | 40 | 5 | 17 | 5 |
| Springer | 22,931 | 3,178 | 933 | 294 |
| Wiley | 165 | 176 | 314 | 104 |
| Google Scholar | ~148,000 | ~19,800 | ~18,700 | ~12,800 |

Due to the general search terms used, the number of results was high for manual primary study selection for ACM, IEEE Xplore, Springer, and Google Scholar databases. Therefore, as others have done (Adams et al., 2016; Garousi and Mäntylä, 2016; Garousi et al., 2018; Godin et al., 2015), I had to restrict the search space by assuming a "search saturation effect." Thus, I checked the first 200 results of each query in each database. I only continued further when the results between 190 and 200 still looked relevant.

I applied the inclusion and exclusion criteria to the papers by reading the title, keywords, and abstract. As the result, I obtained 90 primary studies in total from two searches conducted in January 2020 and March 2021.

### 4.2.3 Backward and forward snowballing

To ensure the inclusion of relevant primary studies as much as possible, I conducted backward and forward snowballing, as recommended by systematic review guidelines (Wohlin, 2014). For backward snowballing, I applied the inclusion and exclusion criteria to each primary study's reference list found via database search. For forward snowballing, I checked the citations listed on Google Scholar to each primary study against the inclusion and exclusion criteria. Snowballing provided 68 primary studies.

### 4.2.4 Manual search

To enrich the primary study pool, I conducted a manual search in two top SE conference proceedings (ICSE and ESEC/FSE) for the years between 2017 and 2020. In addition, I investigated all papers published in AIRE and RAISE, since these workshops directly target the intersection of AI and SE. After applying the inclusion and exclusion criteria to the papers obtained via manual search, I discovered 58 primary studies.

### 4.2.5 De-duplication

I manually entered the metadata of primary studies (e.g., title, abstract, keywords, publication year, venue, etc.) into a spreadsheet. I identified and removed 74 duplicate studies from the list. I scanned the list to find possible different versions of a paper by the same author(s). If most of the content of a paper is repeated in another paper of the author(s), I included the extended version and excluded the shorter version.

### 4.2.6 Quality assessment

To assess the quality of primary studies, I used the quality assessment criteria (Table 5) proposed by Mohanani et al. (2018). I divided the included studies into empirical (127 papers) and non-empirical (15 papers), as suggested by Mohanani et al. (2018). Empirical studies are the ones that analyze primary data; non-empirical studies include both opinion papers and conceptual research (Mohanani et al., 2018).

For each criterion, I scored the primary studies using a 2-point Likert scale (yes = 1, no = 0). To maintain a high-quality input of primary studies for this SLR, I decided to exclude the papers with a score lower than six points out of 12 for empirical studies and three points out of six for non-empirical studies. I excluded one study with a score under the threshold.

Table 5. Quality assessment checklist (Mohanani et al., 2018)

| Quality criteria | Empirical | Non-empirical |
|---|---|---|
| Was a motivation for the study provided? | X | X |
| Was the aim (e.g., objectives, research goal, focus) reported? | X | X |
| Was the study's context (i.e., knowledge areas) mentioned? | X | X |
| Does the paper position itself within the existing literature? | X | X |
| Is relevance (to industry or academia) discussed? | X | X |
| Were the findings or conclusions reported? | X | X |
| Was the research design or method described? | X | |
| Was the sample or sampling strategy described? | X | |
| Was the data collection method(s) reported? | X | |
| Was the data analysis method(s) reported? | X | |
| Were limitations or threats to validity described? | X | |
| Was the relationship between researchers and participants mentioned? | X | |

### 4.2.7 Finalization of primary study pool

After excluding one study (found via DB search), the final primary study pool included 141 individual results. As shown in Figure 4, the DB search yielded 90 results, snowballing 68, and manual search 58. There were 77 studies found via only one search method (DB search: 40, Snowballing: 31, Manual search: 6) and 64 studies via more than search method.

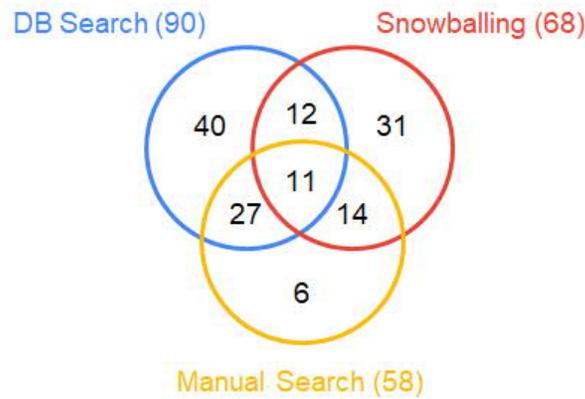

Figure 4. Number of primary studies per search method

I recorded all the metadata of the primary studies into the spreadsheet and checked its completeness. The primary study pool has been finalized with this step and became ready for data extraction.

### 4.3 DATA EXTRACTION

After selecting the primary studies, I started with the data extraction phase. I formed an initial data extraction form (Table 6) based on my RQs. The first six columns (from paper title to affiliation types of study authors) constitute the metadata of the papers. I derived affiliation type simply from affiliation data. I used the list of research methods proposed in Molléri et al. (2019) to identify the primary studies' methods. I used the knowledge areas presented in Wan et al. (2019), which was derived from SWEBOK (Bourque and Fairley, 2014) to classify the challenges and solutions. One hundred twenty one out of 141 primary studies address only one knowledge area. For these studies, the relevant knowledge area is generally mentioned in the title, abstract, or keywords. For the rest of 20 primary studies, I extracted the challenges and solutions by reading the papers and associated them with the knowledge areas. For all primary studies, I recorded the challenges and solutions to the spreadsheet without making any changes as written in the primary studies.

Table 6. Data extraction form

| Field | Categories | Relevant RQ |
|---|---|---|
| Paper title | Free text | - |
| Abstract | Free text | - |
| Keywords | Free text | - |
| Publication year | Number | Demographics |
| Venue (Journal/Conference) | Free text | Demographics |
| Affiliation types of study authors | University, Industry, Collaboration | Demographics |
| SE knowledge area | Requirements Engineering, Design, Software Development and Tools, Testing and Quality, Maintenance and Configuration Management, Software Engineering Process and Management, Organizational Aspects | RQ1, RQ2, RQ3 |
| Research method | Experiment, Interview, Survey, Thematic Analysis, Case Study, Statistical Analysis, Opinion/No Research Method | RQ1 |
| Application scenario | Analysis of ML Framework/Library, Autonomous driving, Classification (except image), Image classification, Mountain car | RQ2 |

| | problem, NLP & Machine translation, No application scenario, Object detection, Prediction of a continuous value, Ranking | |
| --- | --- | --- |
| Traditional ML/DL Algorithm(s) | Traditional ML, DL, Traditional ML & DL, Not mentioned | RQ2 |
| Dataset(s) | Free text | RQ2 |
| Challenge(s) | Free text | RQ3 |
| Solution(s) | Free text | RQ3 |

### 4.4 DATA SYNTHESIS AND REPORTING

Since I managed to categorize the extracted data for most of the RQs, the data extraction phase yielded a set of quantitative data to be synthesized. I reported the frequencies and percentages of each identified category to answer the RQs.

The only RQ that required qualitative analysis was RQ3, that is, the challenges and proposed solutions. I conducted open coding (Miles and Huberman, 1994) to analyze the challenges and solutions. A code symbolically assigns a summative or evocative attribute for a portion of qualitative data (Miles and Huberman, 1994). I conducted open coding in cycles. In the first cycle, I tried to identify any emerging patterns of similarity or contradiction. In the second cycle, I collapsed and expanded codes to understand any patterns. After I extracted the main themes and codes, I revised the codes assigned to each challenge/solution and reported the results.

## 5 RESULTS

In order to obtain an overall impression on the paper pool's content and display which areas the studies concentrate on, I created a bar chart of the most frequently used 50 terms in the primary studies' abstracts (Figure 5). I used unigrams, bigrams, and trigrams when calculating frequencies. Besides, I used lemmatization and removed stopwords when constructing the list of unigrams. Not surprisingly, "system," "software," "application," "DL software," "ML system," and "program" are among the most frequently used terms to refer to traditional software or ML systems. We can observe many terms that represent the population of this study, i.e., "software engineering": "software engineering," "development," "developer," "testing," "test," "bug," "quality," "coverage," "criterion," "requirement," "performance," "robustness," "safety," "analysis," and "failure." Testing" and "test" terms are the sixth and tenth most frequently used terms showing researchers' high interest in testing ML systems, as addressed in the results. There are also many terms representing the intervention of this study, i.e., "machine learning" OR "deep learning": "Deep Learning," "Machine Learning," "learning," "model," "Deep Neural Network," "Artificial Intelligence," "training," "DL model," "neural network," and "ML model." Finally yet importantly, "challenge" and "problem" are among the most used terms by the authors to refer to the difficulties they face.

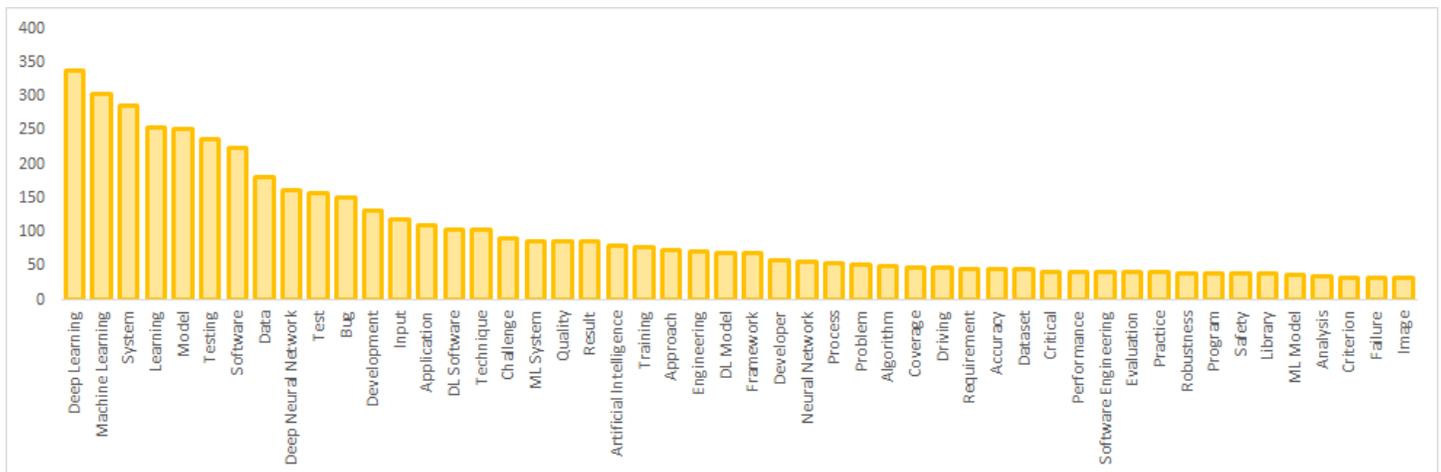

Figure 5. The most frequently used 50 terms in abstracts

In the rest of this section, I address each research question from RQ1 to RQ3.

## 5.1 WHAT RESEARCH METHODS WERE USED?

Figure 6 summarizes the research methods used to investigate SE challenges of engineering ML systems. Most of the empirical studies employed experiments. Besides experiments, the researchers interviewed and surveyed experts to learn about the challenges they face in engineering ML systems. Thematic analysis, case study, and statistical analysis are the other approaches used by the researchers. Fifteen papers either provide the author's opinions or did not report enough details about the research method. Two studies (Amershi et al., 2019; Wan et al., 2019) combined interview and survey methods. Two other studies (Humbatova et al., 2020; Zhang et al., 2020a) used thematic analysis and interview methods together. Alshangiti et al. (2019) performed thematic and statistical analysis together. Pham et al. (2020) conducted experiments to study the variance of DL systems and a survey to understand the awareness of this variance among researchers and practitioners. Due to these three primary studies using multiple research methods, the total number of primary studies equals 147 in Figure 6.

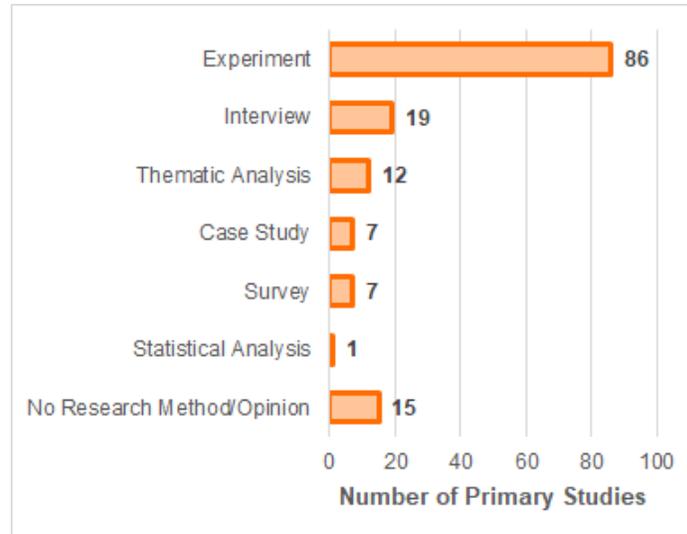

Figure 6. Research methods used

Figure 7 displays the research methods used in the primary studies split across the SE knowledge areas addressed in the papers. The dominant research method, i.e., experiment, is primarily used in studies addressing testing and quality. Researchers also conducted experiments on development & tools and maintenance & configuration management knowledge areas. As expected, experts mentioned many challenges covering all aspects of engineering ML systems in interviews and surveys. Case studies enabled researchers to collect data on activities related to ML systems in the industry. As a result of thematic and statistical analysis, the researchers have identified some difficulties in development & tools, test & quality, and organizational topics.

| SE Knowledge Area / Research Method | Requirements Engineering | Design | Software Development & Tools | Testing & Quality | Maintenance & Configuration Management | Software Eng. Process & Management | Organizational Aspects |
|---|---|---|---|---|---|---|---|
| Experiment | 2 | | 14 | 69 | 4 | 1 | |
| Interview | 6 | 5 | 6 | 4 | 4 | 9 | 3 |
| Thematic Analysis | | | 6 | 7 | 1 | | 1 |
| Case Study | 2 | | 2 | 2 | 4 | 2 | 1 |
| Survey | 5 | 2 | 4 | 3 | 2 | 3 | 3 |
| Statistical Analysis | | | 1 | | | | 1 |
| Opinion/No Research Method | 5 | 2 | 3 | 6 | 3 | 2 | 1 |

Figure 7. Research methods used per SE knowledge area

**5.2 WHAT APPLICATION SCENARIOS AND DATASETS WERE USED FOR IN EXPERIMENTS AND CASE STUDIES?**

The studies that used an experiment or a case study as a research method involved one or more application scenarios to observe SE challenges and validate a solution. Figure 8 shows the distribution of application scenarios used in 93 of the primary studies, the research method of which is either an experiment or a case study. Since one study may involve more than one research method and may cover more than one SE knowledge area, the total of the numbers presented in Figure 8 exceeds 93. One case study (Nakamichi et al., 2020) addressing requirements engineering challenges did not report its application scenario(s).

Image classification is the most popular application scenario used in 57 primary studies. The reason behind this could be the availability of many public datasets, as listed in Table 7. Image classification involves assigning predefined labels to images based on objects they contain (Russakovsky et al., 2015). Related to image classification, but classified as a different type of task, object detection aims to produce a list of object categories present in an image along with an axis-aligned bounding box indicating the position and scale of each instance of each object category (Russakovsky et al., 2015). Three studies involved an ML system for object detection. Twenty studies included a classification scenario other than image classification. Some examples include sentiment classification (Du et al., 2020, Guo et al., 2019, Guo et al., 2020, Li et al., 2020), credit risk category prediction (Aggarwal et al., 2019, Barash et al., 2019), income category classification (Aggarwal et al., 2019), and disease prediction (Moreb et al., 2020).

The third most popular application scenario (used in 13 studies) is autonomous driving, following image and other types of classification scenarios. Autonomous driving includes many ML tasks, including image classification, object (like pedestrian) detection, predicting steering wheel angle (Wu et al., 2021). Eight studies in our primary study pool (Haq et al., 2020; Harel-Canada et al., 2020; Jahangirova and Tonella, 2020; Li et al., 2019b; Stocco and Tonella, 2020; Stocco et al., 2020; Tian et al., 2018; Zhang et al., 2018a) conducted an experiment for predicting steering wheel angle.

Five studies addressed testing and maintenance aspects of ML systems by analyzing ML frameworks or libraries involved in these systems. Liu et al. (2020b) explored technical debts in seven open-source DL frameworks and their impacts on the maintainability of DL systems. Tizpaz-Niari et al. (2020) discovered performance bugs in ML frameworks, such as the implementation of logistic regression in scikit-learn. Ahmed et al. (2020), Dutta et al. (2020), and Wang et al. (2020b) studied problems in DL/ML frameworks related to testing and quality.

Four studies involved an application scenario to predict a numerical value for an observation. Such scenarios fall under regression type of ML problems. Two examples of regression problems are predicting the price of a house and predicting the number of bugs in a source code (Google, 2020). Four studies mention ML systems predicting values for combined sewer overflow (Challa et al., 2020), stock price (Wang et al., 2020b), age (Gao et al., 2020a), real estate value and weather (Arpteg et al., 2018).

Three studies contained application scenarios of Natural Language Processing, two specifically machine translation. Gao et al. (2020b) proposed an estimation tool for GPU memory consumption of DL models to help developers in planning runtime resource requirements. One of their experiments involved BERT (Devlin et al., 2018) model over GLUE (General Language Understanding Evaluation) benchmark (Wang et al., 2018), with various batch sizes and sequence lengths. Sun and Zhou (2018) and Zheng et al. (2019) addressed the challenges of testing an ML system on machine translation.

Only one study (Murphy et al., 2007) is concerned with a ranking problem. Murphy et al. (2007) aim at ranking the probabilities of observing a failure in a device.

Mountain car problem (Moore, 1990) is based on an underpowered car that must drive up a steep hill. The car cannot simply accelerate up the steep slope since gravity is stronger than the car's engine. The solution requires a reinforcement-learning agent to leverage potential energy by learning on two continuous variables: position and velocity. Trujillo et al. (2020) used mountain car problem to investigate whether neuron coverage can be used for deep reinforcement learning systems.

| SE Knowledge Area / Application Scenario | Requirements Engineering | Design | Software Development & Tools | Testing & Quality | Maintenance & Configuration Management | Software Eng. Process & Management | Organizational Aspects |
|---|---|---|---|---|---|---|---|
| Image classification | | | 13 | 44 | 3 | | |
| Classification (except image) | 1 | | 4 | 17 | 2 | 3 | 1 |
| Autonomous driving | 1 | | 1 | 9 | 3 | 1 | |
| Analysis of ML Framework/Library | | | | 4 | 1 | | |
| Prediction of a continuous value | 1 | | 2 | 2 | 1 | 1 | 1 |
| NLP & Machine translation | | | 1 | 2 | | | |
| Object detection | | | 2 | 1 | 1 | 1 | |
| Ranking | | | | 1 | | | |
| Mountain car problem | | | | 1 | | | |
| No application scenario | 1 | | | | | | |

Figure 8. Application scenarios per SE knowledge area

I also classified the algorithms used in the experiments and case studies into two groups: traditional ML and DL algorithms. Among 93 papers, 65 papers (70%) used only DL algorithms. Nineteen papers used traditional ML algorithms, and four papers used both DL and traditional ML algorithms. Five papers did not report the algorithms used.

As briefly explained in Section 2.2, traditional ML algorithms entail a manual feature selection process, which may require domain expertise, statistical analysis of input data, and experiments for building models with different feature sets (Kelleher, 2019). For instance, to detect an object in an image, a data scientist would first need to extract features such as edges, blobs, and textured regions from the image (Zhou et al., 2018). Thereafter, these features can be fed to an ML algorithm to detect objects. On the other hand, DL algorithms are capable of learning more correct features compared to human-designed ones (Zhou et al., 2018). Hence, for scenarios where there is a big gap between the inputs (such as images) and the outputs (such as objects), DL models are able to perform much better than traditional ML models (Zhou et al., 2018). However, large labelled datasets are required to train DL models (Fink et al., 2020). In line with these findings, DL algorithms were used in all primary studies involving autonomous driving, continuous value prediction, object detection, and mountain car problem scenarios. In addition, nearly all of the image classification scenarios were also experimented using DL algorithms. Solving these complex problems utilizing DL models seem rational due to the availability of public labelled datasets. For the classification scenarios excluding image classification, the researchers mostly used traditional ML algorithms. The rationale behind this could be unavailability of sufficient labelled datasets. Another reason could be that traditional ML algorithms are suggested to be used for the problems where features are available (Zhou et al., 2018).

Table 7 shows the list of datasets used more than once in the experiments and case studies in 93 primary studies. Twenty-nine datasets were only used once in an experiment or case study. Since most of the studies are focused on image classification problems, most of the datasets in Table 7 are the datasets for image classification, such as MNIST (LeCun et al., 1998), CIFAR-10 (Krizhevsky and Hinton, 2009), ImageNet (Fei-Fei, 2010), Fashion-MNIST (Xiao et al., 2017), SVHN (Netzer et al., 2011), and IMDB-WIKI dataset. Twenty-four studies include custom datasets constructed by researchers (including synthetic datasets) or privately owned by a company. Eleven studies used Udacity self-driving car challenge dataset for experiments in autonomous driving domain for various tasks such as predicting steering wheel angle (Kim et al., 2019; Li et al., 2019b; Tian et al., 2018).

184 datasets were reported in 93 studies, which means most of the studies used two datasets on average. One reason may be the cost of conducting experiments using various datasets (Zhang et al., 2020c).

Table 7. Datasets used in experiments and case studies

| Dataset | Description | Web Site | # of Primary Studies |
|---|---|---|---|
| MNIST (LeCun et al., 1998) | Images of handwritten digits | http://yann.lecun.com/exdb/mnist | 45 |
| CIFAR-10 (Krizhevsky and Hinton, 2009) | Various color images in 10 classes | https://www.cs.toronto.edu/~kriz/cifar.html | 31 |
| ImageNet (Fei-Fei, 2010) | Visual recognition challenge dataset | http://www.image-net.org | 16 |
| Udacity self-driving car challenge | Udacity Self-Driving Car Challenge images | https://github.com/udacity/self-driving-car | 11 |
| UCI machine learning repository (Dua and Graff, 2019) | Collection of databases, domain theories, and data generators that are used by the ML community for the empirical analysis of ML algorithms | https://archive.ics.uci.edu/ml/index.php | 8 |
| Fashion-MNIST (Xiao et al., 2017) | MNIST-like dataset of fashion images | https://www.kaggle.com/zalando-research/fashionmnist | 7 |
| SVHN (Netzer et al., 2011) | A real-world image dataset of house numbers | http://ufldl.stanford.edu/housenumbers | 7 |
| Adult Census Income (Kohavi, 1996) | Income data from the 1994 Census bureau database with some attributes such as education, hours of work per week, etc. | https://www.kaggle.com/uciml/adult-census-income | 2 |
| IMDB Dataset | Information about films from Internet Movie Database | https://www.imdb.com/interfaces | 2 |
| IMDB-WIKI dataset | Human faces with gender, name, and age information | https://github.com/imdeepmind/processed-imdb-wiki-dataset | 2 |
| Custom | Private dataset constructed by researchers or private dataset of a company | - | 24 |

**5.3 WHICH CHALLENGES AND SOLUTIONS FOR ENGINEERING ML SYSTEMS WERE RAISED BY SE RESEARCHERS?**

This subsection presents the main SE challenges of engineering ML systems identified from the primary studies. The identified SE challenges are classified based on the main SE knowledge areas presented in (Wan et al., 2019). The proposed solutions are also presented in this subsection. Table 8 provides a consolidated list of the challenges with proposed solutions.

### 5.3.1 Requirements Engineering

Requirements engineering involves eliciting, analyzing, specifying, and validating requirements that represent a software system's intended purpose. Many stakeholders may play a role in the RE process with various backgrounds and possibly conflicting objectives. Although a satisfactory RE phase is not sufficient for project success, it is a necessary condition. RE process has numerous potential obstacles, such as incomplete, hidden, inconsistent, underspecified requirements, communication flaws (Fernández et al., 2017), even for traditional software development. The development of ML systems makes the RE process even more challenging (Chechik, 2019; Dalpiaz and Niu, 2020). Wan et al. (2019) interviewed professionals with experience in both traditional and ML system development and reported that nearly every interviewee made a strong statement about differences between the requirements of ML versus non-ML systems.

*Managing expectations of customers:* A potential gap between customers' expectations and the delivered system has been reported as a potential problem in software projects (Wiegers, 1996). Frequent customer engagement is an effective way of reducing this "expectation gap" (Wiegers, 1996). Adding ML capabilities to a software system brings an extra complexity that can increase this gap. Customers generally do not understand the difference between traditional software and ML system (Baier et al., 2019; Vogelsang and Borg, 2019). While it is always possible to develop traditional software that always meets customers' expectations (despite its challenges), this is not always possible for ML systems. Evaluation metrics for an ML model are crucial means of understanding customer expectations. On the other hand, customers face difficulties in selecting proper evaluation metrics and interpreting these for assessing the quality of an ML system (Vogelsang and Borg, 2019). For instance, customers may have to judge on whether 90% prediction accuracy on average is satisfying. Besides, they should assess the impact of incorrect predictions and decide on the accountable for such predictions (Baier et al., 2019). Another potential challenge is convincing customers about the value to be provided by ML capabilities (Kim et al., 2017; Nguyen-Duc et al., 2020).

Nevertheless, customers may ask for "perfect ML systems" and have unrealistic expectations (Fujii et al., 2020; Ishikawa and Yoshioka, 2019; Kim et al., 2017). It is important to inform customers about the benefits of ML systems even without being imperfect (Ishikawa and Yoshioka, 2019) and explain the possibility of change in the level of success of ML systems on production in time (Wan et al., 2019). Customers should be guided to think about possible alternative scenarios in which ML systems are not acting as expected and providing input on how to detect such scenarios on production and how the system should behave in such circumstances.

*Eliciting and analyzing requirements:* Requirements of ML systems are heavily dependent on available data; different data sets may lead to different needs (Rahimi et al., 2019; Wan et al., 2019). Requirements are much more uncertain (Belani et al., 2019; Chechik, 2019; Ishikawa and Yoshioka, 2019; Rahimi et al., 2019; Wan et al., 2019), and hence they are more difficult to elicit and analyze. Some requirements should be stated as hypotheses to be tested via experiments (Wan et al., 2019) to understand what is possible to deliver. In some cases, different ML model prototypes may be built to without concrete business requirements to capture evolving business opportunities (Nguyen-Duc et al., 2020). Conducting such experiments requires additional skills to deal with data (Amershi et al., 2019) while eliciting and analyzing requirements. Besides, new techniques are necessary for analyzing requirements regarding ML capabilities (Rahimi et al., 2019).

*Specifying requirements:* Extensions in the current practices of RE may be needed to specify requirements. For instance, Kaindl and Ferdigg (2020) raised the idea of extending Goal-oriented Requirements Engineering (GORE) to model dynamic goals of intelligence to be realized by an ML component.

Requirements to be fulfilled by ML components are generally specified using quantitative measures (such as accuracy, precision, recall, F measure) (Kim et al., 2017; Vogelsang and Borg, 2019; Wan et al., 2019) that are new for many stakeholders. It may also be difficult to map these measures to business objectives (Baier et al., 2019; de Souza Nascimento et al., 2019). While it is relatively easy to measure a metric such as accuracy, if that metric decouples from business objectives, customers think that the ML system does not deliver what it should provide (Byrne, 2017). In cases where a business or heuristic metric already exists, this metric can be used as a baseline to develop an ML model (Nguyen-Duc et al., 2020). For instance, the output of an existing rule-based customer credit rating algorithm can form a baseline for the performance of an ML component predicting customers' credit scores (Nguyen-Duc et al., 2020). In cases where a business or heuristic metric is not in place, a metric may be defined through building ML model prototypes and experimentation (Nguyen-Duc et al., 2020).

Besides specifying each requirement correctly, all specified requirements should exhibit qualities such as completeness and consistency (Wiegers and Beatty, 2013). Some researchers present recommendations and examples for specifying requirements correctly (Alexander and Stevens, 2002; Kovitz, 1998; Wiegers, 2005). In parallel with the rise of ML systems, new quality metrics to ensure completeness, consistency, etc. should be defined (Alvarez-Rodríguez et al., 2019) and contemporary techniques for specifying types of requirements, such as performance (Kawamoto, 2019), robustness (Kawamoto, 2019), and fairness (Horkoff, 2019; Kawamoto, 2019), should be developed. Sharma and Wehrheim (2020b) provided a specification language for formulating fairness requirements on ML models.

*Dealing with new types of quality attributes:* Quality attributes (also named as non-functional requirements) have been defined and well understood for traditional software (IIBA, 2015; Wiegers and Beatty, 2013). Stakeholders need to elicit requirements on new types of quality attributes, such as explainability (Nguyen-Duc et al., 2020), fairness (Ahuja et al., 2020; Bellamy et al., 2019; Horkoff, 2019), and freshness (Zinkevich, 2017). Nakamichi et al. (2020) proposed a model and a measurement method to be used for agreeing with customers regarding quality requirements when enhancing enterprise systems with ML capabilities.

Explainability of an ML system measures how a human observer can understand the reasons behind a decision (such as a prediction) made by that ML system (Dam et al., 2018). Explainability is quite essential in building trust between an ML system and its users (Nguyen-Duc et al., 2020; Ribeiro et al., 2016). Therefore, requirements on explainability should be elicited. Users may need to have explanations on a failure (Baier et al., 2019; Ishikawa and Yoshioka, 2019), a single decision (Vogelsang and Borg, 2019), or a model used in an ML system (Vogelsang and Borg, 2019). In some domains, such as software analytics, making models explainable to users is as important as making accurate decisions (Dam et al., 2018). For instance, Jiarpakdee et al. (2020) argued that the lack of explainability of defect prediction models could hinder the adoption of these models in practice. Developers will not tend to use such models unless they are told why a piece of code is proposed to be defective (Jiarpakdee et al., 2020). An explainable defect prediction model should be able to present the important factors that led to a prediction, such as high number of class and method declaration lines, high number of distinct developers contributed to a file, and low proportion of code ownership (Tantithamthavorn et al., 2020). Since explainability is an important requirement, Defense Advanced Research Projects Agency (DARPA) has started a program named "Explainable AI (XAI)," which aimed at developing more explainable models (Meacham et al., 2019; Nushi et al., 2018).

As ML systems are increasingly being used in domains sensitive to discrimination (such as education, employment, healthcare) due to protected characteristics (such as gender and race), it becomes crucial to avoid making decisions biased by protected characteristics. Various definitions of this new type of quality attribute, fairness, have been presented (Corbett-Davies and Goel, 2018). It is essential to elicit requirements on fairness and identify protected characteristics (Vogelsang and Borg, 2019). For example, a software system assessing loan applications can be attributed to be fair with respect to age and race, if it approves or rejects the loan for all pairs of individuals having identical names, income, savings, employment status and requested loan amount but different age or race (Galhotra et al., 2017). If protected characteristics are directly identifiable, excluding them from the training data can fulfill the fairness requirement. On the other hand, some characteristics, which do not seem to be protected, may highly correlate with a protected characteristic (Kamishima et al., 2011). For instance, address data, which may be correlated with race, may be used to train a model for credit scoring. It is crucial to find out such red-lining effects (Calders and Verwer, 2010) or indirect discrimination (Pedreshi et al., 2008).

Freshness requirement (Zinkevich, 2017) refers to the conditions for updating ML models. ML models' performance may change over time due to some reasons, such as changing patterns in input data (data drift). It is crucial to determine the tolerance to performance degradation and the conditions that will trigger ML model updates. Freshness requirements can be defined as a period (daily, weekly, monthly update of ML models). A set of rules can also be defined for ML model updates. In such a case, the performance of ML models should be monitored, and required actions should be taken. Analyzing freshness requirements is essential to get prepared for ensuring a certain level of satisfaction for users continuously.

Robustness requirements are vital, especially for safety-critical systems, such as those used in autonomous driving. Therefore, Hu et al. (2020) proposed a method for specifying requirements on robustness. Human performance in safety-critical tasks can be obtained and used as a baseline to specify robustness requirements (Hu et al., 2020). For example, an ML component should detect a pedestrian even if a small amount of noise is added to the image or it is rotated by a limited amount (Hu et al., 2020).

There are also efforts to identify possible risks and negative impacts of ML systems (Floridi, 2019). For instance, "Ethics guidelines for trustworthy AI" define the ethical principles that an ML system should adhere to (European Commission, 2019). These principles should be continuously considered while engineering ML systems to ensure Trustworthy AI (European Commission, 2019).

*Dealing with new types of conflicts between requirements:* It is crucial to find the right balance of quality attributes for delivering successful software systems (Boehm and In, 1996). To be able to do this, conflicts among desired quality attributes should be identified (Boehm and In, 1996). Possible conflicts for traditional software have been identified and documented a long time ago (Boehm and In, 1996; Bosch, 2000; Kotonya and Sommerville, 1998; Wiegers and Beatty, 2013). ML systems pose new challenges for conflict resolution (Horkoff, 2019). Recently, researchers started to explore conflicts between specific quality attributes, such as the trade-off between fairness and performance or accuracy (Kamishima et al., 2011).

*Dealing with changing emphasis on different requirements:* Stakeholders often focus on functional and non-functional requirements during the RE process. On the other hand, data requirements analysis is an integral part of RE and focuses on information needs, providing a set of procedures for identifying, analyzing, and validating data requirements (Loshin, 2011). Stakeholders aim at understanding data entities to be used in traditional software. Description, data type, length, and value range of data elements are analyzed to understand the data entities' structure (Wiegers and Beatty, 2013). To develop ML components, stakeholders also have to analyze what data instances of these data entities are present and what they can get out of these (Vogelsang and Borg, 2019). In addition, data quality is an important aspect to be considered during RE since it directly affects the performance of an ML model (Challa et al., 2020).

Business rules include government regulations, corporate policies, and industry standards and form the rules that a software system must comply with (Wiegers and Beatty, 2013). Business rules are not software requirements themselves, but they sometimes originate new requirements or apply constraints on requirements. With the digitalization of high-volume data, especially personal data, data privacy regulations have been put in place. For instance, the EU's new data privacy rules, the General Data Protection Regulation (GDPR), significantly impact the development and use of ML systems in Europe. Besides, companies also form their data privacy policies (Tata et al., 2017). For instance, Google's data privacy policy only allows working with anonymized and aggregated summary statistics, making exploratory analysis extremely difficult (Tata et al., 2017). Data privacy rules should be considered carefully while analyzing requirements (Baier et al., 2019; Belani et al., 2019; Vogelsang and Borg, 2019). Although this should also be done for traditional software, data privacy rules can significantly affect what an ML system can perform. A team can allocate some effort to anonymize data (Fung et al., 2007; Muntés-Mulero and Nin, 2009) by assessing its cost and benefit (Li and Li, 2009). New approaches are being developed to provide a better understanding of various privacy-related requirements for improving privacy policy enforcement when developing systems integrated with social networks (Caramujo and da Silva, 2015). In projects where data are obtained from outside sources, ensuring data privacy and compliance to GDPR may involve cross-organizational collaboration (Nguyen-Duc et al., 2020).

### 5.3.2 Design

Design activities aim at producing logical descriptions of how a system will work (Larman, 2005). These descriptions can be at a high-level, i.e., architectural design and lower levels. Many best practices for designing traditional software have been presented in the literature. Design patterns for low-level design (Gamma et al., 1995), architectural patterns and styles (Fowler, 2002; Microsoft, 2009) for high-level design, and some other design principles, such as GRASP (Gamma et al., 1995) and SOLID (Martin, 2002), are already in place to be used by software engineers. While all of these are valid for engineering many software systems components, developing ML capabilities requires some changes and additions to these best practices (Tripakis, 2018; Washizaki et al., 2019b).

There is some literature on designing the internals of ML components, such as design patterns for convolutional neural networks (Smith and Topin, 2016) and architecture for computer vision (Szegedy et al., 2016). As I pointed out before (in Section 4.1), I focus on design from a SE perspective.

*Designing for monitoring performance degradation on production:* A degradation in how well an ML system meets users' expectations should be expected on production in time (Wan et al., 2019). Known as "concept drift," underlying training data distribution may change, and these changes make the model built on old data inconsistent with the new data (Tsymbal, 2004). This change poses two crucial challenges: detecting when concept drift occurs and keeping the models up-to-date (Maimon and Rokach, 2010).

Yokoyama (2019) proposes an architectural pattern for locating performance problems and rolling back in case of a failure on production. Defense Advanced Research Projects Agency (DARPA) started a program named "Assured Autonomy" (DARPA, 2020), who's one of the aims is developing design toolchains to achieve continual assurance. Continual assurance is defined as an assurance of the safety and functional correctness of a system provided provisionally at design time and continually monitored, updated, and evaluated on production as the system and its environment evolves.

*Using new solution patterns for solving problems:* Engineers do not have a mature set of solution patterns for solving the issues posed by ML. Sculley et al. (2014) reflected upon their own experiences in a position paper recounting their views on the difficulties of developing ML systems. They reported one of the complexities posed by ML systems as the "change anything, changes everything" principle, which refers to the dependencies among all the parts of an ML system, i.e., application code, "glue code," ML libraries, and external data (Sculley et al., 2014). This principle prevents the use of standard techniques (such as abstraction and information hiding) for reducing coupling (Wan et al., 2019), which is a fundamental design principle for traditional software (Sculley et al., 2014). Not being able to isolate the impact of a specific change anywhere in the system of dependencies could be why our practitioners so often resorted to ad hoc practices like trial and error and rules of thumb.

Researchers started to publish the lessons learned from failures as anti-patterns (Sculley et al., 2014). Belani et al. (2019) state avoiding anti-patterns as a design challenge and warns engineers for staying away from producing glue code, pipeline jungles, dead experimental code paths, and configuration debt. Washizaki et al. (2019b) state that ad-hoc solutions are being used for solving everyday problems in developing ML systems. Washizaki et al. (2019b) list some of the patterns and anti-patterns for designing ML systems by stating that these patterns do not cover all the common problems. There is a need for a catalog of patterns dedicated to ML-specific problems (Washizaki et al., 2019b). A recent study (Washizaki et al., 2019a) has surveyed architectural and design (anti-)patterns for ML systems to bridge the gap between traditional and ML systems concerning design.

Jahić and Roitsch (2020) pointed out that organizations face challenges in identifying required changes in software architectures when they enhance their systems with ML capabilities. Some organizations are reusing internal design patterns used and validated in their past projects (Washizaki et al., 2020). Washizaki et al. (2020) identified 15 design patterns for ML systems and plan to expand the list via further empirical studies. John et al. (2020a; 2020b) proposed a framework involving five architectural alternatives ranging from centralized cloud to fully decentralized edge architectures for designing ML systems. Trade-offs among device cost, ML model performance, and data privacy requirements guide the selection of an architectural alternative (John et al., 2020b). Scheerer et al. (2020) proposed classes of dependability in which ML systems may be categorized. Such a categorization can lead to a proper architecture to ensure dependability of an ML system, i.e., its ability to deliver service that can justifiably be trusted (Avizienis et al., 2001).

*Dealing with high-volume data:* Distributed architectural patterns are widely used in designing ML systems to cope with high-volume data (Wan et al., 2019), which leads to additional complexity in architectural and detailed design (Wan et al., 2019). While Anderson (2015) emphasizes the paramount significance of data modeling in engineering data-intensive systems that are scalable, robust, and efficient, de Souza Nascimento et al. (2019) state the difficulty of succeeding in doing this. Benton et al. (2020) proposes an architecture for an ML system that deals with high-volume data. This architecture includes a component for data federation to deal with structured, unstructured, and streaming data.

### 5.3.3 Software Development and Tools

While developers code the solution in traditional software components, they infer the solution using data and ML algorithms in ML components (Pruss, 2017). Therefore, developing ML systems necessitates rethinking current development practices, tools, and infrastructures (Pruss, 2017). ML model development practice differs from traditional software development due to its data dependency, uncertainty, and experimentation requirements. Organizations, which continuously develop and deploy ML systems, must have a proper process to support the highly iterative development, testing, and deployment of ML models (Sapp, 2017). Based on the final paper pool, most challenges in developing ML systems manifest in issues related to data; models; dependencies; infrastructure and tools; ML algorithms, techniques, and libraries; and reuse.

*Dealing with data:* Data preparation is a vital and inevitable group of activities for developing ML systems (Sapp, 2017). Discovering, accessing, collecting, cleaning, and transforming data is challenging and time-consuming (Amershi et al., 2019;

Baier et al., 2019; Correia et al., 2020; Fredriksson et al., 2020; Fujii et al., 2020; Hill et al., 2016; Kim et al., 2017; Lwakatare et al., 2019; Sankaran et al., 2017). There may be various types of data sources, such as transactional systems, data warehouses, data lakes, data meshes, and real-time data streams (Sato et al., 2019). Developers may build data pipelines to have data ready for model development. Developing data pipelines may involve dealing with structured and unstructured data (Alshangiti et al., 2019). In addition, combining data is not always straightforward (Kim et al., 2017; Lwakatare et al., 2020) since different systems can have other objectives for storing data, hence leading to data semantics heterogeneity and data integration problems. Data pipelines are essential artifacts, which should be version-controlled, tested, deployed, and maintained (Sato et al., 2019). A frequently seen type of bug in ML systems is a "data bug" (Islam, 2019). Handling exceptions in data is a serious problem in developing ML models (Zhang et al., 2020a). For instance, imbalanced datasets will most likely lead to inaccurate prediction models (He and Garcia, 2009). Therefore, it is important to prepare data for analysis, analyze the distribution of them using some metrics (Tantithamthavorn et al., 2018). Data verification tools may help engineers catch data bugs, such as improper format or encoding, missing data, etc. (Islam, 2019; Zhang et al., 2020a).

*Understanding ML algorithms, techniques, and libraries:* Once data is ready, developers use algorithms, techniques, and libraries to produce models (Sato et al., 2019). They face challenges in understanding these (Alshangiti et al., 2019; Zhang et al., 2019a; Zhang et al., 2020a), especially the ones related to DL (Alshangiti et al., 2019), and identify a proper model (Islam, 2019). Use of automated model and parameter recommendation tools (Islam, 2019), better documentation (Alshangiti et al., 2019), or having ML engineers in a team may help in overcoming this challenge. To provide more comprehensive knowledge at different level of expertise, Hashemi et al. (2020) automatically generated ML system documentation from StackOverflow Q&As.

*Dealing with models:* After having requirements and data ready, the team develops models. This development endeavor involves some sub-activities, such as feature engineering, model training, model evaluation, and model deployment (Amershi et al., 2019). Developing models may require many iterations and experimentation, including many feedback loops between sub-activities (Amershi et al., 2019; Pham et al., 2020). Even the model requirements may not be apparent before starting with model development and may be clarified after many experiment iterations. Developers work on extracting selecting informative features for models (Amershi et al., 2019; Hill et al., 2016; Kim et al., 2017). They may need to optimize a possible tradeoff between the best features and model complexity (Sapp, 2017). For DL models, feature engineering is performed implicitly during model training (Amershi et al., 2019). Nevertheless, DL models should be evaluated using various important metrics such as robustness and neuron coverage, besides the widely-used prediction accuracy metric (Tian et al., 2020). Tian et al. (2020) developed EvalDNN tool to assess a pre-trained model with respect to different metrics by writing a few lines of code. The tool supports multiple frameworks and metrics and can be used via an API (Tian et al., 2020). Ghamizi et al. (2020) developed a tool, called FeatureNET, for generating a large variety of DL models and performing experiments with these diverse models. FeatureNET aims to help engineers to evaluate more DL models with less effort and deploy better performing models (Ghamizi et al., 2020).

Developers usually reported that analyzing and understanding the structure and behavior of ML models (mainly neural networks) is very difficult (Amershi et al., 2019; Arpteg et al., 2018; Hill et al., 2016; Xie et al., 2019a; Zhang et al., 2019b). This difficulty may lead to problems in evaluating and debugging models. Researchers attempt to provide a solution for this challenge via visualization. DeepVisual is a visual programming tool for designing and developing DL systems (Xie et al., 2019a). DARVIZ tool supports a visual model-driven development environment (Sankaran et al., 2017). Another type of ML model analysis encompasses verification of a compressed model against an original model. Compression techniques are used for deploying trained models on computationally- and energy-constraint devices (Paulsen et al., 2020). Paulsen et al. (2020) proposed the NEURODIFF technique to verify compressed models and to enable developing robust and resource-efficient ML systems.

Several ML models are vulnerable to adversarial examples, which are only slightly different from correctly classified examples drawn from the training data distribution (Szegedy et al., 2014). Developing fault-tolerant ML systems is a vital objective, especially in safety-critical settings (Protzel et al., 1993). One approach for this is the identification of out-of-distribution (OoD) data, which are quite different from training data. Zhou et al. (2020b) found out that OoD-aware ML models are more robust according to their experiments. Wang et al. (2020a) proposed the Dissector approach for identifying adversarial inputs that are still within the model's handling capabilities, i.e., within-range (Wang et al., 2020a). This approach tries to uncover the maximum potential of an ML model and reduce the need for re-training (Wang et al., 2020a). Another approach for dealing with adversarial inputs and developing robust ML models is enriching training data via data augmentation. Gao et al. (2020a) and Yokoyama et al. (2020) utilized genetic algorithms to generate realistic transformations of training data points and increase the amount of training data. Their approaches skip unsuitable variants in order not to increase training time unless a more robust ML model is obtained (Gao et al., 2020a; Yokoyama et al., 2020).

Another essential concern in developing ML models is fairness (Corbett-Davies and Goel, 2018). Biased decisions breaking ethical rules should be removed from ML models. Chakraborty et al. (2020a) proposed a method named Fairway, which combines pre-processing and in-processing approaches to remove ethical bias from training data and trained model.

Model deployment involves some challenges (Chen et al., 2020b). In some cases, a lot of manual work may be needed for deployment (Baier et al., 2019). If deployments are widespread, some automated deployment mechanism may be considered (Amershi et al., 2019). Moreover, developers may have to scale up models to deployment architectures using code parallelization (Anderson, 2015; Baier et al., 2019).

*Dealing with dependencies:* Building highly cohesive and loosely coupled components is a powerful way of managing complexity in traditional software development (Larman, 2005). It is possible to encapsulate behavior and associated data in modular components. On the other hand, ML components generally depend on external data (Sculley et al., 2015). Although there are some tools for exploring such data dependencies (McMahan et al., 2013), these are not as capable and common as the static analysis tools for traditional software. Another type of dependency is named "undeclared consumers" (Sculley et al., 2015). The outputs of an ML model may be used by many other components implicitly. In such cases, the changes in the ML model may negatively affect the dependent components (Belani et al., 2019; Hill et al., 2016; Lwakatare et al., 2019). Therefore, it is vital to consider the potential negative impacts of such dependencies during development.

*Reusing models:* Reuse in traditional software development is vital to increase productivity and quality and decrease development time and cost (Ravichandran and Rothenberger, 2003). While reusing ML libraries is very common, effective, and efficient, reusing ML models in different domains or systems is not straightforward (Alvarez-Rodríguez et al., 2019; Amershi et al., 2019; Baier et al., 2019; Guo et al., 2019; Sankaran et al., 2017). Adapting the implementation of a neural network for a different task (Zhang et al., 2019a) or transferring a built solution to another domain (Baier et al., 2019) remains a challenge. Using training data from other domains is another reuse strategy in ML. Transfer learning refers to using training data from another domain to develop an ML model in a domain of interest (Pan and Yang, 2009; Weiss et al., 2016). Transfer learning is mostly preferred when it is not possible or too costly to obtain training data. Although transfer learning is an attractive topic in ML, it has not been addressed by the studies in the final paper pool.

*Dealing with the development environment, tools, and infrastructure:* The presence of diverse and incompatible programming and data tools is a severe concern for engineering ML systems (Kim et al., 2017). A heterogeneous tool set challenges engineers in many tasks, such as data processing, environment setup, and model deployment (Alshangiti et al., 2019). This heterogeneity causes potential discrepancies among development, quality assurance, and production environments (Zhang et al., 2020a). Although Docker images with all desired software pre-installed may help to some extent (Zhang et al., 2020a), researchers recognized the importance of having a proper ML system development infrastructure (Baier et al., 2019), preferably by integrating ML development support into traditional software development infrastructure (Alvarez-Rodríguez et al., 2019; Amershi et al., 2019). On the other hand, such integration may be challenging due to the different characteristics of ML and traditional software components (Amershi et al., 2019). There are some attempts to bring ML system development capabilities to existing Integrated Development Environments, such as Azure ML for Visual Studio Code (Amershi et al., 2019). Due to the highly iterative nature of ML model development, the infrastructure should enable experiment management (Arpteg et al., 2018; Gharibi et al., 2019; Lwakatare et al., 2019). ModelKB is a tool prototype to manage experiments involving DL models (Gharibi et al., 2019). The Simple-ML project aims at enabling engineers without significant ML expertise to efficiently create ML systems (Reimann and Kniesel-Wünsche, 2020). Some components are being developed to handle data and ML workflows to develop explainable, efficient, and scalable ML systems. For instance, a unified ML API generalize the functionality of scikit-learn, Keras, and PyTorch libraries along with an enhanced documentation (Reimann and Kniesel-Wünsche, 2020).

Tripakis (2018) and Menzies (2019) points out the heterogeneity of DL platforms that leads to interoperability problems. The model-driven development approach may increase the level of abstraction and facilitate the intuitive development of DL models in a platform-agnostic fashion (Sankaran et al., 2017).

Working with a large volume of data that requires a distributed system brings additional complexity to development (Arpteg et al., 2018). Distributed systems require further knowledge and experience and additional cost and management of associated hardware and software (Arpteg et al., 2018). For instance, improper decision making during ML model development may cause an overflow error in GPU memory. Gao et al. (2020b) developed a tool, named DNNMem, to estimate GPU memory consumption of DL models. This tool can be used to prevent failures due to memory consumption and hence waste of computing resources, as well as reduction in developer productivity.

### 5.3.4 Testing and Quality

Testing ML systems poses challenges that arise from the fundamentally different nature and development of ML systems (as briefly explained in Section 2.2), compared to relatively more deterministic traditional software systems (Zhang et al., 2020c). In 2007, Murphy et al. (2007) mentioned the idea of testing ML systems and classified such systems as "non-testable". Although non-testability is attributed to a lack of reliable test oracle (Weyuker, 1982), there are several challenges in testing ML systems. I classified these challenges into eight categories.

*Designing test cases:* Test case design refers to identifying test inputs and outputs that allow tests to be executed. Well-designed test cases can help in finding many types of defects in ML systems (Humbatova et al., 2020). Many researchers mentioned the difficulty of test case generation for ML systems (Ahmed et al., 2020; Chechik et al., 2019; Wan et al., 2019). As early as in 2007, Murphy et al. (2007) proposed creating test cases to test ML systems, particularly those that implement ranking algorithms. Groce et al. (2013) proposed a method for effective test case selection. Ma et al. (2019) proposed a test generation technique based on combinatorial testing for DL systems.

The discovery of adversarial examples and evaluation of model robustness are fundamental problems in engineering ML systems (Du et al., 2020; Fujii et al., 2020; Sun et al., 2018). Adversarial examples cause ML models to produce an error output with high confidence, even though their difference from natural inputs are subtle (Szegedy et al., 2014). For instance, an image classifier may label two very similar images, which cannot be differentiated by a human, incorrectly. As a result of this, Sun et al. (2019a) presented DeepConcolic, a tool for testing deep neural networks. DeepConcolic adopted concolic analysis approach and generates test cases guided by neuron coverage and MC/DC variants for deep neural networks (Sun et al., 2019a). Zhang et al. (2020b) developed a tool named KuK (Know the UnKnown), which generates uncommon input samples via genetic algorithms. Some researchers focused on generating test inputs as naturally as possible for testing autonomous cars, such as DeepTest (Tian et al., 2018), DeepHunter (Xie et al., 2019b), and DeepRoad (Zhang et al., 2018a). On the other hand, some of the generated images (test inputs) were still unnatural and could not be recognized by humans (Zhang et al., 2020c). Gambi et al. (2019) followed a different approach and proposed an approach to generate test cases from police reports of car crashes to cope with the scarcity of sensory data collected during real car crashes. Riccio and Tonella (2020) model-based input generation approach to produce realistic inputs and developed a tool named DeepJanus. Byun and Rayadurgam (2020) proposed a manifold-based test generation framework for automatic generation of realistic test cases for testing ML systems involving image classification components. Berend et al. (2020) generated test cases by considering input data distribution and managed to increase model robustness.

Metamorphic testing (MeT) is one of the test case selection techniques proposed by Chen et al. (1998). MeT relies on metamorphic relations that refers to the relationships among the inputs and outputs of multiple executions of a software system under test (Zhou et al., 2015). For instance, a metamorphic relation for a search engine can be *Search_Result_Count(keyword1) >= Search_Result_Count(keyword1 + keyword2)*, where + denotes the concatenation of two keywords (Segura et al., 2016). In other words, the number of search results for *keyword1* should be greater or equal than that obtained for the phrase consisted of *keyword1* and *keyword2* (Segura et al., 2016). Violation of a metamorphic relation points to a defect (Zhou et al., 2015). Braiek and Khomh (2019) proposed DeepEvolution, a search-based approach for testing DL models that relies on metaheuristics to ensure the maximum diversity in generated test cases. They leveraged metamorphic transformations to limit the search space for inputs (Braiek and Khomh, 2019). Wang and Su (2020) applied MeT to validate DNN-based object detectors. Their tool MetaOD generated new images by inserting extra objects to existing images (Wang and Su, 2020). These new images were used to detect failures in object detectors and helped in improving the performance of DL models (Wang and Su, 2020). Santos et al. (2020) applied MeT to test image classifiers and obtained promising results. On the contrary, Kang et al. (2020) did not consider small variations created from existing inputs via MeT. They introduced a method called SINVAD to search for fault-revealing inputs in a wider search space and produced test cases capable of validating the robustness of DNNs (Kang et al., 2020).

Fairness can be an essential quality attribute of ML systems (Corbett-Davies and Goel, 2018), as pointed out in Section 5.3.1. Galhotra et al. (2017) proposed a technique to measure causal discrimination and managed to detect significant ML model discrimination towards gender, marital status, or race as many as 77.2% of the individuals in datasets. Causal discrimination suggests that a software system is attributed to be fair with respect to a set of characteristics, if it produces the same output for every two individuals who differ only in those characteristics (Galhotra et al., 2017). Udeshi et al. (2018) proposed an approach to systematically generate discriminatory test inputs and reveal fairness violations in ML models. They first performed random sampling on the input space and then derived further test cases based on these random samples (Udeshi et al., 2018). Aggarwal et al. (2019) performed systematic searching to generate test cases for detecting problems in fairness and aimed at eliminating the redundancies produced by random sampling. Sharma and Wehrheim (2020b) suggested a technique to verify ML model fairness against a set of fairness requirements specified in a language designed by the authors. In another study, Sharma and Wehrheim (2019) proposed a metamorphic testing framework for testing balancedness (which complements fairness) of an ML classifier. They mutated training datasets in various ways to generate new datasets (Sharma and Wehrheim, 2019). Sharma and Wehrheim (2020a) applied verification-based testing to check monotonicity of generated models.

Some researchers aimed at reducing testing costs by selecting a subset of test cases and at the same time keeping model quality estimation at a satisfactory level. Zhou et al. (2020a) presented their approach named DeepReduce, which helped them reduce testing data 7.5% on average and still reliably estimate the performance of DL models. Chen et al. (2020a) proposed PACE approach (Practical ACcuracy Estimation), which selects a small subset of test inputs that can represent a larger set test inputs. To assist an input subset selection, Shen et al. (2020) proposed a technique named Multiple-Boundary Clustering and Prioritization (MCP). This technique clusters test samples into the boundary areas and specifies the priority to select samples evenly from all boundary areas (Shen et al., 2020). Yan et al. (2019) designed the Adaptive Random Testing for DL systems (ARTDL) method targeting to improve testing efficiency. They managed to detect defects spending less effort by selecting test cases having the furthest distance from test cases not causing any failure. All of these approaches try to reduce input data labeling and testing effort.

*Evaluating test cases:* Test case evaluation is a complex task since it is hard to formalize and measure test cases' characteristics influence quality (Durelli et al., 2019). Coverage of a test suite is used for test case evaluation when precise quality measures are not present (Durelli et al., 2019). The higher coverage a test suite achieves, the more defects hidden in code are expected to be detected (Zhang et al., 2020c). Unlike traditional software, code coverage cannot be used for ML components since the decision logic of ML capability is obtained via training an ML model, not via explicit coding (Ishikawa and Yoshioka, 2019; Khomh et al., 2018; Kim et al., 2019; Ma et al., 2018b; Sekhon and Fleming, 2019; Sun et al., 2018; Zhang et al., 2020c). The size of test case sets is enormous, and missing test cases is a more challenging problem for ML systems (Khomh et al., 2018). Kim et al. (2019) and Sekhon and Fleming (2019) state that existing coverage criteria are not sufficiently fine-grained to capture subtle behaviors exhibited by ML systems. Ma et al. (2018a) designed a set of coverage criteria for deep neural networks to assess the testing adequacy. Sun et al. (2018) formalized the coverage criteria for deep neural networks studied in the literature and used it to increase the coverage. In their more recent study, Sun et al. (2019b) proposed four coverage criteria tailored to deep neural networks' structural features and semantics. Du et al. proposed new coverage criteria for DL systems based on recurrent neural networks and utilized these criteria to generate more adversarial samples and reveal more failures (Du et al., 2019). Although researchers try to develop new coverage criteria appropriate for DL models, Li et al. (2019) could not find a strong correlation between the number of misclassified inputs in a test set and its structural coverage. Yan et al. (2020) claimed that DNN coverage criteria do not have monotonic relations with model quality measured by model accuracy in the presence of adversarial examples. Based on the experiments conducted by Harel-Canada et al. (2020), higher neuron coverage led to fewer defects detected, less natural inputs, and more biased prediction preferences. Feng et al. (2020) also criticized coverage-based methods for being not effective as expected. They proposed a test prioritization technique called DeepGini, especially for image classification tasks. They adopted a statistical approach, which is more effective and less time-consuming based on their experiments (Feng et al., 2020). Trujillo et al. (2020) studied the reliability of the use of neuron coverage in testing Deep Reinforcement Learning (DeepRL) models, which is a particular type of DNN models. Based on their preliminary results, neuron coverage is not sufficient to reach reliable conclusions about the quality DeepRL models (Trujillo et al., 2020). Gerasimou et al. (2020) proposed an approach and a tool, called DeepImportance, to determine more influential neurons in a DL model. Their hypothesis is that some neurons are more important than the others in affecting the performance of a DL model. To this end, they designed an importance-driven coverage criterion, which can establish the adequacy of a test set to trigger different combinations of important neurons' behaviors (Gerasimou et al., 2020).

Mutation Testing (MuT) is a fault-based testing technique that claims to address the shortcomings of coverage criteria (Jia and Harman, 2010; Zhang et al., 2018c). MuT involves a mutation adequacy score, which is used to measure the effectiveness of a set of test cases in terms of its ability to detect faults (Jia and Harman, 2010). Ma et al. (2018b), Jahangirova and Tonella (2020), Klampfl et al. (2020), and Guo et al. (2020) investigated the applicability of MuT for DL systems. Ma et al. (2018b) proposed a MuT framework, called DeepMutation, to inject faults that could be potentially introduced during DL development. Jahangirova and Tonella (2020) identified some mutation operators and revised mutation killing by taking the stochastic nature of DL systems into account. Klampfl et al. (2020) proposed a testing approach to detect various types of mutants. Guo et al. (2020) proposed a search-based approach named Audee. This approach involves three different mutation strategies to generate diverse test cases by exploring combinations of model structures, parameters, weights and inputs (Guo et al., 2020). All of these researchers reported that they planned for further research to adapt well-established MuT techniques used for engineering traditional software to ML systems testing.

Flaky tests are observed frequently and may cause problems in software development process (Luo et al., 2014). Such tests can run non-deterministically pass or fail when run on the same version of the code (Bell et al., 2018). Dutta et al. (2020) proposed a technique, called FLASH, for systematically running tests with different random number generator seeds to detect flaky tests designed to test an ML system.

*Preparing test data:* Test data preparation has been identified as a challenge for traditional software development decades ago (Korel, 1990). Access to high quality test data is a concern for developing ML systems (Arpteg et al., 2018; Guo et al., 2018; Knauss et al., 2017; Li et al., 2019b; Ma et al., 2018a; Ma et al., 2018b; Wan et al., 2019). In some cases, collecting testing datasets may require manual labeling, which is labor-intensive (Li et al., 2019b; Wan et al., 2019). DLFuzz, proposed by Guo et al. (2018), aims to generate adversarial examples without manual labeling effort. By finding new and especially rare inputs that improve neuron coverage in DL models, DLFuzz tries to ensure the reliability and robustness of DL systems (Guo et al., 2018). Besides, since ML systems are expected to cope with the external world's dynamic nature to some degree, datasets should be updated (Arpteg et al., 2018). Therefore, an infrastructure for automated data collection and labeling may be developed (Knauss et al., 2017). Simulators can be used to generate data for testing in some domains. For instance, according to the experiments conducted by Haq et al. (2020), simulator-generated datasets yielded DNN prediction errors that are similar to those obtained via real-world datasets in autonomous driving domain.

*Executing tests:* Test execution refers to running the code (including an ML model) and comparing the expected and observed results. One needs a test environment involving trained ML models to execute the tests (Gharibi et al., 2019). ModelKB tool allows a tester to test a specific experiment model via a user interface (Gharibi et al., 2019). On the other hand, ModelKB tool prototype does not support all kinds of ML algorithms and ML problems (Gharibi et al., 2019).

Haq et al. (2020) executed tests on DNNs for autonomous vehicles in offline and online mode. For offline testing, DNNs were tested as individual units based on test datasets obtained independently from the DNNs under test. For online testing, DNNs were embedded into a simulator and tested in a closed-loop environment in the simulator. As a result, they found offline testing more optimistic as many safety violations identified through online testing could not be identified through online testing (Haq et al., 2020). In addition, large prediction errors generated by offline testing always led to severe safety violations detectable by online testing (Haq et al., 2020).

Cross-framework and cross-platform support become more critical as ML models are deployed on various hardware types, such as cloud platforms, mobile devices, and edge computing devices (Zhang et al., 2019a). To this respect, Zhang et al. (2019a) proposed a differential testing framework to test ML models against potential inconsistent behavior in different settings.

*Evaluating test results:* Test evaluation refers to assessing the testing results using test oracles and giving pass or fail decisions for test scenarios (Garousi et al., 2018). Identifying test oracle ("oracle problem" (Barr et al., 2014)) is one of the challenges in ML systems testing (Liem and Panichella, 2020; Murphy et al., 2007; Stocco et al., 2020). Dwarakanath et al. (2018) applied metamorphic relations to image classifications with SVM and DL systems to tackle the oracle problem. Similarly, Xie et al. (2011) proposed an approach based on metamorphic testing to alleviate the oracle problem in testing ML classification algorithms. Nakajima (2018) proposed a behavioral oracle that monitors changes in certain statistical indicators during the training process and forms a basis for metamorphic relations to be checked. Based on these metamorphic relations, Nakajima (2018) generated test inputs for testing neural network models. Cheng et al. (2018) conducted some experiments and found out that metamorphic relations are not effective. Qin et al. (2018) proposed a program synthesis technique to systematically construct oracle-alike mirror programs to alleviate the oracle problem. Chen et al. (2019) applied the variable strength combinatorial testing technique to measure the adequacy of deep neural network testing. Barash et al. (2019) proposed generating test cases using the modeling process of combinatorial testing. They tried to utilize business requirements and detect weak sides of an ML system from a business perspective (Barash et al., 2019).

Zheng et al. (2019) proposed a method for identifying translation failures in neural machine translation systems without reference translations, i.e., test oracle. Sun and Zhou (2018) used a metamorphic testing technique to test machine translation services without a human assessor or reference translation.

The non-deterministic nature of ML systems makes test evaluation more challenging (Arpteg et al., 2018; Ishikawa, 2018; Ishikawa and Yoshioka, 2019; Nishi et al., 2018; Tao et al., 2019). Unlike testing traditional software, finding one or a few incorrect results does not necessarily indicate the presence of a bug (Dwarakanath et al., 2018). Since ML systems are validated via probabilistic ML performance metrics (like accuracy, F-measure), observing an unexpected output resulting from a single test case execution does not necessarily mean a bug (Zhang et al., 2018b). Moreover, good performance during testing cannot guarantee the satisfactory performance of ML systems on production (Wan et al., 2019). Barash et al. introduced a method to bridge the gap between business requirements and ML performance metrics (Barash et al., 2019). By doing so, they were able to evaluate test results against business requirements (Barash et al., 2019).

*Debugging and fixing:* It is very critical to detect and remove bugs as much as possible before launching an ML system. Different from traditional software, data bugs are also very vital for ML systems besides bugs in code. The difficulties of debugging ML systems were pointed out by several studies (Khomh et al., 2018; Wan et al., 2019; Zhang et al., 2018b; Zhang et al., 2019a). One reason for this is the prevalence of non-determinism in the training process, which leads to hard-to-reproduce bugs (Wan et al., 2019; Zhang et al., 2018b). Another pain point is the lack of debugging and profiling support for ML systems (Zhang et al., 2019a; Zhang et al., 2020a). To this end, some approaches and tools have been developed for finding bugs in training data, ML/DL models, and code, such as DeepFault (Eniser et al., 2019), LAMP (Ma et al., 2017), MODE (Ma et al., 2018c), and DARVIZ (Sankaran et al., 2017). DeepFault is a whitebox deep neural network testing approach to identify suspicious neurons that may lead to inadequate performance (Eniser et al., 2019). Ma et al. (2017) proposed LAMP technique and developed a prototype tool for finding bugs in input data, graph models, and graph-based ML algorithm implementations. MODE identifies faulty neurons in neural networks (Ma et al., 2018c). DARVIZ tool helps in debugging the training process of DL models (Sankaran et al., 2017). Chakraborty et al. (2020b) focused on detecting, explaining, and visualizing bias in ML models.

ML systems heavily depend on ML libraries, frameworks, and platforms. The bugs in these components impair ML systems (Guo et al., 2019). With this, Pham et al. (2019) proposed an approach, named CRADLE, to find and localize bugs in DL libraries. CRADLE utilizes publicly available DL models as inputs to invoke DL libraries and adopt differential testing to capture the triggered defects (Wang et al., 2020b). Wang et al. (2020b) proposed another approach (LEMON) for DL library testing by generating effective DL models via guided mutation. They claimed that publicly available DL models used by CRADLE usually focus on popular tasks and only invoke a limited portion of library code (Wang et al., 2020b). To overcome this shortcoming of CRADLE, they used model-level mutations to generate models to test DL libraries by invoking relatively rarely used portions of library code (Wang et al., 2020b). Tizpaz-Niari et al. (2020) presented a method and a tool (DPFuzz) for discovering and explaining performance bugs in ML frameworks . They used fuzzing to generate inputs that uncover performance bugs and discriminant analysis to explain differences in performance (Tizpaz-Niari et al., 2020). Humbatova et al. (2020) analyzed the issues in some GitHub projects using DL frameworks Tensorflow, Keras, and PyTorch and built a taxonomy of the defects. They aimed at identifying defect types and guide developers to both avoid and locate such defects (Humbatova et al., 2020).

There are also domain-specific open-source software libraries. Apollo Baidu and Autoware are two of these in autonomous driving domain. Garcia et al. (2020) identified 499 bugs in these two software systems. They classified these bugs into 13 root causes, 20 bug symptoms, and 18 categories of software components these bugs often affect (Garcia et al., 2020). These findings can be used to develop bug detection tools, localize defects or repair defects automatically in autonomous driving domain (Garcia et al., 2020).

Islam et al. (2020) studied bug repairs on Deep Neural Networks (DNN) obtained from StackOverflow and GitHub. They found that developers are mostly fixing data dimensions and neural network connectivity (Islam et al., 2020). While developers fix bugs, they may introduce adversarial vulnerabilities and new bugs (Islam et al., 2020). In another study, Pan (2020) observed that bug fixes may cause both an increase and decrease in robustness of DL models. Li et al. (2020) focused on debugging and fixing confidence errors in DNNs. By doing so, they intend to understand correctly when a model works well and when not (Li et al., 2020).

*Managing tests:* Test management refers to planning, controlling, and monitoring testing activities (Garousi et al., 2018). In a case study, Rivero et al. (2020) utilized test management tools and spreadsheets to manage test process in which various teams were involved. Alvarez-Rodríguez et al. (2019) states the need for certification and qualification activities to manage tests, especially for safety-critical software systems.

*Automating tests:* Software test automation has already moved beyond a luxury to become a necessity to cope with large complex systems (Mosley and Posey, 2002) and decrease testing costs (Ramler and Wolfmaier, 2006). The cost and required resources to test ML components, especially for autonomous driving, is steadily increasing (Knauss et al., 2017). Therefore, new tools and techniques are required to automate testing for ML systems for providing more robust systems, especially in safety-sensitive domains (Knauss et al., 2017). Tian et al. (2018) automated test case generation for safety-critical DL-based systems like autonomous cars. Rivero et al. (2020) reported that they automated unit, integration, and ML model testing in an industrial project. They also decided to automate functional tests; however, they decided to switch to manual exploratory testing due to the dramatic changes in the interfaces during the project (Rivero et al., 2020).

### 5.3.5 Maintenance and Configuration Management

With the emergence of many general-purpose ML libraries, services, platforms combined with the accumulation of ML experience, developing ML systems has become relatively easy and cheap, whereas maintaining them over time is difficult and expensive (Sculley et al., 2015). Maintaining ML systems involves keeping track of additional configuration items, models, and data, besides code (Sato et al., 2019). Moreover, ML systems should be monitored continuously since their performance is subject to change due to their non-deterministic nature. The papers in my pool mainly identify three groups of challenges, which are interrelated:

*Dealing with configuration management of data and ML models:* ML systems enlarged the scope of tracking and controlling change. Data and ML models are essential configuration items besides code (Belani et al., 2019; Sato et al., 2019). While versioning data is a vital need (Alvarez-Rodríguez et al., 2019; Amershi et al., 2019; Belani et al., 2019; Wan et al., 2019; Yokoyama, 2019), the methods and tools to support this is not mature yet (Amershi et al., 2019). Versioning data does not only involve keeping track of datasets but also keeping track of metadata of datasets (Gebru et al., 2018) and their relationships to ML models (Amershi et al., 2019). Another requirement can be comparing datasets and exploring differences between datasets (Sutton et al., 2018). All of these aspects make configuration management of the data component complicated (Morgenthaler et al., 2012). ML models derived from data constitute the other type of essential configuration item (Alvarez-Rodríguez et al., 2019; Belani et al., 2019; Gharibi et al., 2019; Wan et al., 2019; Yokoyama, 2019). Models are mainly made up of algorithms, hyperparameters, and their dependencies on datasets (Amershi et al., 2019; Yokoyama, 2019). Wu et al. (2020) used the existing clone analysis techniques to identify similar ML models. Model management, which refers to tracking, storing, query, comparing, reproducing, and sharing models (Gharibi et al., 2019), has been identified as one of the most challenging and time-consuming activities (Schelter et al., 2018; Vartak, 2018). There are some attempts to address the challenges in model management, such as ModelHub (Miao et al., 2017), ModelDB (Vartak et al., 2016), and MLflow (Zaharia et al., 2018). Open source tools, such as DVC (dvc.org), were launched for version control of ML components. These methods and tools should be integrated with the overall engineering process and the other tools used to develop and maintain ML systems.

ML models are also dependent on ML frameworks. It is a challenge to keep pace with the latest versions of ML libraries and their dependencies (Han et al., 2020). Software Composition Analysis (SCA) tools can help in discovering all related components, their supporting libraries, and their direct and indirect dependencies (Han et al., 2020). Another critical maintenance issue is the technical debts present in ML frameworks (Liu et al., 2020b). Liu et al. (2020b) identified design, defect, documentation, test, requirement, compatibility, and algorithm debts in ML frameworks, which may cause maintenance problems in ML systems.

Peng et al. (2020) observed that as ML models evolve, it poses maintenance challenges on the code logic that should co-evolve. Failing to update the code logic accordingly as ML models evolve may introduce bugs and fail to achieve the optimal performance of ML models, and further negatively influence the entire ML-powered system (Peng et al., 2020).

Benton (2020) suggests using a single platform to manage all components (data pipelines, ML models, code logic, etc.) of an ML system. For instance, Kubernetes ecosystem may ease maintenance and problem localization (Benton, 2020).

*Dealing with the history of experiments:* While it is necessary to keep track of changes in data and ML models, some large-scale ML systems involve configuration management at a higher level, i.e., keeping track of experiments. Each experiment may have many components that affect the outcome, i.e., the ML model. These components can include hardware (GPU models primarily), platform (operating system and installed packages), source code (model training and pre-processing), configuration (model configuration and pre-processing settings), training data (input signals and target values), and model state (versions of trained models) (Arpteg et al., 2018; Wan et al., 2019; Yokoyama, 2019). The development of ML components is an iterative and experimental process (Gharibi et al., 2019), and hence, it is prevalent to perform a vast number of experiments to identify the optimal ML model (Arpteg et al., 2018). Also, it is possible to use automated meta-optimization methods and conduct lots of experiments without human intervention. Whether done automatically or manually, experimentation activities generate many artifacts (datasets, hyperparameters, models, libraries, etc.), which should be versioned (Golovin et al., 2017). Therefore, experiment management is a recent and complicated challenge in engineering ML systems (Hazelwood et al., 2018; Zaharia et al., 2018). Teams may require to compare, reproduce, and share experiments (Gharibi et al., 2019; Lwakatare et al., 2019). Improper experiment management may result in some problems: (1) experiments may not be reproduced when needed (Hill et al., 2016; Zaharia et al., 2018); (2) some experiments may be repeated unintentionally (Hill et al., 2016); (3) experiments and their results may not be analyzed, reported, and shared appropriately.

*Dealing with re-training and re-deployment:* One of the hardest parts of maintaining ML systems is to keep the performance (such as accuracy) at a certain level or improve it if needed. Unforeseen changes in the external world may cause changes in input data patterns and negatively affect the performance of ML components (known as "concept drift" (Tsymbal, 2004), see Section 5.3.2). The type of concept drift, i.e., sudden or gradual (Tsymbal, 2004), and its impact should be considered when dealing with re-training and re-deployment. In some cases, concept drifts in data streams should be detected promptly (Wang and Abraham, 2015). Žliobaitė (2010) provides a framework for thinking about decision points when addressing concept drift. For instance, Stocco and Tonella (2020) suggest using drift detectors to identify input data deviating from the distribution of training data. Newly identified data can be used directly for adaptive retraining (Stocco and Tonella, 2020). Another option is to augment training data with the guidance of newly identified input data and retrain ML model (Ren et al., 2020).

Therefore, it is critical to monitor the performance and take the necessary actions, if required, (Baier et al., 2019; Carleton et al., 2020; Wan et al., 2019). Automatic maintenance mechanisms can be built to keep the performance at a certain level (Wan et al., 2019). Such mechanisms can detect performance degradation and take the required actions by re-training ML models with new data. On the other hand, it is questionable to what extent this can be achieved in all domains without human intervention, especially in risky domains, such as health. In risky domains, automated checks can be used to trigger notifications if predefined thresholds are violated (Baier et al., 2019).

In some cases, it may be required to deploy an updated ML model to the production environment to get users' feedback. Canary release approach may limit the possible negative impact of the new ML model and help to cope with the risk of deploying an updated ML model (Yokoyama, 2019). Canary release refers to deploying a new ML model (or any other component) for a restricted number of users and rollback the deployment if a negative impact is observed (Sato, 2014).

## 5.3.6 Software Engineering Process and Management

There are some process models, phases, activities that document steps to develop and maintain ML models. For instance, Hulten (2018) breaks down the ML process into five stages: (1) getting data to model, (2) feature engineering, (3) modeling, (4) deployment, and (5) maintenance. Rao (2019) shows the typical steps involved in the model development lifecycle: (1) define the problem, (2) collect, cleanse, and prepare data, (3) build and train model, (4) tune hyperparameters and validate model, and (5) deploy to production. The steps are not executed as a waterfall process; there is feedback from deployment to problem definition and an iteration from building a model to deployment (Rao, 2019). Shams (2018) demonstrates a three-phase ML product development process to shorten the delivery time from idea generation to deployment. On the other hand, processes that guide teams in engineering ML systems with ML and traditional software components are needed.

*Harmonizing the activities for developing ML components with software process:* Adding ML components to a software system involves additional actions to be performed. Teams developing ML systems need more guidance in the form of a software process that can be tailored according to specific project needs (Alvarez-Rodríguez et al., 2019; Belani et al., 2019; Correia et al., 2020; de Souza Nascimento et al., 2019; Liu et al., 2020a; Ozkaya, 2020; Serban et al., 2020). Data lifecycle activities form a critical and compelling subgroup of activities, which profoundly affect ML projects' success (Munappy et al., 2020; Polyzotis et al., 2018; Wan et al., 2019). Therefore, SE process standards should be enriched to handle data lifecycle activities (Huang, 2017). Besides, new agile approaches, such as DataOps (Atwal, 2019; Ereth, 2018; Munappy et al., 2020), are needed to meet increasingly demanding and evolving business requirements via ML systems. ML model lifecycle activities are another vital subgroup of activities integrated with the SE process (Alvarez-Rodríguez et al., 2019; Byrne, 2017). Besides data and ML model lifecycles, teams need guidance on various aspects, such as engineering requirements (Alvarez-Rodríguez et al., 2019; Belani et al., 2019), handling data (de Souza Nascimento et al., 2019), testing (Alvarez-Rodríguez et al., 2019; Fujii et al., 2020), and maintenance (Baier et al., 2019). Moreb et al. (2020) proposed a process for engineering ML systems in health informatics domain.

According to Sensemaking-Coevolution-Implementation Theory, sensemaking is one of the core activities in engineering software systems (Ralph, 2015). Simply put, sensemaking refers to understanding the context that surrounds the problems and software requirements against which a software system will be developed (Ralph, 2015). Wolf and Paine (2020) emphasizes the difficulty of sensemaking for business context, ML algorithms, libraries, and models in ML systems engineering projects.

*Assessing the ML process:* Teams may need to evaluate their ML process to understand and improve their capabilities. There are some early attempts to determine the ML process' maturity (Amershi et al., 2019) and how ML components are used in larger software systems (Lwakatare et al., 2019). A mature process assessment is required to identify possible smells regarding the ML process, especially in large-scale ML systems (Sculley et al., 2015).

*Estimating effort:* Due to the uncertainties in the ML process, it is hard to estimate effort and do planning (Wan et al., 2019). In some cases, the team cannot know whether the quantitative targets are achievable until final ML models are obtained (Wan et al., 2019). As a result of not being able to estimate effort, business owners act impatiently and cancel ML component development despite promising intermediate results (Arpteg et al., 2018).

### 5.3.7 Organizational Aspects

Engineering complex systems, such as ML systems, is a complicated endeavor with organizational and technical aspects (Clarke et al., 2016). Part of the organizational challenges involves identifying skill sets and roles required to engineer ML systems (Flaounas, 2017). Seven papers in the final pool address the challenges in two main categories from an organizational perspective: Having the required skill sets and having a team working harmoniously.

*Dealing with various skill sets:* Engineering an ML system involves skill sets from multiple fields, i.e., machine learning, deep learning, data science, mathematics, algorithms, software engineering, and statistics (Alshangiti et al., 2019; Alvarez-Rodríguez et al., 2019; Belani et al., 2019; Correia et al., 2020; Wan et al., 2019; Workera, 2020). A proliferation of roles, including specialists, in particular, SE phases, such as design (Northrop et al., 2019) or specific types of algorithms such as natural language processing (Byrne, 2017). Practitioners and academicians already started to identify emerging roles to classify skill sets and have proper people in teams to engineer ML systems (Kim et al., 2017; Workera, 2020). While there are generic roles such as data scientist, ML engineer, data analyst, ML software engineer, ML researcher, and software engineer, there are more specific roles, such as data evangelist, data preparer, data shaper, data analyzer (Kim et al., 2017). A data preparer requires a good data slicing and dicing skills for proper data cleaning and feature preprocessing (Alshangiti et al., 2019). An ML engineer need to understand the mathematical and statistical background of models for proper model fitting and tuning (Alshangiti et al., 2019). A software engineer may have to deal with distributed systems, web services, and multi-threaded programming (Alshangiti et al., 2019). Alvarez-Rodríguez et al. (2019) identified the formation of a workforce equipped with the required skills to tackle different types of problems in various domains as a challenge.

*Building harmony among isolated roles:* Having the required people with proper skill sets is necessary but not sufficient to engineer ML systems. For instance, the synergy between data scientists, domain experts, and customers during data processing is essential for proper data collection, data enrichment and labeling (Correia et al., 2020). People having different backgrounds have cultural (Arpteg et al., 2018) and language differences. For example, while data scientists may be pragmatic when coding as long as they achieve the desired results, software engineers may care about code quality and maintainability (Arpteg et al., 2018). Therefore, it is challenging to get all these people working together towards achieving a common objective (Amershi et al., 2019). It is even challenging to have all these people on the same page (Kim et al., 2017).

## 5.4 DEMOGRAPHICS

Figure 9 shows the number of primary studies published each year. The earliest paper (Murphy et al., 2007) was published at the International Conference on Software Engineering & Knowledge Engineering in 2007. There was no paper or one paper per year until 2017. Three papers were published in 2017, after which we see a sharp increase starting from 2018. Since I conducted the search and primary study selection in January 2020 and March 2021, I did not include the papers that have been published in 2021 to see the trend over the years.

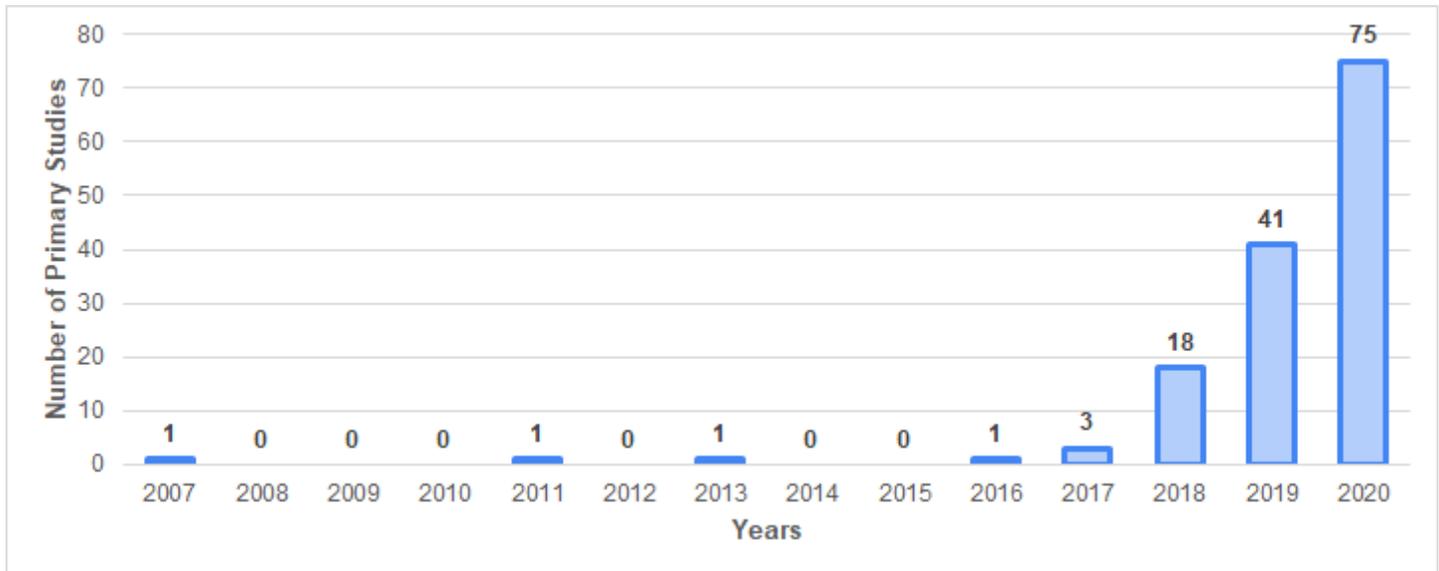

Figure 9. Number of primary studies over the years

91% of the studies were presented in conferences and workshops (129 papers in conferences/workshops and 12 papers in journals). The conference with the most papers (20) is the ACM Joint Meeting on European Software Engineering Conference and Symposium on the Foundations of Software Engineering (ESEC/FSE). The journals with the most papers (three in each) are IEEE Access, IEEE Software, and IEEE Transactions on Software Engineering. The appendix includes the list of the SE venues in which the primary studies were presented and published.

I classified the papers based on the researchers' affiliations: university, industry (research organizations and companies), and collaboration (for papers whose authors are from both university and industry). One hundred and one papers (72%) were authored solely by academic researchers, 11 papers (8%) by researchers from research organizations or companies, and 29 papers (20%) were written jointly by universities and other organizations.

Figure 10 shows the affiliation types per knowledge area. Since a paper can address more than one knowledge area, a paper may have been classified under more than one area. Testing & Quality is the most focused knowledge area by academia and industry. Also, researchers collaborated the most within the scope of testing and quality. Development & tools and requirements engineering areas the second and third areas in which researchers collaborated.

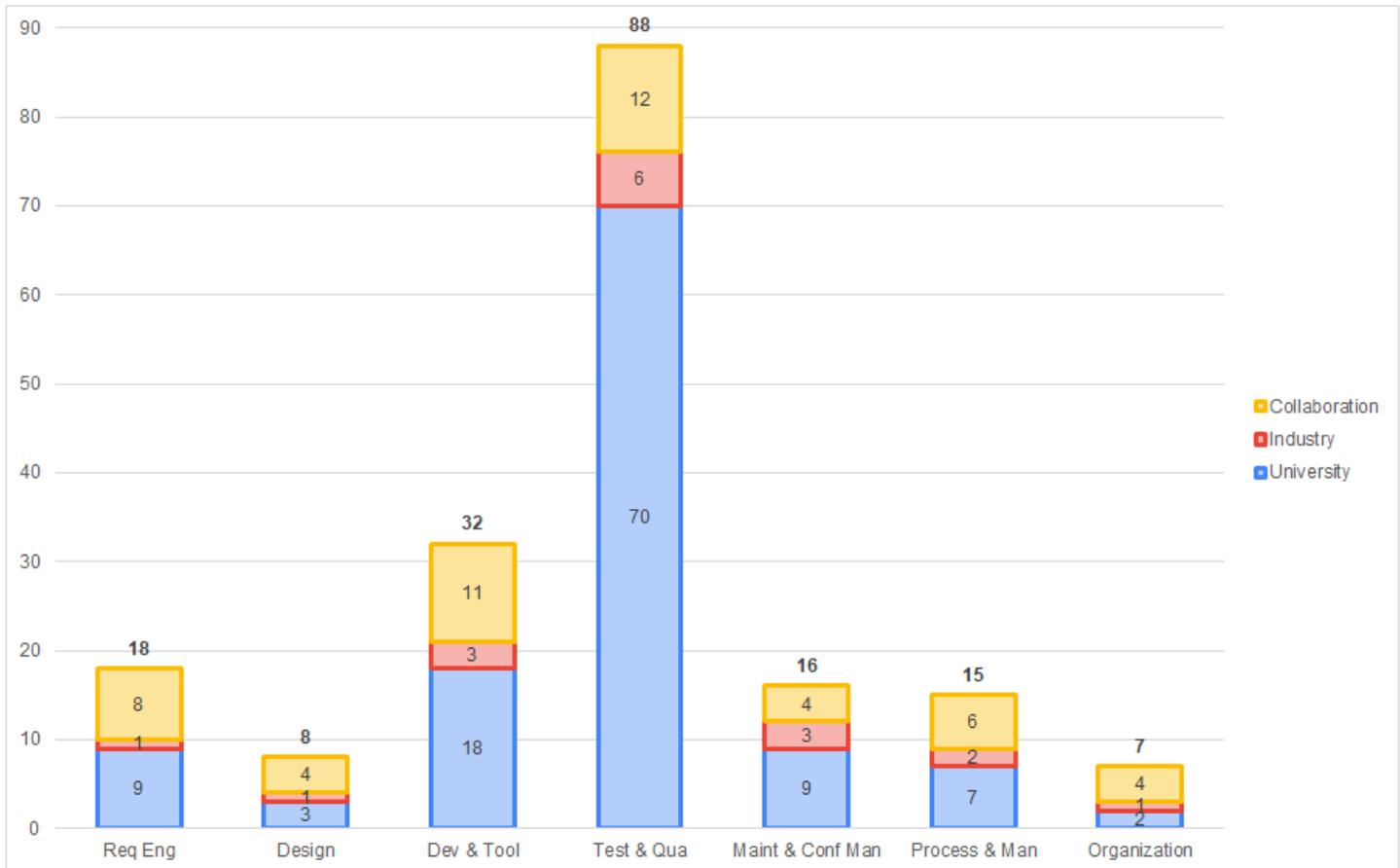

Figure 10. Affiliation types per knowledge area

## 6 DISCUSSION

The number of primary studies in SE venues on engineering ML systems has approximately doubled annually every year since 2018. This trend clearly shows the increasing interest of the SE community in engineering ML systems. In addition, the top SE conferences, i.e., ESEC/FSE and ICSE, receive submissions on this emerging topic. The reason for having much fewer papers in journals may be due to the fact that this topic is still in its infancy. There are many gaps that should be addressed by researchers, ideally from academia and industry, to develop production-ready ML systems.

This section presents some potential research directions from the SE perspective (Section 6.1), other implications independent from SE aspects (Section 6.2), the possible benefits of this review (Section 6.3), and then limitations and potential threats to validity of this review (Section 6.4).

### 6.1 POTENTIAL RESEARCH DIRECTIONS FROM THE SE PERSPECTIVE

In this subsection, I summarize the main research directions from an SE perspective. Researchers may refer to Table 8 to form more detailed research questions.

*Requirements engineering for ML systems:* Specifying ML components' requirements requires a different perspective than traditional software components. It is difficult and mostly impossible to characterize all the behaviors of ML components under all circumstances (Tuncali et al., 2019). Rather than specifying precise requirements, hypothesizing what outcomes can be obtained from data should be the starting point (Wan et al., 2019), and requirements shall be refined via experimentation (Ishikawa and Yoshioka, 2019). Equally important, more guidance and examples are needed on how to map ML performance metrics to business objectives and metrics.

Besides, ML systems include traditional software components as well. Therefore, integrating proper practices to specify ML and traditional software components' requirements is a research problem (Vogelsang and Borg, 2019). Specification of new types of quality attributes, such as explainability (Gade et al., 2019), fairness (Ahuja et al., 2020; Bellamy et al., 2019; Horkoff, 2019), and freshness (Zinkevich, 2017), for ML components, may initiate new research questions.

Another research direction can be building a risk assessment framework for ML systems to understand the potential impacts of false predictions. For instance, Schwerdtner et al. (2020) proposed a risk assessment framework for ML systems developed for classification problems.

Data privacy regulations are vital inputs for the development of ML systems. Privacy impact assessments and rules on data collection to prevent potential privacy violations may be required. Data stewardship requirements may be specified to prevent the use of personal information that is adverse or unfair to the individuals the data relates to. Guidelines on privacy impact assessment, data collection rules, and data stewardship may help requirements engineers involving in ML systems development.

*Designing ML systems:* Components with ML capabilities are becoming an architectural part of software systems, sharing cross-cutting functional and non-functional concerns (Nguyen-Duc et al., 2020). Monitoring for potential performance degradation on production (Wan et al., 2019; Yokoyama, 2019) and high-volume data processing (de Souza Nascimento et al., 2019; Wan et al., 2019) are two important design considerations for ML systems. New architectural styles and patterns for ML systems are required for robust systems to address these concerns (Lin and Kolcz, 2012; Wan et al., 2019).

Another challenge arises when organizations decide to enhance their existing systems with ML capabilities (Jahić and Roitsch, 2020). Although there are some design patterns proposed in the literature to address these challenges (Washizaki et al., 2020), further guidance is needed on this topic.

*SE tooling for developing ML systems:* ML systems' development involves a diverse set of frameworks, tools, libraries, and programming languages (Rahman et al., 2019). The popularity of ML has brought many tools, mostly open-source (Louridas and Ebert, 2016). However, this diverse set of tools have emerged with challenges, such as compatibility (Rahman et al., 2019), integration of various components (Rahman et al., 2019), required knowledge and experience, etc. Lack of developer support tools for data-intensive systems is a challenge for developing ML systems (Anderson, 2015). Some tools emerged from academia, such as DARVIZ (Sankaran et al., 2017), DeepVisual (Xie et al., 2019a), NeuralVis (Zhang et al., 2019b), ModelKB (Gharibi et al., 2019), to solve specific problems in developing ML systems. There are some efforts to bring various tools together to support the effective development of ML systems (Smith et al., 2020). There are some platforms, such as MLflow (Zaharia et al., 2018), focusing only on the ML lifecycle. To support other SE aspects, such as generating Python components ready for deployment, version control conflict handling, and continuous integration, some tools, such as nbdev, have been released (Howard and Gugger, 2020). On the other hand, compared to tool support for traditional software development, there is room for improvement to have SE tooling for developing ML systems.

Developers deal with additional complications while developing ML systems. They need guidance and tool support to cope with exceptions in data, adversarial examples, and potentially biased decisions breaking ethical rules. Another challenge is the estimation of resource requirements for production environment. Tools to estimate computation and memory requirements, such as DNNMem (Gao et al., 2020b), may help developers. The development of ML models to run on energy- and resource-constrained devices requires new techniques for model compression and assessment of their performance.

*Testing ML systems:* SE researchers focused mostly on testing ML systems. Despite many studies, we do not have mature testing techniques and tools to develop ML systems in industrial settings. Designing and evaluating test cases, preparing test data, executing tests, and assessing test results contain many research problems. To name a few of the research problems, generating test inputs, ensuring adequate test coverage, coping with oracle problem, and ensuring test data quality.

As in testing scientific software (Kanewala and Bieman, 2014), perfect oracles may not be available for ML systems. Therefore, developing test case design and selection techniques that consider the imperfect characteristics of the oracle used for testing may be useful. Two promising techniques for designing test cases are metamorphic testing techniques and leveraging adversarial techniques to detect erroneous behavior (Mjeda and Botterweck, 2019) on which further research is needed. One of the challenges to be addressed is the identification of metamorphic relations that should be satisfied by an ML system (Kanewala and Bieman, 2014).

Coverage criteria designed for ML/DL models have been used to evaluate test cases (Ma et al., 2018a; Sun et al., 2018; Sun et al., 2019b; Du et al., 2019). However, the success of these coverage criteria is questionable (Li et al., 2019). Mutation testing (MuT) has been used to address the shortcomings of coverage criteria (Jia and Harman, 2010; Zhang et al., 2018c). Both areas, i.e., design of coverage criteria and use of MuT for ML systems, involve further research opportunities.

The economic aspect of testing ML systems should also be addressed. Techniques for designing optimum test case sets are needed to reduce the resource requirements for testing.

Regression testing of ML systems has not been addressed in the primary studies. Automation of regression testing and techniques for test case prioritization are potential essential issues for testing ML systems.

*Debugging tool support:* Debugging and profiling tool support is an essential deficiency for engineering ML systems (Zhang et al., 2019a). Although there are some approaches and prototype tools, such as DeepFault (Eniser et al., 2019), LAMP (Ma et al., 2017), MODE (Ma et al., 2018c), and DARVIZ (Sankaran et al., 2017), more mature debugging tools integrated with ML system development environments are required.

Open-source ML libraries include many bugs (Garcia et al., Pham et al., 2019; 2020; Wang et al., 2020b). Researchers analyzed and classified these bugs (Garcia et al., 2020). It may be worth developing bug detection tools using these bug classifications to increase the quality of ML libraries.

*Test management:* Planning, controlling, and monitoring of testing activities for ML systems have not been addressed by researchers. Some domains involving safety-critical ML systems, such as healthcare and autonomous driving, may be in urgent need of certification to reduce harmful risks.

*Test automation:* Automating tests has become a necessity to cope with large-scale complex systems (Mosley and Posey, 2002). More testing can be conducted via test automation to reduce the risks of ML systems. Therefore, new tools and techniques are required to automate testing for ML systems.

*Quality models for ML systems:* Current systems and software quality models (such as ISO/IEC 25010) should be revisited to address the different characteristics of ML systems, such as trustworthiness (Siebert et al., 2020) and fairness. As an example of the fulfillment of this need, the QA4AI Consortium was established in 2018 in Japan to discuss the quality assurance of ML systems (Fujii et al., 2020). A set of guidelines was published to ensure ML systems' quality (QA4AI, 2020). Rudraraju and Boyanapally, (2019) focused on developing a data quality model for ML systems. They identified new data quality attributes that are not present in the Data Quality model defined in the standard ISO/IEC 25012.

*Maintaining ML systems:* Data, ML models, and experiments have become essential configuration items in maintaining ML systems. Keeping versions of data and ML models require tools integrated with the development environment. Keeping track of experiments, reproducing, and comparing them on demand are open research areas for software engineers and tool developers.

Keeping pace with the latest versions of ML libraries and their dependencies is a challenge (Han et al., 2020). Tool support is required to maintain the ML libraries used by an ML system. Another maintainability issue related to ML libraries is the technical debt present in these libraries (Liu et al., 2020b). More research is required to address and decrease the technical debt in ML libraries.

To maintain the performance of an ML system above a level that meets customer expectations, drift detectors may be developed to identify data deviating from training data distribution (Stocco and Tonella, 2020). The identified data can be used for adaptive retraining.

*Tailoring and assessing SE process:* While there are many processes for ML lifecycle (Hulten, 2018; Rao, 2019; Shams, 2018) and traditional software development (Klünder et al., 2019), a set of harmonized practices for developing ML systems is required (Alvarez-Rodríguez et al., 2019; Belani et al., 2019; de Souza Nascimento et al., 2019). Blending the knowledge and experience in ML and SE areas and applying harmonized practices in industrial case studies seem to be a research direction. Researchers should develop DevOps practices for data and ML model components (like DataOps (Atwal, 2019; Ereth, 2018) and ModelOps (Hummer et al., 2019)) to keep ML systems up-to-date against rapidly changing external factors. Moreover, frameworks for process assessment can be another open area for research (Sculley et al., 2015).

*Managing ML systems development projects:* While there are some examples of using ML algorithms for project management (two examples in ML for SE by Nassif et al. (2016) and Pospieszny et al. (2018)), our primary study pool does not include any practices for managing ML systems development projects. Wan et al. only mentioned the difficulty of effort estimation in ML projects (Wan et al., 2019). Planning, controlling, and monitoring ML projects involve open research questions, such as effort estimation by Nemecek and Bemley (1993).

*Forming coherent teams:* Engineering an ML system involves skill sets from various fields, i.e., machine learning, deep learning, data science, mathematics, algorithms, software engineering, and statistics (Alshangiti et al., 2019; Alvarez-Rodríguez et al., 2019; Belani et al., 2019; Wan et al., 2019; Workera, 2020). Organizations need to determine the roles and responsibilities and associated required skillsets for engineering ML systems. Forming a coherent team from people with different backgrounds may bring new problems to tackle.

Table 8. Challenges and proposed solutions identified by researchers classified by knowledge area

| Knowledge Area | Challenge | Primary Study ID | Proposed Solution(s) |
|---|---|---|---|
| Requirements Engineering | Managing expectations of customers | Baier et al., 2019; Fujii et al., 2020; Ishikawa and Yoshioka, 2019; Kim et al., 2017; Nguyen-Duc et al., 2020; Vogelsang and Borg, 2019; Wan et al., 2019 | • inform customers about the benefits of ML systems even without being imperfect (Ishikawa and Yoshioka, 2019)<br>• explain the possibility of change in the level of success of ML systems on production in time (Wan et al., 2019) |
| | Eliciting and analyzing requirements | Amershi et al., 2019; Ishikawa and Yoshioka, 2019; Rahimi et al., 2019; Wan et al., 2019 | • state some requirements as hypotheses to be tested via experiments (Wan et al., 2019)<br>• refine requirements by means of experiments (Ishikawa and Yoshioka, 2019; Nguyen-Duc et al., 2020) |
| | Specifying requirements | Alvarez-Rodríguez et al., 2019; Baier et al., 2019; de Souza Nascimento et al., 2019; Horkoff, 2019; Kaindl and Ferdigg, 2020; Kawamoto, 2019; Kim et al., 2017; Nguyen-Duc et al., 2020; Sharma and Wehrheim, 2020b; Vogelsang and Borg, 2019; Wan et al., 2019 | • use a checklist to identify business metrics (de Souza Nascimento et al., 2019)<br>• use logical formulas to express classification performance, robustness, and fairness of ML classifiers (Kawamoto, 2019)<br>• extend Goal-oriented Requirements Engineering (GORE) (Kaindl and Ferdigg, 2020)<br>• use a specification language for specifying fairness requirements (Sharma and Wehrheim, 2020b) |
| | Dealing with new types of quality attributes | Baier et al., 2019; Horkoff, 2019; Hu et al., 2020; Ishikawa and Yoshioka, 2019; Nakamichi et al., 2020; Nguyen-Duc et al., 2020; Vogelsang and Borg, 2019 | • use a model and a measurement method to agree with customers on quality attributes (Nakamichi et al., 2020)<br>• use baselines for specifying quality attributes when possible (such as human performance for a safety-critical task) (Hu et al., 2020) |
| | Dealing with new types of conflicts between requirements | Horkoff, 2019 | - |
| | Dealing with changing emphasis on different types of requirements | Baier et al., 2019; Challa et al., 2020; Nguyen-Duc et al., 2020; Vogelsang and Borg, 2019 | • assess data quality during RE |
| Design | Designing for monitoring performance degradation on production | Wan et al., 2019; Yokoyama, 2019 | • use a specific software architectural pattern to deal with operational problems such as problem localization and rollback at failure (Yokoyama, 2019) |
| | Using new solution patterns for solving problems | Jahić and Roitsch, 2020; John et al., 2020a; John et al., 2020b; Scheerer et al., 2020; Wan et al., 2019; Washizaki et al., 2020 | • design effective solutions via experiments (Wan et al., 2019)<br>• develop and use a design patterns catalog (Washizaki et al., 2020)<br>• develop and use a catalog of architectural patterns (John et al., 2020a; John et al., 2020b) |

| | Dealing with high-volume data | de Souza Nascimento et al., 2019; Wan et al., 2019 | • use a checklist to validate a design model against high-volume data (de Souza Nascimento et al., 2019) |
|---|---|---|---|
| Software Development and Tools | Dealing with data | Alshangiti et al., 2019; Amershi et al., 2019; Baier et al., 2019; Correia et al., 2020; Fredriksson et al., 2020; Fujii et al., 2020; Hill et al., 2016; Islam, 2019; Kim et al., 2017; Lwakatare et al., 2019; Sankaran et al., 2017; Zhang et al., 2020a | • use data verification tools (Islam, 2019; Zhang et al., 2020a) |
| | Understanding ML algorithms, techniques, and libraries | Alshangiti et al., 2019; Hashemi et al., 2020; Islam, 2019; Zhang et al., 2019a; Zhang et al., 2020a | • produce better documentation for ML libraries (better explanations for methods and parameters, more tutorials covering various use cases, FAQ sections) (Alshangiti et al., 2019)<br>• use automated model and parameter recommendation tools (Islam, 2019)<br>• utilize resources such as Stack Overflow for knowledge acquisition via automatic documentation generation (Hashemi et al., 2020) |
| | Dealing with models | Amershi et al., 2019; Arpteg et al., 2018; Baier et al., 2019; Chakraborty et al., 2020a; Chen et al., 2020b; Gao et al., 2020a; Ghamizi et al., 2020; Hill et al., 2016; Kim et al., 2017; Paulsen et al., 2020; Pham et al., 2020; Sankaran et al., 2017; Tian et al., 2020; Wang et al., 2020a; Xie et al., 2019a; Yokoyama et al., 2020; Zhang et al., 2019b; Zhou et al., 2020b | • use a model-driven development based and platform agnostic framework to generate DL library specific (for TensorFlow, CAFFE, Theano, Torch, etc.) code (such as DARVIZ) (Sankaran et al., 2017)<br>• use a visual tool to develop DL models (such as DeepVisual) (Xie et al., 2019a)<br>• use a verification tool (such as NEURODIFF) to develop robust and resource-efficient ML systems involving compressed ML models (Paulsen et al., 2020)<br>• use a visual tool for supporting engineers in the understanding structure of neural network models (such as NeuralVis) (Zhang et al., 2019b)<br>• use a tool (such as EvalDNN) to assess a DL model based on various metrics (Tian et al., 2020)<br>• use a tool (such as FeatureNET) to generate and evaluate DL models (Ghamizi et al., 2020)<br>• consider an out-of-distribution module to develop more robust ML systems (Zhou et al., 2020b)<br>• use an approach (such as Dissector) for developing fault-tolerant ML systems by minimizing re-training effort (Wang et al., 2020a) |
| | | | • use data augmentation to enrich training data and obtain more robust ML models (Gao et al., 2020a; Yokoyama et al., 2020)<br>• use a method (such as Fairway) to remove ethical bias from training data and trained model (Chakraborty et al., 2020a) |
| | Dealing with dependencies | Hill et al., 2016; Lwakatare et al., 2019 | - |
| | Reusing models | Alvarez-Rodríguez et al., 2019; Amershi et al., 2019; Baier et al., 2019; Guo et al., 2019; Sankaran et al., 2017; Zhang et al., 2019a | - |
| | Dealing with the development environment, tools, and infrastructure | Alshangiti et al., 2019; Alvarez-Rodríguez et al., 2019; Amershi et al., 2019; Arpteg et al., 2018; Baier et al., 2019; Gao et al., 2020b; Gharibi et al., 2019; Guo et al., 2019; Kim et al., 2017; Lwakatare et al., 2019; Reimann and Kniesel-Wünsche, 2020; Sankaran et al., 2017; Zhang et al., 2020a | • use Docker images with all desired software pre-installed to prevent discrepancies among development, quality assurance, and production environments (Zhang et al., 2020a)<br>• develop and use extensions on current IDEs for ML system development, such as Azure ML for Visual Studio Code (Amershi et al., 2019)<br>• use a tool (such as DARVIZ) for model abstraction to provide interoperability across platforms (Sankaran et al., 2017)<br>• use unified APIs (ML API) hiding the complexity of ML libraries (Reimann and Kniesel-Wünsche, 2020)<br>• use a tool to visualize and compare experiments (such as ModelKB) (Gharibi et al., 2019)<br>• use a tool (such as DNNMem) to estimate GPU memory consumption to prevent waste of computing resources due to failures during model training (Gao et al., 2020b) |
| Testing and Quality | Designing test cases | Aggarwal et al., 2019; Ahmed et al., 2020; Barash et al., 2019; Berend et al., 2020; Braiek and Khomh, 2019; Byun and Rayadurgam, 2020; Chechik et al., 2019; Chen et al., 2020a; Du et al., 2020; Fujii et al., 2020; Gambi et al., 2019; Groce et al., 2013; Humbatova et al., 2020; Kang et al., 2020; Ma et al., 2019; Murphy et al., 2007; Riccio and Tonella, 2020; Santos et al., 2020; Sharma and Wehrheim, 2019; Sharma and Wehrheim, 2020a; Sharma and Wehrheim, 2020b; Shen et al., 2020; Sun et al., 2018; Sun et al., 2019a; Tian et al., 2018; Udeshi et al., 2018; Wan et al., 2019; Wang and Su, 2020; Xie et al., 2019b; Yan et al., 2019; Zhang et al., 2018a; Zhang et al., 2020b; Zhou et al., 2020a | • use a method/tool for testing DL models (Groce et al., 2013; Ma et al., 2019; Murphy et al., 2007), such as DeepEvolution (Braiek and Khomh, 2019), DeepCT (Ma et al., 2019), DeepConcolic (Sun et al., 2019a)<br>• utilize Metamorphic Testing to test DL systems (Santos et al., 2020; Wang and Su, 2020)<br>• use a method/tool to generate fault-revealing inputs (Berend et al., 2020; Byun and Rayadurgam, 2020; Kang et al., 2020), such as DeepJanus (Riccio and Tonella, 2020), Know the UnKnown (Zhang et al., 2020b)<br>• use a method/tool for generating test inputs for autonomous cars (Gambi et al., 2019; Tian et al., 2018; Xie et al., 2019b; Zhang et al., 2018a) |

| | | | • use an approach to generate discriminary test inputs to reveal fairness violations (Aggarwal et al., 2019; Sharma and Wehrheim, 2019; Udeshi et al., 2018)<br>• use a technique to verify ML model fairness (Sharma and Wehrheim, 2020b)<br>• apply verification-based testing to check monotonicity of generated models (Sharma and Wehrheim, 2020a)<br>• reduce test data volume using an approach, such as DeepReduce (Zhou et al., 2020a), PACE Chen et al., 2020a), Multiple-Boundary Clustering and Prioritization (Shen et al., 2020)<br>• improve testing efficiency by applying a method, such as Adaptive Random Testing for DL systems (ARTDL) (Yan et al., 2019) |
|---|---|---|---|
| | Evaluating test cases | Du et al., 2019; Dutta et al., 2020; Feng et al., 2020; Gerasimou et al., 2020; Guo et al., 2020; Harel-Canada et al., 2020; Ishikawa and Yoshioka, 2019; Jahangirova and Tonella, 2020; Khomh et al., 2018; Kim et al., 2019; Klampfl et al., 2020; Li et al., 2019a; Ma et al., 2018a; Ma et al., 2018b; Sekhon and Fleming, 2019; Sun et al., 2018; Sun et al., 2019b; Trujillo et al., 2020; Yan et al., 2020 | • use a coverage criterion appropriate for ML/DL models (Du et al., 2019; Kim et al., 2019; Ma et al., 2018a; Sekhon and Fleming, 2019; Sun et al., 2018; Sun et al., 2019b)<br>• consider other criteria when neuron coverage criterion is not sufficient, such as proposed in DeepGini (Feng et al., 2020), DeepImportance (Gerasimou et al., 2020)<br>• use a mutation adequacy score (Guo et al., 2020 Jahangirova and Tonella, 2020; Klampfl et al., 2020; Ma et al., 2018b)<br>• detect flaky tests using a technique such as FLASH (Dutta et al., 2020)<br>• measure the quality of test data (Ma et al., 2018b) |
| | Preparing test data | Arpteg et al., 2018; Guo et al., 2018; Haq et al., 2020; Knauss et al., 2017; Li et al., 2019b; Ma et al., 2018a; Ma et al., 2018b; Wan et al., 2019 | • use a tool (such as DLFuzz) to generate adversarial examples without manual labeling effort (Guo et al., 2018)<br>• develop an infrastructure for automated data collection and labeling (Knauss et al., 2017)<br>• consider using simulators to generate test data when appropriate (Haq et al., 2020) |
| | Executing tests | Gharibi et al., 2019; Haq et al., 2020; Zhang et al., 2019a | • use a tool to execute tests (such as ModelKB) (Gharibi et al., 2019)<br>• use a differential testing framework to detect potential inconsistent behavior of ML models on different settings (Zhang et al., 2019a) |
| | Evaluating test results | Arpteg et al., 2018; Barash et al., 2019; Chen et al., 2019; Cheng et al., 2018; Dwarakanath et al., | • use metamorphic relations to tackle with oracle problem (Dwarakanath et al., 2018; Nakajima, 2018; Sun and Zhou, |
| | | 2018; Ishikawa, 2018; Ishikawa and Yoshioka, 2019; Liem and Panichella, 2020; Murphy et al., 2007; Nakajima, 2018; Nishi et al., 2018; Qin et al., 2018; Stocco et al., 2020; Sun and Zhou, 2018; Tao et al., 2019; Wan et al., 2019; Xie et al., 2011; Zhang et al., 2018b; Zheng et al., 2019 | 2018; Xie et al., 2011); counter-evidence to use metamorphic relations (Cheng et al., 2018)<br>• use combinatorial testing to tackle with oracle problem (Barash et al., 2019; Chen et al., 2019)<br>• use a program synthesis technique to tackle with oracle problem (Qin et al., 2018) |
| | Debugging and fixing | Chakraborty et al., 2020b; Eniser et al., 2019; Garcia et al., 2020; Guo et al., 2019; Humbatova et al., 2020; Islam et al., 2020; Khomh et al., 2018; Li et al., 2020; Ma et al., 2017; Ma et al., 2018c; Pan, 2020; Pham et al., 2019; Sankaran et al., 2017; Tizpaz-Niari et al., 2020; Wan et al., 2019; Wang et al., 2020b; Zhang et al., 2018b; Zhang et al., 2019a; Zhang et al., 2020a | • use an approach/tool to debug and fix DL models, such as DeepFault (Eniser et al., 2019), LAMP (Ma et al., 2017), MODE (Ma et al., 2018c), DARVIZ (Sankaran et al., 2017)<br>• use an approach to find and localize bugs in DL libraries, such as CRADLE (Pham et al., 2019), LEMON (Wang et al., 2020b)<br>• use a tool (such as DPFuzz) to uncover performance bugs in ML frameworks (Tizpaz-Niari et al., 2020)<br>• compare the ML model with a model produced by a simple ML algorithm as a baseline and use the result as an indication for possible bugs in code and data (Wan et al., 2019) |
| | Managing tests | Alvarez-Rodríguez et al., 2019; Rivero et al., 2020 | • use test management tools and spreadsheets for managing test process in which various teams involve (Rivero et al., 2020) |
| | Automating tests | Knauss et al., 2017; Rivero et al., 2020; Tian et al., 2018 | • use a systematic technique to automate test case generation, such as DeepTest (Tian et al., 2018) |
| Maintenance and Configuration Management | Dealing with configuration management of data and ML models | Alvarez-Rodríguez et al., 2019; Amershi et al., 2019; Benton, 2020; Gharibi et al., 2019; Han et al., 2020; Liu et al., 2020b; Peng et al., 2020; Wan et al., 2019; Wu et al., 2020; Yokoyama, 2019 | • use a tool for data and ML model configuration management (such as ModelKB) (Gharibi et al., 2019)<br>• use Software Composition Analysis (SCA) tools to discover all related components of ML frameworks (Han et al., 2020)<br>• use a single platform (like Kubernetes) to manage all components (data pipelines, ML models, code logic, etc.) of an ML system (Benton, 2020) |
| | Dealing with the history of experiments | Arpteg et al., 2018; Hill et al., 2016; Lwakatare et al., 2019; Wan et al., 2019; Yokoyama, 2019 | • use a proper software architecture suitable for troubleshooting (Yokoyama, 2019) |
| | Dealing with re-training and re-deployment | Baier et al., 2019; Ren et al., 2020; Stocco and Tonella, 2020; Wan et al., 2019; Yokoyama, 2019 | • apply Canary release (Sato, 2014) approach for risky ML model re-deployments (Yokoyama, 2019)<br>• retrain ML models using input data deviating from the distribution of training data (Ren et al., 2020; Stocco and Tonella, 2020) |
| Software Engineering Process | Harmonizing the activities for developing ML components with software process | Alvarez-Rodríguez et al., 2019; Baier et al., 2019; Correia et al., 2020; de Souza Nascimento et al., 2019; Fujii et al., 2020; Liu et al., 2020a; Moreb et al., 2020; Munappy et al., 2020; Ozkaya, 2020; | • use checklists to perform ML-related activities in a standard way (de Souza Nascimento et al., 2019)<br>• integrate DataOps practices with software process (Munappy et al., 2020) |
| | | Serban et al., 2020; Wan et al., 2019; Wolf and Paine, 2020 | • benefit from ML systems engineering processes, an example from health informatics domain (Moreb et al., 2020) |
| | Assessing the ML process | Amershi et al., 2019; Lwakatare et al., 2019 | - |
| | Estimating effort | Arpteg et al., 2018; Wan et al., 2019 | - |
| Organiza-tional Aspects | Dealing with various skillsets | Alshangiti et al., 2019; Alvarez-Rodríguez et al., 2019; Correia et al., 2020; Kim et al., 2017; Wan et al., 2019 | - |
| | Building harmony among isolated roles | Amershi et al., 2019; Arpteg et al., 2018; Kim et al., 2017 | - |

## 6.2 OTHER IMPLICATIONS

*Industry-academia collaboration:* As we can see in Figure 10, this research topic is mostly dominated by academicians, although engineering ML systems is a highly practical area. The industrial involvement is of critical importance to enrich the literature and help mature this field. Therefore, encouraging researchers from the industry to collaborate with academia and share their experience in SE venues may empower structured information on this research topic.

*Formation of research groups:* ML systems engineering brings together two challenging research areas, i.e., SE and ML, and requires various skillsets. The formation of research groups with relevant skillsets could lead to better and more relevant results.

*Use of various research methods for all SE aspects:* As pointed out under RQ1, experiments are mostly used for testing and quality aspects. There are a few studies targeting development and maintenance. The paper pool does not include any study that uses experiment to address requirements- and design-related challenges. Experiments and case studies should be undertaken to examine requirements- and design-related challenges identified via interviews and surveys. Process, management, and organizational aspects should also be addressed using various research methods.

*Use of various application scenarios:* The answer to RQ2 indicates that researchers have mainly used classification problems for their experiments and case studies. It may be useful to use different ML problems, such as clustering, to find further challenges. Some of the concerns, such as some non-functional requirements, may be domain-specific (Nguyen-Duc et al., 2020) and can be discovered by exploring various ML problems.

## 6.3 POTENTIAL BENEFITS OF THIS REVIEW

Next, I will explore the potential benefits of this SLR for practitioners, researchers and academicians, and educators.

*For practitioners:* More and more software systems are involving ML capabilities. Practitioners should be aware of the new challenges that emerged with ML components since ML's influences on software systems are expected to affect their role significantly (Meade et al., 2019). This study provides an overview of these challenges and some suggestions for solutions (mostly not directly applicable but providing guidance).

*For researchers and academicians:* This SLR can be a valuable resource for future research on engineering ML systems' challenges. As Table 8 reveals, many challenges have not been addressed. Moreover, even for the difficulties addressed, the proposed solutions are mostly conceptual or implemented as a prototype tool without industrial use. Researchers and academicians can use the challenges reported in this study for initiating research projects. In addition, I believe that the bibliography provides a good base for understanding the current situation. The appendix section provides the list of the SE venues that involve papers on engineering ML systems.

*For educators:* The 2020 version of the Future of Jobs Survey reveals that the top increasingly strategic job roles are data analysts and scientists, AI and ML specialists, robotics engineers, software and application developers, and digital transformation specialists (World Economic Forum, 2020). The adaptation of university curricula based on this industrial need is therefore essential. Educators may use this study to either design new courses or adapt existing SE courses. One of the pioneer courses is the "Software Engineering for AI-Enabled Systems" course offered at Carnegie Mellon University (Kästner and Kang, 2020). The course takes a SE perspective on building software systems with a significant AI/ML component. It discusses how to take an idea and an ML model developed by data scientists and deploy it as part of a scalable and maintainable system.

## 6.4 LIMITATIONS AND POTENTIAL THREATS TO VALIDITY

The scope of this study is limited to the following parameters:

- Date: This study covers primary studies published until the end of 2020, i.e., 31 December 2020.
- Type of Literature: This study comprises studies published in peer-reviewed academic venues. Grey literature, e.g., papers only published in arxiv.org, blogs, videos, etc., have been excluded.
- Perspective: The primary studies have been selected from SE venues to reflect an SE perspective on ML systems engineering. Therefore, papers published in AI, ML, DL, Data Science, Data Management venues have been excluded.

Validity considerations are applicable for SMS and SLR studies, like empirical studies (Petersen et al., 2008; Petersen et al., 2015). The threats to this SLR's validity are mainly related to the specification of the candidate pool of papers, primary study selection bias, data extraction, and data synthesis.

The candidate pool of papers was specified by searching online databases using keywords. I used broad terms to form search keywords to reduce the risk of excluding potentially relevant studies. With this approach, I decreased precision and increased recall found more candidate papers to be assessed for specifying the final set of primary studies. I also searched for five widely used online databases in SMS and SLR studies in computer science and software engineering. Besides these five databases, I also searched Google Scholar to enrich the paper pool. To mitigate the risk of missing some relevant studies, I also carried out both backward and forward snowballing along with a manual search. On the other hand, selection of primary studies only from SE venues forms a threat to having a complete set of primary studies. There may be other relevant studies published in AI, ML, DL, Data Science, and Data Management venues. However, I believe that SE venues can provide a representative set to identify state of the art and challenges of engineering ML systems.

Personal bias might have been introduced during the application of inclusion and exclusion criteria. To minimize this type of bias and errors whenever I could not decide on inclusion/exclusion by reading the abstract, I scanned the full text for the final decision.

The validity of data extraction is another essential aspect that directly affects this study's results. In order to ensure the accuracy of the extracted data, I used existing categories in the literature where possible (RQ1, RQ2, and SE knowledge areas in RQ3). I aimed to decrease the risk of researcher bias by mapping the relevant data in primary studies to the specified groups. I applied the open coding technique iteratively and incrementally to identify the challenges and solutions (RQ3). This coding process potentially entails some researcher bias.

In general, primary study selection, data extraction, and synthesis are subject to researcher bias, and this bias may be higher than those SLRs with multiple authors. For instance, there is empirical evidence showing that single data extraction results in more errors than double data extraction (Buscemi et al., 2006). Despite its disadvantages, we can find SLRs with single authors in SE (Malhotra, 2015) and other research areas (Mitchell, 2011).

## 7 CONCLUSIONS AND FUTURE WORK

Advances in machine learning lead to a transition from the conventional view of software development, where algorithms are hard-coded by humans, to ML systems that are materialized through learning from data. Therefore, we need to rethink our ways of developing software systems and consider the particularities required by these new types of applications.

This study aims to systematically identify, analyze, summarize, and synthesize SE research's current state for engineering ML systems. Researchers have been showing an increasing interest in this research area since 2018 (Figure 9). Although there are a lot of primary studies (Section 9.1) and secondary studies (Section 3) on this topic, many research questions remain unanswered (Section 6.1 and Table 8). More cooperation between industry and academia and conducting more experiments using real-world problems would help to extend the SE body of knowledge for engineering ML systems. Moreover, reporting on lessons learned from action research in the industry can provide valuable insights for answering research questions.

I plan to conduct a multi-vocal literature review (Garousi et al., 2016) to identify more challenges and solution proposals reported by the industry in future work. Papers from AI, ML, data science, and data management venues can also be included in such a review. I plan to involve other researchers to decrease researcher bias and to cope with the increasing number of primary sources. Another future research plan is to elaborate on state-of-the-art in a specific dimension, such as requirements engineering for ML systems and organizational aspects of engineering ML systems.

## 8 APPENDIX: DISTRIBUTION OF THE PRIMARY STUDIES PER VENUE

| Venue | Number of Primary Studies | Reference(s) |
|---|---|---|
| ACM Joint Meeting on European Software Engineering Conference and Symposium on the Foundations of Software Engineering (ESEC/FSE) | 20 | (Aggarwal et al., 2019; Ahmed et al., 2020; Barash et al., 2019; Chakraborty et al., 2020a; Chen et al., 2020b; Du et al., 2019; Gambi et al., 2019; Gao et al., 2020b; Guo et al., 2018; Harel-Canada et al., 2020; Islam, 2019; Li et al., 2019b; Li et al., 2020; Ma et al., 2017; Ma et al., 2018c; Peng et al., 2020; Riccio and Tonella, 2020; Wang et al., 2020b; Yan et al., 2020; Zhang et al., 2019b) |
| IEEE/ACM International Conference on Software Engineering (ICSE) | 18 | (Amershi et al., 2019; Byun and Rayadurgam, 2020; Gao et al., 2020a; Garcia et al., 2020; Gerasimou et al., 2020; Humbatova et al., 2020; Islam et al., 2020; Kim et al., 2019; Li et al., 2019a; Liu et al., 2020b; Pham et al., 2019; Sankaran et al., 2017; Sekhon and Fleming, 2019; Stocco et al., 2020; Tian et al., 2018; Wang et al., 2020a; Zhang et al., 2020a; Zhang et al., 2020b) |
| ACM/IEEE International Conference on Automated Software Engineering (ASE) | 14 | (Berend et al., 2020; Chakraborty et al., 2020b; Du et al., 2020; Guo et al., 2019; Guo et al., 2019; Ma et al., 2018a; Paulsen et al., 2020; Pham et al., 2020; Shen et al., 2020; Sun et al., 2018; Udeshi et al., 2018; Wang and Su, 2020; Yokoyama et al., 2020; Zhang et al., 2018a) |
| ACM SIGSOFT International Symposium on Software Testing and Analysis | 7 | (Dutta et al., 2018; Dwarakanath et al., 2018; Feng et al., 2020; Sharma and Wehrheim, 2020a; Tizpaz-Niari et al., 2020; Xie et al., 2019b; Zhang et al., 2018b) |
| IEEE/ACM International Conference on Software Engineering - Companion (ICSE-Companion) | 7 | (Ghamizi et al., 2020; Knauss et al., 2017; Pan, 2020; Sun et al., 2019a; Sun et al., 2019b; Tian et al., 2020; Zheng et al., 2019) |
| IEEE International Conference on Software Testing, Verification and Validation (ICST) | 5 | (Chen et al., 2019; Haq et al., 2020; Jahangirova and Tonella, 2020; Nishi et al., 2018; Sharma and Wehrheim, 2019) |
| IEEE International Conference on Software Maintenance and Evolution (ICSME) | 4 | (Braiek and Khomh, 2019; Han et al., 2020; Ren et al., 2020; Washizaki et al., 2020) |
| IEEE/ACM International Conference on Software Engineering Workshops (ICSEW) | 4 | (Kang et al., 2020; Liem and Panichella, 2020; Trujillo et al., 2020; Wolf and Paine, 2020) |
| International Workshop on Artificial Intelligence for Requirements Engineering (AIRE) | 4 | (Challa et al., 2020; Hu et al., 2020; Kaindl and Ferdigg, 2020; Vogelsang and Borg, 2019) |
| ACM/IEEE International Symposium on Empirical Software Engineering and Measurement (ESEM) | 3 | (Alshangiti et al., 2019; de Souza Nascimento et al., 2019; Serban et al., 2020) |
| IEEE Access | 3 | (Moreb et al., 2020; Tao et al., 2019; Yan et al., 2019) |
| IEEE International Conference on Software Quality, Reliability and Security (QRS) | 3 | (Cheng et al., 2018; Klampfl et al., 2020; Qin et al., 2018) |
| IEEE Software | 3 | (Benton, 2020; Khomh et al., 2018; Ozkaya, 2020) |
| IEEE Transactions on Software Engineering | 3 | (Groce et al., 2013; Kim et al., 2017; Wan et al., 2019) |
| International Symposium on Software Reliability Engineering (ISSRE) | 3 | (Ma et al., 2018b; Zhang et al., 2019a; Zhou et al., 2020a) |
| Asia-Pacific Software Engineering Conference (APSEC) | 2 | (John et al., 2020b; Zhou et al., 2020b) |
| Brazilian Symposium on Software Quality (SBQS) | 2 | (Correia et al., 2020; Rivero et al., 2020) |
| Euromicro Conference on Software Engineering and Advanced Applications (SEAA) | 2 | (Arpteg et al., 2018; John et al., 2020a) |
| IEEE International Conference on Software Analysis, Evolution and Reengineering (SANER) | 2 | (Hashemi et al., 2020; Ma et al., 2019) |
| IEEE International Requirements Engineering Conference (RE) | 2 | (Horkoff, 2019; Nakamichi et al., 2020) |
| IEEE/ACM International Conference on Program Comprehension (ICPC) | 2 | (Wu et al., 2020; Xie et al., 2019a) |
| International Conference on Fundamental Approaches to Software Engineering (FASE) | 2 | (Chechik et al., 2019; Eniser et al., 2019) |
| International Conference on Product-Focused Software Process Improvement (PROFES) | 2 | (Fredriksson et al., 2020; Munappy et al., 2020) |
| ACM Transactions on Software Engineering and Methodology (TOSEM) | 1 | (Chen et al., 2020a) |
| Australasian Software Engineering Conference (ASWEC) | 1 | (Sun and Zhou, 2018) |
| Brazilian Symposium on Systematic and Automated Software Testing (SAST) | 1 | (Santos et al., 2020) |
| European Conference on Information Systems (ECIS) | 1 | (Baier et al., 2019) |
| European Conference on Software Architecture (ECSA) | 1 | (Jahić and Roitsch, 2020) |
| Evaluation and Assessment in Software Engineering (EASE) | 1 | (Nguyen-Duc et al., 2020) |
| IEEE International Conference on Software Architecture Companion (ICSA-C) | 1 | (Yokoyama, 2019) |
| IEEE International Requirements Engineering Conference Workshops (REW) | 1 | (Rahimi et al., 2019) |
| IEEE Symposium on Visual Languages and Human-Centric Computing (VL/HCC) | 1 | (Hill et al., 2016) |
| IFIP International Conference on Testing Software and Systems (ICTSS) | 1 | (Sharma and Wehrheim, 2020b) |
| INCOSE International Symposium | 1 | (Alvarez-Rodríguez et al., 2019) |
| International Conference on Agile Software Development (XP) | 1 | (Lwakatare et al., 2019) |
| International Conference on Art, Science, and Engineering of Programming | 1 | (Reimann and Kniesel-Wünsche, 2020) |
| International Conference on Conceptual Modeling (ER) | 1 | (Ishikawa, 2018) |
| International Conference on Software and System Processes (ICSSP) | 1 | (Liu et al., 2020a) |
| International Conference on Software Engineering & Knowledge Engineering (SEKE) | 1 | (Murphy et al., 2007) |
| International Conference on Software Engineering and Formal Methods (SEFM) | 1 | (Kawamoto, 2019) |
| International Journal of Software Engineering and Knowledge Engineering | 1 | (Fujii et al., 2020) |
| International Symposium on Software Engineering for Adaptive and Self-Managing Systems (SEAMS) | 1 | (Scheerer et al., 2020) |
| International Symposium on Software Reliability Engineering Workshops (ISSREW) | 1 | (Stocco and Tonella, 2020) |
| International Workshop on Realizing Artificial Intelligence Synergies in Software Engineering (RAISE) | 1 | (Gharibi et al., 2019) |
| International Workshop on Structured Object-Oriented Formal Language and Method (SOFL) | 1 | (Nakajima, 2018) |
| Joint International Workshop on Conducting Empirical Studies in Industry | 1 | (Ishikawa and Yoshioka, 2019) |

| and International Workshop on Software Engineering Research and Industrial Practice (CESSER-IP) | | |
|---|---|---|
| Journal of Systems and Software | 1 | (Xie et al., 2011) |

# 9 REFERENCES

This section is divided into two parts: (1) Citations to the primary studies reviewed in the SLR; and (2) Other (regular) references cited throughout the paper.

## 9.1 PRIMARY STUDIES (SOURCES REVIEWED IN THE SLR)

Aggarwal et al., 2019 Aggarwal, A., Lohia, P., Nagar, S., Dey, K., & Saha, D. (2019, August). Black box fairness testing of machine learning models. In Proceedings of the 2019 27th ACM Joint Meeting on European Software Engineering Conference and Symposium on the Foundations of Software Engineering (pp. 625-635).

Ahmed et al., 2020 Ahmed, M. S., Ishikawa, F., & Sugiyama, M. (2020, November). Testing machine learning code using polyhedral region. In Proceedings of the 28th ACM Joint Meeting on European Software Engineering Conference and Symposium on the Foundations of Software Engineering (pp. 1533-1536).

Alshangiti et al., 2019 Alshangiti, M., Sapkota, H., Murukannaiah, P. K., Liu, X., & Yu, Q. (2019, September). Why is Developing Machine Learning Applications Challenging? A Study on Stack Overflow Posts. In 2019 ACM/IEEE International Symposium on Empirical Software Engineering and Measurement (ESEM) (pp. 1-11). IEEE.

Alvarez-Rodríguez et al., 2019 Alvarez-Rodríguez, J. M., Zuñiga, R. M., Pelayo, V. M., & Llorens, J. (2019, July). Challenges and opportunities in the integration of the Systems Engineering process and the AI/ML model lifecycle. In INCOSE International Symposium (Vol. 29, No. 1, pp. 560-575).

Amershi et al., 2019 Amershi, S., Begel, A., Bird, C., DeLine, R., Gall, H., Kamar, E., ... & Zimmermann, T. (2019, May). Software engineering for machine learning: A case study. In 2019 IEEE/ACM 41st International Conference on Software Engineering: Software Engineering in Practice (ICSE-SEIP) (pp. 291-300). IEEE.

Arpteg et al., 2018 Arpteg, A., Brinne, B., Crnkovic-Friis, L., & Bosch, J. (2018, August). Software engineering challenges of deep learning. In 2018 44th Euromicro Conference on Software Engineering and Advanced Applications (SEAA) (pp. 50-59). IEEE.

Baier et al., 2019 Baier, L., Jöhren, F., & Seebacher, S. (2019). Challenges in the deployment and operation of machine learning in practice. In 2019 27th European Conference on Information Systems (ECIS).

Barash et al., 2019 Barash, G., Farchi, E., Jayaraman, I., Raz, O., Tzoref-Brill, R., & Zalmanovici, M. (2019, August). Bridging the gap between ML solutions and their business requirements using feature interactions. In Proceedings of the 2019 27th ACM Joint Meeting on European Software Engineering Conference and Symposium on the Foundations of Software Engineering (pp. 1048-1058).

Benton, 2020 Benton, W. C. (2020). Machine learning systems and intelligent applications. IEEE Software, 37(4), 43-49.

Berend et al., 2020 Berend, D., Xie, X., Ma, L., Zhou, L., Liu, Y., Xu, C., & Zhao, J. (2020, September). Cats are not fish: Deep learning testing calls for out-of-distribution awareness. In 2020 35th IEEE/ACM International Conference on Automated Software Engineering (ASE) (pp. 1041-1052). IEEE.

Braiek and Khomh, 2019 Braiek, H. B., & Khomh, F. (2019). DeepEvolution: A Search-Based Testing Approach for Deep Neural Networks. In 2019 IEEE International Conference on Software Maintenance and Evolution (ICSME) (pp. 454-458). IEEE.

Byun and Rayadurgam, 2020 Byun, T., & Rayadurgam, S. (2020, June). Manifold for machine learning assurance. In Proceedings of the ACM/IEEE 42nd International Conference on Software Engineering: New Ideas and Emerging Results (pp. 97-100).

Chakraborty et al., 2020a Chakraborty, J., Majumder, S., Yu, Z., & Menzies, T. (2020, November). Fairway: a way to build fair ML software. In Proceedings of the 28th ACM Joint Meeting on European Software Engineering Conference and Symposium on the Foundations of Software Engineering (pp. 654-665).

Chakraborty et al., 2020b Chakraborty, J., Peng, K., & Menzies, T. (2020, September). Making fair ML software using trustworthy explanation. In 2020 35th IEEE/ACM International Conference on Automated Software Engineering (ASE) (pp. 1229-1233). IEEE.

Challa et al., 2020 Challa, H., Niu, N., & Johnson, R. (2020, September). Faulty Requirements Made Valuable: On the Role of Data Quality in Deep Learning. In 2020 IEEE Seventh International Workshop on Artificial Intelligence for Requirements Engineering (AIRE) (pp. 61-69). IEEE.

Chechik et al., 2019 Chechik, M., Salay, R., Viger, T., Kokaly, S., & Rahimi, M. (2019, April). Software assurance in an uncertain world. In International Conference on Fundamental Approaches to Software Engineering (pp. 3-21). Springer, Cham.

Chen et al., 2019 Chen, Y., Wang, Z., Wang, D., Fang, C., & Chen, Z. (2019, April). Variable strength combinatorial testing for deep neural networks. In 2019 IEEE International Conference on Software Testing, Verification and Validation Workshops (ICSTW) (pp. 281-284). IEEE.

Chen et al., 2020a Chen, J., Wu, Z., Wang, Z., You, H., Zhang, L., & Yan, M. (2020). Practical Accuracy Estimation for Efficient Deep Neural Network Testing. ACM Transactions on Software Engineering and Methodology (TOSEM), 29(4), 1-35.

Chen et al., 2020b Chen, Z., Cao, Y., Liu, Y., Wang, H., Xie, T., & Liu, X. (2020, November). A comprehensive study on challenges in deploying deep learning based software. In Proceedings of the 28th ACM Joint Meeting on European Software Engineering Conference and Symposium on the Foundations of Software Engineering (pp. 750-762).

Cheng et al., 2018 Cheng, D., Cao, C., Xu, C., & Ma, X. (2018, July). Manifesting bugs in machine learning code: An explorative study with mutation testing. In 2018 IEEE International Conference on Software Quality, Reliability and Security (QRS) (pp. 313-324). IEEE.

Correia et al., 2020 Correia, J. L., Pereira, J. A., Mello, R., Garcia, A., Fonseca, B., Ribeiro, M., ... & Tiengo, W. (2020, December). Brazilian Data Scientists: Revealing their Challenges and Practices on Machine Learning Model Development. In 19th Brazilian Symposium on Software Quality (pp. 1-10).

de Souza Nascimento et al., 2019 de Souza Nascimento, E., Ahmed, I., Oliveira, E., Palheta, M. P., Steinmacher, I., & Conte, T. (2019, September). Understanding Development Process of Machine Learning Systems: Challenges and Solutions. In 2019 ACM/IEEE International Symposium on Empirical Software Engineering and Measurement (ESEM) (pp. 1-6). IEEE Computer Society.

Du et al., 2019 Du, X., Xie, X., Li, Y., Ma, L., Liu, Y., & Zhao, J. (2019, August). Deepstellar: Model-based quantitative analysis of stateful deep learning systems. In Proceedings of the 2019 27th ACM Joint Meeting on European Software Engineering Conference and Symposium on the Foundations of Software Engineering (pp. 477-487).

Du et al., 2020 Du, X., Li, Y., Xie, X., Ma, L., Liu, Y., & Zhao, J. (2020, September). Marble: Model-based Robustness Analysis of Stateful Deep Learning Systems. In 2020 35th IEEE/ACM International Conference on Automated Software Engineering (ASE) (pp. 423-435). IEEE.

Dutta et al., 2020 Dutta, S., Shi, A., Choudhary, R., Zhang, Z., Jain, A., & Misailovic, S. (2020, July). Detecting flaky tests in probabilistic and machine learning applications. In Proceedings of the 29th ACM SIGSOFT International Symposium on Software Testing and Analysis (pp. 211-224).

Dwarakanath et al., 2018 Dwarakanath, A., Ahuja, M., Sikand, S., Rao, R. M., Bose, R. J. C., Dubash, N., & Podder, S. (2018, July). Identifying implementation bugs in machine learning based image classifiers using metamorphic testing. In Proceedings of the 27th ACM SIGSOFT International Symposium on Software Testing and Analysis (pp. 118-128).

Eniser et al., 2019 Eniser, H. F., Gerasimou, S., & Sen, A. (2019, April). Deepfault: Fault localization for deep neural networks. In International Conference on Fundamental Approaches to Software Engineering (pp. 171-191). Springer, Cham.

Feng et al., 2020 Feng, Y., Shi, Q., Gao, X., Wan, J., Fang, C., & Chen, Z. (2020, July). DeepGini: prioritizing massive tests to enhance the robustness of deep neural networks. In Proceedings of the 29th ACM SIGSOFT International Symposium on Software Testing and Analysis (pp. 177-188).

Fredriksson et al., 2020 Fredriksson, T., Mattos, D. I., Bosch, J., & Olsson, H. H. (2020, November). Data Labeling: An Empirical Investigation into Industrial Challenges and Mitigation Strategies. In International Conference on Product-Focused Software Process Improvement (pp. 202-216). Springer, Cham.

Fujii et al., 2020 Fujii, G., Hamada, K., Ishikawa, F., Masuda, S., Matsuya, M., Myojin, T., Nishi, Y., Ogawa, H., Toku, T., Tokumoto, S., Kazunori, T. & Ujita, Y. (2020). Guidelines for Quality Assurance of Machine Learning-Based Artificial Intelligence. International Journal of Software Engineering and Knowledge Engineering, 1-18.

Gambi et al., 2019 Gambi, A., Huynh, T., & Fraser, G. (2019, August). Generating effective test cases for self-driving cars from police reports. In Proceedings of the 2019 27th ACM Joint Meeting on European Software Engineering Conference and Symposium on the Foundations of Software Engineering (pp. 257-267).

Gao et al., 2020a Gao, X., Saha, R. K., Prasad, M. R., & Roychoudhury, A. (2020, October). Fuzz testing based data augmentation to improve robustness of deep neural networks. In 2020 IEEE/ACM 42nd International Conference on Software Engineering (ICSE) (pp. 1147-1158). IEEE.

Gao et al., 2020b Gao, Y., Liu, Y., Zhang, H., Li, Z., Zhu, Y., Lin, H., & Yang, M. (2020, November). Estimating gpu memory consumption of deep learning models. In Proceedings of the 28th ACM Joint Meeting on European Software Engineering Conference and Symposium on the Foundations of Software Engineering (pp. 1342-1352).

Garcia et al., 2020 Garcia, Joshua, Yang Feng, Junjie Shen, Sumaya Almanee, Yuan Xia, and and Qi Alfred Chen. "A comprehensive study of autonomous vehicle bugs." In Proceedings of the ACM/IEEE 42nd International Conference on Software Engineering, pp. 385-396. 2020.

Gerasimou et al., 2020 Gerasimou, S., Eniser, H. F., Sen, A., & Cakan, A. (2020, October). Importance-driven deep learning system testing. In 2020 IEEE/ACM 42nd International Conference on Software Engineering (ICSE) (pp. 702-713). IEEE.

Ghamizi et al., 2020 Ghamizi, S., Cordy, M., Papadakis, M., & Traon, Y. L. (2020, June). FeatureNET: diversity-driven generation of deep learning models. In Proceedings of the ACM/IEEE 42nd International Conference on Software Engineering: Companion Proceedings (pp. 41-44).

Gharibi et al., 2019 Gharibi, G., Walunj, V., Rella, S., & Lee, Y. (2019, May). ModelKB: towards automated management of the modeling lifecycle in deep learning. In 2019 IEEE/ACM 7th International Workshop on Realizing Artificial Intelligence Synergies in Software Engineering (RAISE) (pp. 28-34). IEEE.

Groce et al., 2013 Groce, A., Kulesza, T., Zhang, C., Shamasunder, S., Burnett, M., Wong, W. K., Stumpf, S., Das, S., Shinsel, A., Bice, F. & McIntosh, K. (2013). You are the only possible oracle: Effective test selection for end users of interactive machine learning systems. IEEE Transactions on Software Engineering, 40(3), 307-323.

Guo et al., 2018 Guo, J., Jiang, Y., Zhao, Y., Chen, Q., & Sun, J. (2018, October). DIfuzz: Differential fuzzing testing of deep learning systems. In Proceedings of the 2018 26th ACM Joint Meeting on European Software Engineering Conference and Symposium on the Foundations of Software Engineering (pp. 739-743).

Guo et al., 2019 Guo, Q., Chen, S., Xie, X., Ma, L., Hu, Q., Liu, H., Liu, Y., Zhao, J. & Li, X. (2019, November). An empirical study towards characterizing deep learning development and deployment across different frameworks and platforms. In 2019 34th IEEE/ACM International Conference on Automated Software Engineering (ASE) (pp. 810-822). IEEE.

Guo et al., 2020 Guo, Q., Xie, X., Li, Y., Zhang, X., Liu, Y., Li, X., & Shen, C. (2020, September). Audee: Automated Testing for Deep Learning Frameworks. In 2020 35th IEEE/ACM International Conference on Automated Software Engineering (ASE) (pp. 486-498). IEEE.

Han et al., 2020 Han, J., Deng, S., Lo, D., Zhi, C., Yin, J., & Xia, X. (2020, September). An empirical study of the dependency networks of deep learning libraries. In 2020 IEEE International Conference on Software Maintenance and Evolution (ICSME) (pp. 868-878). IEEE.

Haq et al., 2020 Haq, F. U., Shin, D., Nejati, S., & Briand, L. C. (2020, October). Comparing offline and online testing of deep neural networks: An autonomous car case study. In 2020 IEEE 13th International Conference on Software Testing, Validation and Verification (ICST) (pp. 85-95). IEEE.

Harel-Canada et al., 2020 Harel-Canada, F., Wang, L., Gulzar, M. A., Gu, Q., & Kim, M. (2020, November). Is neuron coverage a meaningful measure for testing deep neural networks?. In Proceedings of the 28th ACM Joint Meeting on European Software Engineering Conference and Symposium on the Foundations of Software Engineering (pp. 851-862).

Hashemi et al., 2020 Hashemi, Y., Nayebi, M., & Antoniol, G. (2020, February). Documentation of Machine Learning Software. In 2020 IEEE 27th International Conference on Software Analysis, Evolution and Reengineering (SANER) (pp. 666-667). IEEE.

Hill et al., 2016 Hill, C., Bellamy, R., Erickson, T., & Burnett, M. (2016, September). Trials and tribulations of developers of intelligent systems: A field study. In 2016 IEEE Symposium on Visual Languages and Human-Centric Computing (VL/HCC) (pp. 162-170). IEEE.

Horkoff, 2019 Horkoff, J. (2019, September). Non-functional requirements for machine learning: Challenges and new directions. In 2019 IEEE 27th International Requirements Engineering Conference (RE) (pp. 386-391). IEEE.

Hu et al., 2020 Hu, B. C., Salay, R., Czarnecki, K., Rahimi, M., Selim, G., & Chechik, M. (2020, September). Towards Requirements Specification for Machine-learned Perception Based on Human Performance. In 2020 IEEE Seventh International Workshop on Artificial Intelligence for Requirements Engineering (AIRE) (pp. 48-51). IEEE.

Humbatova et al., 2020 Humbatova, N., Jahangirova, G., Bavota, G., Riccio, V., Stocco, A., & Tonella, P. (2020, June). Taxonomy of real faults in deep learning systems. In Proceedings of the ACM/IEEE 42nd International Conference on Software Engineering (pp. 1110-1121).

Ishikawa and Yoshioka, 2019 Ishikawa, F., & Yoshioka, N. (2019, May). How do engineers perceive difficulties in engineering of machine-learning systems?-Questionnaire survey. In 2019 IEEE/ACM Joint 7th International Workshop on Conducting Empirical Studies in Industry (CESI) and 6th International Workshop on Software Engineering Research and Industrial Practice (SER&IP) (pp. 2-9). IEEE.

Ishikawa, 2018 Ishikawa, F. (2018, October). Concepts in quality assessment for machine learning-from test data to arguments. In International Conference on Conceptual Modeling (pp. 536-544). Springer, Cham.

Islam et al., 2020 Islam, M. J., Pan, R., Nguyen, G., & Rajan, H. (2020, October). Repairing deep neural networks: Fix patterns and challenges. In 2020 IEEE/ACM 42nd International Conference on Software Engineering (ICSE) (pp. 1135-1146). IEEE.

Islam, 2019 Islam, M. J., Nguyen, G., Pan, R., & Rajan, H. (2019, August). A comprehensive study on deep learning bug characteristics. In Proceedings of the 2019 27th ACM Joint Meeting on European Software Engineering Conference and Symposium on the Foundations of Software Engineering (pp. 510-520).

Jahangirova and Tonella, 2020 Jahangirova, G., & Tonella, P. (2020, October). An empirical evaluation of mutation operators for deep learning systems. In 2020 IEEE 13th International Conference on Software Testing, Validation and Verification (ICST) (pp. 74-84). IEEE.

Jahić and Roitsch, 2020 Jahić, J., & Roitsch, R. (2020, September). State of the Practice Survey: Predicting the Influence of AI Adoption on System Software Architecture in Traditional Embedded Systems. In European Conference on Software Architecture (pp. 155-169). Springer, Cham.

John et al., 2020a John, M. M., Olsson, H. H., & Bosch, J. (2020, August). AI on the Edge: Architectural Alternatives. In 2020 46th Euromicro Conference on Software Engineering and Advanced Applications (SEAA) (pp. 21-28). IEEE.

John et al., 2020b John, M. M., Olsson, H. H., & Bosch, J. (2020, December). AI Deployment Architecture: Multi-Case Study for Key Factor Identification. In 2020 27th Asia-Pacific Software Engineering Conference (APSEC) (pp. 395-404). IEEE.

Kaindl and Ferdigg, 2020 Kaindl, H., & Ferdigg, J. (2020, September). Towards an Extended Requirements Problem Formulation for Superintelligence Safety. In 2020 IEEE Seventh International Workshop on Artificial Intelligence for Requirements Engineering (AIRE) (pp. 33-38). IEEE.

Kang et al., 2020 Kang, S., Feldt, R., & Yoo, S. (2020, June). SINVAD: Search-based Image Space Navigation for DNN Image Classifier Test Input Generation. In Proceedings of the IEEE/ACM 42nd International Conference on Software Engineering Workshops (pp. 521-528).

Kawamoto, 2019 Kawamoto, Y. (2019, September). Towards logical specification of statistical machine learning. In International Conference on Software Engineering and Formal Methods (pp. 293-311). Springer, Cham.

Khomh et al., 2018 Khomh, F., Adams, B., Cheng, J., Fokaefs, M., & Antoniol, G. (2018). Software engineering for machine-learning applications: The road ahead. IEEE Software, 35(5), 81-84.

Kim et al., 2017 Kim, M., Zimmermann, T., DeLine, R., & Begel, A. (2017). Data scientists in software teams: State of the art and challenges. IEEE Transactions on Software Engineering, 44(11), 1024-1038.

Kim et al., 2019 Kim, J., Feldt, R., & Yoo, S. (2019, May). Guiding deep learning system testing using surprise adequacy. In 2019 IEEE/ACM 41st International Conference on Software Engineering (ICSE) (pp. 1039-1049). IEEE.

Klampfl et al., 2020 Klampfl, L., Chetouane, N., & Wotawa, F. (2020, December). Mutation Testing for Artificial Neural Networks: An Empirical Evaluation. In 2020 IEEE 20th International Conference on Software Quality, Reliability and Security (QRS) (pp. 356-365). IEEE.

Knauss et al., 2017 Knauss, A., Schroder, J., Berger, C., & Eriksson, H. (2017, May). Software-related challenges of testing automated vehicles. In 2017 IEEE/ACM 39th International Conference on Software Engineering Companion (ICSE-C) (pp. 328-330). IEEE.

Li et al., 2019a Li, Z., Ma, X., Xu, C., & Cao, C. (2019, May). Structural coverage criteria for neural networks could be misleading. In 2019 IEEE/ACM 41st International Conference on Software Engineering: New Ideas and Emerging Results (ICSE-NIER) (pp. 89-92). IEEE.

Li et al., 2019b Li, Z., Ma, X., Xu, C., Cao, C., Xu, J., & Lü, J. (2019, August). Boosting operational dnn testing efficiency through conditioning. In Proceedings of the 2019 27th ACM Joint Meeting on European Software Engineering Conference and Symposium on the Foundations of Software Engineering (pp. 499-509).

Li et al., 2020 Li, Z., Ma, X., Xu, C., Xu, J., Cao, C., & Lü, J. (2020, November). Operational calibration: Debugging confidence errors for DNNs in the field. In Proceedings of the 28th ACM Joint Meeting on European Software Engineering Conference and Symposium on the Foundations of Software Engineering (pp. 901-913).

Liem and Panichella, 2020 Liem, C. C., & Panichella, A. (2020, June). Oracle Issues in Machine Learning and Where to Find Them. In Proceedings of the IEEE/ACM 42nd International Conference on Software Engineering Workshops (pp. 483-488).

Liu et al., 2020a Liu, H., Eksmo, S., Risberg, J., & Hebig, R. (2020, June). Emerging and changing tasks in the development process for machine learning systems. In Proceedings of the international conference on software and system processes (pp. 125-134).

Liu et al., 2020b Liu, J., Huang, Q., Xia, X., Shihab, E., Lo, D., & Li, S. (2020, June). Is using deep learning frameworks free? characterizing technical debt in deep learning frameworks. In Proceedings of the ACM/IEEE 42nd International Conference on Software Engineering: Software Engineering in Society (pp. 1-10).

Lwakatare et al., 2019 Lwakatare, L. E., Raj, A., Bosch, J., Olsson, H. H., & Crnkovic, I. (2019, May). A taxonomy of software engineering challenges for machine learning systems: An empirical investigation. In International Conference on Agile Software Development (pp. 227-243). Springer, Cham.

Ma et al., 2017 Ma, S., Aafer, Y., Xu, Z., Lee, W. C., Zhai, J., Liu, Y., & Zhang, X. (2017, August). LAMP: data provenance for graph based machine learning algorithms through derivative computation. In Proceedings of the 2017 11th Joint Meeting on Foundations of Software Engineering (pp. 786-797).

Ma et al., 2018a Ma, L., Juefei-Xu, F., Zhang, F., Sun, J., Xue, M., Li, B., Chen, C., Su, T., Li, L., Liu, Y. & Zhao, J. (2018, September). Deepgauge: Multi-granularity testing criteria for deep learning systems. In Proceedings of the 33rd ACM/IEEE International Conference on Automated Software Engineering (pp. 120-131).

Ma et al., 2018b Ma, L., Zhang, F., Sun, J., Xue, M., Li, B., Juefei-Xu, F., Xie, C., Li, L., Liu, Y., Zhao, J. & Wang, Y. (2018, October). Deepmutation: Mutation testing of deep learning systems. In 2018 IEEE 29th International Symposium on Software Reliability Engineering (ISSRE) (pp. 100-111). IEEE.

Ma et al., 2018c Ma, S., Liu, Y., Lee, W. C., Zhang, X., & Grama, A. (2018, October). MODE: automated neural network model debugging via state differential analysis and input selection. In Proceedings of the 2018 26th ACM Joint Meeting on European Software Engineering Conference and Symposium on the Foundations of Software Engineering (pp. 175-186).

Ma et al., 2019 Ma, L., Juefei-Xu, F., Xue, M., Li, B., Li, L., Liu, Y., & Zhao, J. (2019, February). Deepct: Tomographic combinatorial testing for deep learning systems. In 2019 IEEE 26th International Conference on Software Analysis, Evolution and Reengineering (SANER) (pp. 614-618). IEEE.

Moreb et al., 2020 Moreb, M., Mohammed, T. A., & Bayat, O. (2020). A novel software engineering approach toward using machine learning for improving the efficiency of health systems. IEEE Access, 8, 23169-23178.

Munappy et al., 2020 Munappy, A. R., Bosch, J., & Olsson, H. H. (2020, November). Data Pipeline Management in Practice: Challenges and Opportunities. In International Conference on Product-Focused Software Process Improvement (pp. 168-184). Springer, Cham.

Murphy et al., 2007 Murphy, C., Kaiser, G. E., & Arias, M. (2007, July). An Approach to Software Testing of Machine Learning Applications. In 2007 Proceedings of the Nineteenth International Conference on Software Engineering & Knowledge Engineering (SEKE'2007) (p. 167).

Nakajima, 2018 Nakajima, S. (2018, November). Dataset diversity for metamorphic testing of machine learning software. In International Workshop on Structured Object-Oriented Formal Language and Method (pp. 21-38). Springer, Cham.

Nakamichi et al., 2020 Nakamichi, K., Ohashi, K., Namba, I., Yamamoto, R., Aoyama, M., Joeckel, L., Siebert, J. & Heidrich, J. (2020, August). Requirements-driven method to determine quality characteristics and measurements for machine learning software and its evaluation. In 2020 IEEE 28th International Requirements Engineering Conference (RE) (pp. 260-270). IEEE.

Nguyen-Duc et al., 2020 Nguyen-Duc, A., Sundbø, I., Nascimento, E., Conte, T., Ahmed, I., & Abrahamsson, P. (2020). A Multiple Case Study of Artificial Intelligent System Development in Industry. In Proceedings of the Evaluation and Assessment in Software Engineering (pp. 1-10).

Nishi et al., 2018 Nishi, Y., Masuda, S., Ogawa, H., & Uetsuki, K. (2018, April). A test architecture for machine learning product. In 2018 IEEE International Conference on Software Testing, Verification and Validation Workshops (ICSTW) (pp. 273-278). IEEE.

Ozkaya, 2020 Ozkaya, I. (2020). What Is Really Different in Engineering AI-Enabled Systems?. IEEE Software, 37(4), 3-6.

Pan, 2020 Pan, R. (2020, October). Does fixing bug increase robustness in deep learning?. In 2020 IEEE/ACM 42nd International Conference on Software Engineering: Companion Proceedings (ICSE-Companion) (pp. 146-148). IEEE.

Paulsen et al., 2020 Paulsen, B., Wang, J., Wang, J., & Wang, C. (2020, September). NeuroDiff: Scalable Differential Verification of Neural Networks using Fine-Grained Approximation. In 2020 35th IEEE/ACM International Conference on Automated Software Engineering (ASE) (pp. 784-796). IEEE.

Peng et al., 2020 Peng, Z., Yang, J., Chen, T. H., & Ma, L. (2020, November). A first look at the integration of machine learning models in complex autonomous driving systems: a case study on Apollo. In Proceedings of the 28th ACM Joint Meeting on European Software Engineering Conference and Symposium on the Foundations of Software Engineering (pp. 1240-1250).

Pham et al., 2019 Pham, H. V., Lutellier, T., Qi, W., & Tan, L. (2019, May). CRADLE: cross-backend validation to detect and localize bugs in deep learning libraries. In 2019 IEEE/ACM 41st International Conference on Software Engineering (ICSE) (pp. 1027-1038). IEEE.

Pham et al., 2020 Pham, H. V., Qian, S., Wang, J., Lutellier, T., Rosenthal, J., Tan, L., Yu Y & Nagappan, N. (2020, September). Problems and Opportunities in Training Deep Learning Software Systems: An Analysis of Variance. In 2020 35th IEEE/ACM International Conference on Automated Software Engineering (ASE) (pp. 771-783). IEEE.

Qin et al., 2018 Qin, Y., Wang, H., Xu, C., Ma, X., & Lu, J. (2018, July). SynEva: Evaluating ml programs by mirror program synthesis. In 2018 IEEE International Conference on Software Quality, Reliability and Security (QRS) (pp. 171-182). IEEE.

Rahimi et al., 2019 Rahimi, M., Guo, J. L., Kokaly, S., & Chechik, M. (2019, September). Toward Requirements Specification for Machine-Learned Components. In 2019 IEEE 27th International Requirements Engineering Conference Workshops (REW) (pp. 241-244). IEEE.

Reimann and Kniesel-Wünsche, 2020 Reimann, L., & Kniesel-Wünsche, G. (2020, March). Achieving guidance in applied machine learning through software engineering techniques. In Conference Companion of the 4th International Conference on Art, Science, and Engineering of Programming (pp. 7-12).

Ren et al., 2020 Ren, X., Yu, B., Qi, H., Juefei-Xu, F., Li, Z., Xue, W., Ma, L. & Zhao, J. (2020, September). Few-Shot Guided Mix for DNN Repairing. In 2020 IEEE International Conference on Software Maintenance and Evolution (ICSME) (pp. 717-721). IEEE.

Riccio and Tonella, 2020 Riccio, V., & Tonella, P. (2020, November). Model-based exploration of the frontier of behaviours for deep learning system testing. In Proceedings of the 28th ACM Joint Meeting on European Software Engineering Conference and Symposium on the Foundations of Software Engineering (pp. 876-888).

Rivero et al., 2020 Rivero, L., Diniz, J., Silva, G., Borralho, G., Braz Junior, G., Paiva, A., ... & Oliveira, M. (2020, December). Deployment of a Machine Learning System for Predicting Lawsuits Against Power Companies: Lessons Learned from an Agile Testing Experience for Improving Software Quality. In 19th Brazilian Symposium on Software Quality (pp. 1-10).

Sankaran et al., 2017 Sankaran, A., Aralikatte, R., Mani, S., Khare, S., Panwar, N., & Gantayat, N. (2017, May). DARVIZ: deep abstract representation, visualization, and verification of deep learning models. In 2017 IEEE/ACM 39th International Conference on Software Engineering: New Ideas and Emerging Technologies Results Track (ICSE-NIER) (pp. 47-50). IEEE.

Santos et al., 2020 Santos, S. H., da Silveira, B. N. C., Andrade, S. A., Delamaro, M., & Souza, S. R. (2020, October). An Experimental Study on Applying Metamorphic Testing in Machine Learning Applications. In Proceedings of the 5th Brazilian Symposium on Systematic and Automated Software Testing (pp. 98-106).

Scheerer et al., 2020 Scheerer, M., Klamroth, J., Reussner, R., & Beckert, B. (2020, June). Towards classes of architectural dependability assurance for machine-learning-based systems. In Proceedings of the IEEE/ACM 15th International Symposium on Software Engineering for Adaptive and Self-Managing Systems (pp. 31-37).

Sekhon and Fleming, 2019 Sekhon, J., & Fleming, C. (2019, May). Towards improved testing for deep learning. In 2019 IEEE/ACM 41st International Conference on Software Engineering: New Ideas and Emerging Results (ICSE-NIER) (pp. 85-88). IEEE.

Serban et al., 2020 Serban, A., van der Blom, K., Hoos, H., & Visser, J. (2020, October). Adoption and effects of software engineering best practices in machine learning. In Proceedings of the 14th ACM/IEEE International Symposium on Empirical Software Engineering and Measurement (ESEM) (pp. 1-12).

Sharma and Wehrheim, 2019 Sharma, A., & Wehrheim, H. (2019, April). Testing machine learning algorithms for balanced data usage. In 2019 12th IEEE Conference on Software Testing, Validation and Verification (ICST) (pp. 125-135). IEEE.

Sharma and Wehrheim, 2020a Sharma, A., & Wehrheim, H. (2020, July). Higher income, larger loan? Monotonicity testing of machine learning models. In Proceedings of the 29th ACM SIGSOFT International Symposium on Software Testing and Analysis (pp. 200-210).

Sharma and Wehrheim, 2020b Sharma, A., & Wehrheim, H. (2020, December). Automatic Fairness Testing of Machine Learning Models. In IFIP International Conference on Testing Software and Systems (pp. 255-271). Springer, Cham.

Shen et al., 2020 Shen, W., Li, Y., Chen, L., Han, Y., Zhou, Y., & Xu, B. (2020, September). Multiple-Boundary Clustering and Prioritization to Promote Neural Network Retraining. In 2020 35th IEEE/ACM International Conference on Automated Software Engineering (ASE) (pp. 410-422). IEEE.

Stocco and Tonella, 2020 Stocco, A., & Tonella, P. (2020, October). Towards Anomaly Detectors that Learn Continuously. In 2020 IEEE International Symposium on Software Reliability Engineering Workshops (ISSREW) (pp. 201-208). IEEE.

Stocco et al., 2020 Stocco, A., Weiss, M., Calzana, M., & Tonella, P. (2020, June). Misbehaviour prediction for autonomous driving systems. In Proceedings of the ACM/IEEE 42nd International Conference on Software Engineering (pp. 359-371).

Sun and Zhou, 2018 Sun, L., & Zhou, Z. Q. (2018, November). Metamorphic testing for machine translations: MT4MT. In 2018 25th Australasian Software Engineering Conference (ASWEC) (pp. 96-100). IEEE.

Sun et al., 2018 Sun, Y., Wu, M., Ruan, W., Huang, X., Kwiatkowska, M., & Kroening, D. (2018, September). Concolic testing for deep neural networks. In Proceedings of the 33rd ACM/IEEE International Conference on Automated Software Engineering (pp. 109-119).

Sun et al., 2019a Sun, Y., Huang, X., Kroening, D., Sharp, J., Hill, M., & Ashmore, R. (2019, May). DeepConcolic: testing and debugging deep neural networks. In 2019 IEEE/ACM 41st International Conference on Software Engineering: Companion Proceedings (ICSE-Companion) (pp. 111-114). IEEE.

Sun et al., 2019b Sun, Y., Huang, X., Kroening, D., Sharp, J., Hill, M., & Ashmore, R. (2019, May). Structural Test Coverage Criteria for Deep Neural Networks. In 2019 IEEE/ACM 41st International Conference on Software Engineering: Companion Proceedings (ICSE-Companion) (pp. 320-321). IEEE.

Tao et al., 2019 Tao, C., Gao, J., & Wang, T. (2019). Testing and Quality Validation for AI Software–Perspectives, Issues, and Practices. IEEE Access, 7, 120164-120175.

Tian et al., 2018 Tian, Y., Pei, K., Jana, S., & Ray, B. (2018, May). Deeptest: Automated testing of deep-neural-network-driven autonomous cars. In Proceedings of the 40th international conference on software engineering (pp. 303-314).

Tian et al., 2020 Tian, Y., Zeng, Z., Wen, M., Liu, Y., Kuo, T. Y., & Cheung, S. C. (2020, October). EvalDNN: a toolbox for evaluating deep neural network models. In 2020 IEEE/ACM 42nd International Conference on Software Engineering: Companion Proceedings (ICSE-Companion) (pp. 45-48). IEEE.

Tizpaz-Niari et al., 2020 Tizpaz-Niari S., Černý, P., & Trivedi, A. (2020, July). Detecting and understanding real-world differential performance bugs in machine learning libraries. In Proceedings of the 29th ACM SIGSOFT International Symposium on Software Testing and Analysis (pp. 189-199).

Trujillo et al., 2020 Trujillo, M., Linares-Vásquez, M., Escobar-Velásquez, C., Dusparic, I., & Cardozo, N. (2020, June). Does Neuron Coverage Matter for Deep Reinforcement Learning? A Preliminary Study. In Proceedings of the IEEE/ACM 42nd International Conference on Software Engineering Workshops (pp. 215-220).

Udeshi et al., 2018 Udeshi, S., Arora, P., & Chattopadhyay, S. (2018, September). Automated directed fairness testing. In Proceedings of the 33rd ACM/IEEE International Conference on Automated Software Engineering (pp. 98-108).

Vogelsang and Borg, 2019 Vogelsang, A., & Borg, M. (2019, September). Requirements Engineering for Machine Learning: Perspectives from Data Scientists. In 2019 IEEE 27th International Requirements Engineering Conference Workshops (REW) (pp. 245-251). IEEE.

Wan et al., 2019 Wan, Z., Xia, X., Lo, D., & Murphy, G. C. (2019). How does Machine Learning Change Software Development Practices?. IEEE Transactions on Software Engineering.

Wang and Su, 2020 Wang, S., & Su, Z. (2020, September). Metamorphic Object Insertion for Testing Object Detection Systems. In 2020 35th IEEE/ACM International Conference on Automated Software Engineering (ASE) (pp. 1053-1065). IEEE.

Wang et al., 2020a Wang, H., Xu, J., Xu, C., Ma, X., & Lu, J. (2020, October). Dissector: Input validation for deep learning applications by crossing-layer dissection. In 2020 IEEE/ACM 42nd International Conference on Software Engineering (ICSE) (pp. 727-738). IEEE.

Wang et al., 2020b Wang, Z., Yan, M., Chen, J., Liu, S., & Zhang, D. (2020, November). Deep learning library testing via effective model generation. In Proceedings of the 28th ACM Joint Meeting on European Software Engineering Conference and Symposium on the Foundations of Software Engineering (pp. 788-799).

Washizaki et al., 2020 Washizaki, H., Takeuchi, H., Khomh, F., Natori, N., Doi, T., & Okuda, S. (2020, September). Practitioners' insights on machine-learning software engineering design patterns: a preliminary study. In 2020 IEEE International Conference on Software Maintenance and Evolution (ICSME) (pp. 797-799). IEEE.

Wolf and Paine, 2020 Wolf, C. T., & Paine, D. (2020, June). Sensemaking Practices in the Everyday Work of AI/ML Software Engineering. In Proceedings of the IEEE/ACM 42nd International Conference on Software Engineering Workshops (pp. 86-92).

Wu et al., 2020 Wu, X., Qin, L., Yu, B., Xie, X., Ma, L., Xue, Y., ... & Zhao, J. (2020, July). How are Deep Learning Models Similar? An Empirical Study on Clone Analysis of Deep Learning Software. In Proceedings of the 28th International Conference on Program Comprehension (pp. 172-183).

Xie et al., 2011 Xie, X., Ho, J. W., Murphy, C., Kaiser, G., Xu, B., & Chen, T. Y. (2011). Testing and validating machine learning classifiers by metamorphic testing. Journal of Systems and Software, 84(4), 544-558.

Xie et al., 2019a Xie, C., Qi, H., Ma, L., & Zhao, J. (2019, May). DeepVisual: a visual programming tool for deep learning systems. In 2019 IEEE/ACM 27th International Conference on Program Comprehension (ICPC) (pp. 130-134). IEEE.

Xie et al., 2019b Xie, X., Ma, L., Juefei-Xu, F., Xue, M., Chen, H., Liu, Y., Zhao, J., Li, B., Yin, J. & See, S. (2019, July). Deephunter: A coverage-guided fuzz testing framework for deep neural networks. In Proceedings of the 28th ACM SIGSOFT International Symposium on Software Testing and Analysis (pp. 146-157).

Yan et al., 2019 Yan, M., Wang, L., & Fei, A. (2019). ARTDL: Adaptive random testing for deep learning systems. IEEE Access, 8, 3055-3064.

Yan et al., 2020 Yan, S., Tao, G., Liu, X., Zhai, J., Ma, S., Xu, L., & Zhang, X. (2020, November). Correlations between deep neural network model coverage criteria and model quality. In Proceedings of the 28th ACM Joint Meeting on European Software Engineering Conference and Symposium on the Foundations of Software Engineering (pp. 775-787).

Yokoyama et al., 2020 Yokoyama, H., Onoue, S., & Kikuchi, S. (2020, September). Towards Building Robust DNN Applications: An Industrial Case Study of Evolutionary Data Augmentation. In 2020 35th IEEE/ACM International Conference on Automated Software Engineering (ASE) (pp. 1184-1188). IEEE.

Yokoyama, 2019 Yokoyama, H. (2019, March). Machine learning system architectural pattern for improving operational stability. In 2019 IEEE International Conference on Software Architecture Companion (ICSA-C) (pp. 267-274). IEEE.

Zhang et al., 2018a Zhang, M., Zhang, Y., Zhang, L., Liu, C., & Khurshid, S. (2018, September). DeepRoad: GAN-based metamorphic testing and input validation framework for autonomous driving systems. In 2018 33rd IEEE/ACM International Conference on Automated Software Engineering (ASE) (pp. 132-142). IEEE.

Zhang et al., 2018b Zhang, Y., Chen, Y., Cheung, S. C., Xiong, Y., & Zhang, L. (2018, July). An empirical study on TensorFlow program bugs. In Proceedings of the 27th ACM SIGSOFT International Symposium on Software Testing and Analysis (pp. 129-140).

Zhang et al., 2019a Zhang, T., Gao, C., Ma, L., Lyu, M., & Kim, M. (2019, October). An empirical study of common challenges in developing deep learning applications. In 2019 IEEE 30th International Symposium on Software Reliability Engineering (ISSRE) (pp. 104-115). IEEE.

Zhang et al., 2019b Zhang, X., Yin, Z., Feng, Y., Shi, Q., Liu, J., & Chen, Z. (2019, November). NeuralVis: Visualizing and Interpreting Deep Learning Models. In 2019 34th IEEE/ACM International Conference on Automated Software Engineering (ASE) (pp. 1106-1109). IEEE.

Zhang et al., 2020a Zhang, R., Xiao, W., Zhang, H., Liu, Y., Lin, H., & Yang, M. (2020, October). An empirical study on program failures of deep learning jobs. In 2020 IEEE/ACM 42nd International Conference on Software Engineering (ICSE) (pp. 1159-1170). IEEE.

Zhang et al., 2020b Zhang, X., Xie, X., Ma, L., Du, X., Hu, Q., Liu, Y., Zhao, J. & Sun, M. (2020, October). Towards characterizing adversarial defects of deep learning software from the lens of uncertainty. In 2020 IEEE/ACM 42nd International Conference on Software Engineering (ICSE) (pp. 739-751). IEEE.

Zheng et al., 2019 Zheng, W., Wang, W., Liu, D., Zhang, C., Zeng, Q., Deng, Y., Yang, W., He, P. & Xie, T. (2019, May). Testing untestable neural machine translation: An industrial case. In 2019 IEEE/ACM 41st International Conference on Software Engineering: Companion Proceedings (ICSE-Companion) (pp. 314-315). IEEE.

Zhou et al., 2020a Zhou, J., Li, F., Dong, J., Zhang, H., & Hao, D. (2020, October). Cost-Effective Testing of a Deep Learning Model through Input Reduction. In 2020 IEEE 31st International Symposium on Software Reliability Engineering (ISSRE) (pp. 289-300). IEEE.

Zhou et al., 2020b Zhou, L., Yu, B., Berend, D., Xie, X., Li, X., Zhao, J. & Liu, X. (2020, December). An Empirical Study on Robustness of DNNs with Out-of-Distribution Awareness. In 2020 27th Asia-Pacific Software Engineering Conference (APSEC) (pp. 266-275). IEEE.

## 9.2 BIBLIOGRAPHY

Adams et al., 2016 Adams, J., Hillier-Brown, F. C., Moore, H. J., Lake, A. A., Araujo-Soares, V., White, M., & Summerbell, C. (2016). Searching and synthesising 'grey literature' and 'grey information' in public health: critical reflections on three case studies. Systematic reviews, 5(1), 164.

Ahuja et al., 2020 Ahuja, M. K., Belaid, M. B., Bernabé, P., Collet, M., Gotlieb, A., Lal, C., Marijan, D., Sen, S., Sharif, A. & Spieker, H. (2020). Opening the Software Engineering Toolbox for the Assessment of Trustworthy AI. arXiv preprint arXiv:2007.07768.

Alexander and Stevens, 2002 Alexander, I. F., & Stevens, R. (2002). Writing better requirements. London: Addison Wesley.

Anderson, 2015 Anderson, K. M. (2015, May). Embrace the challenges: software engineering in a big data world. In 2015 IEEE/ACM 1st International Workshop on Big Data Software Engineering (pp. 19-25). IEEE.

Andreessen, 2011 Andreessen, M. (2011). Why software is eating the world. Wall Street Journal, 20(2011), C2.

Ashmore et al., 2021 Ashmore, R., Calinescu, R., & Paterson, C. (2021). Assuring the Machine Learning Lifecycle: Desiderata, Methods, and Challenges. ACM Computing Surveys.

Atwal, 2019 Atwal, H. (2019). Practical DataOps: Delivering Agile Data Science at Scale. Apress.

Avizienis, A., Laprie, J. C., & Randell, B. (2001). Fundamental concepts of dependability (pp. 7-12). University of Newcastle upon Tyne, Computing Science.

Azeem et al., 2019 Azeem, M. I., Palomba, F., Shi, L., & Wang, Q. (2019). Machine learning techniques for code smell detection: A systematic literature review and meta-analysis. Information and Software Technology, 108, 115-138.

Barr et al., 2014 Barr, E. T., Harman, M., McMinn, P., Shahbaz, M., & Yoo, S. (2014). The oracle problem in software testing: A survey. IEEE transactions on software engineering, 41(5), 507-525.

Basili et al., 1994 Basili, V. R., Caldiera, G., & Rombach, H. D. (1994). The Goal Question Metric Approach, Chapter in Encyclopedia of Software Engineering. Wiley.

Belani et al., 2019 Belani, H., Vukovic, M., & Car, Ž. (2019, September). Requirements Engineering Challenges in Building AI-Based Complex Systems. In 2019 IEEE 27th International Requirements Engineering Conference Workshops (REW) (pp. 252-255). IEEE.

Bell, J., Legunsen, O., Hilton, M., Eloussi, L., Yung, T., & Marinov, D. (2018, May). DeFlaker: Automatically detecting flaky tests. In 2018 IEEE/ACM 40th International Conference on Software Engineering (ICSE) (pp. 433-444). IEEE.

Bellamy et al., 2019 Bellamy, R. K., Dey, K., Hind, M., Hoffman, S. C., Houde, S., Kannan, K., Lohia, P., Mehta, S., Mojsilovic, A., Nagar, S. & Zhang, Y. (2019). Think your artificial intelligence software is fair? Think again. IEEE Software, 36(4), 76-80.

Benchettara et al., 2010 Benchettara, N., Kanawati, R., & Rouveirol, C. (2010, August). Supervised machine learning applied to link prediction in bipartite social networks. In 2010 International Conference on Advances in Social Networks Analysis and Mining (pp. 326-330). IEEE.

Boehm and In, 1996 Boehm, B., & In, H. (1996). Identifying quality-requirement conflicts. IEEE software, 13(2), 25-35.

Booch, 2019 Booch, G. (2019) Grady Booch on the Future of AI. Available at https://www.infoq.com/news/2019/02/Grady-Booch-Future-AI/, https://www.infoq.com/presentations/ai-best-practices/. Last accessed on 23 April 2021.

Bosch et al., 2020 Bosch, J., Crnkovic, I., & Olsson, H. H. (2020). Engineering AI Systems: A Research Agenda. arXiv preprint arXiv:2001.07522.

Bosch, 2000 Bosch, J. (2000). Design and use of software architectures: adopting and evolving a product-line approach. Addison-Wesley Professional.

Bourque and Fairley, 2014 Bourque, P., & Fairley, R. E. (2014). Guide to the software engineering body of knowledge (SWEBOK (R)): Version 3.0. IEEE Computer Society Press.

Braiek and Khomh, 2020 Braiek, H. B., & Khomh, F. (2020). On testing machine learning programs. Journal of Systems and Software, 164, 110542.

Buscemi, N., Hartling, L., Vandermeer, B., Tjosvold, L., & Klassen, T. P. (2006). Single data extraction generated more errors than double data extraction in systematic reviews. Journal of clinical epidemiology, 59(7), 697-703.

Byrne, 2017 Byrne, C. (2017). Development Workflows for Data Scientists. O'Reilly Media.

Calders and Verwer, 2010 Calders, T., & Verwer, S. (2010). Three naive Bayes approaches for discrimination-free classification. Data Mining and Knowledge Discovery, 21(2), 277-292.

Caramujo and da Silva, 2015 Caramujo, J., & da Silva, A. M. R. (2015, July). Analyzing privacy policies based on a privacy-aware profile: The Facebook and LinkedIn case studies. In 2015 IEEE 17th Conference on Business Informatics (Vol. 1, pp. 77-84). IEEE.

Carleton et al., 2020 Carleton, A. D., Harper, E., Lyu, M. R., Eldh, S., Xie, T., & Menzies, T. (2020). Expert Perspectives on AI. IEEE Software, 37(4), 87-94.

Chechik, 2019 Chechik, M. (2019, September). Uncertain Requirements, Assurance and Machine Learning. In 2019 IEEE 27th International Requirements Engineering Conference (RE) (pp. 2-3). IEEE.

Chen et al., 1998 Chen, T.Y., Cheung, S.C. & Yiu, S.M. (1998). Metamorphic testing: A new approach for generating next test cases, Technical Report HKUST-CS98-01 (PDF), Department of Computer Science, The Hong Kong University of Science and Technology, Hong Kong, arXiv:2002.12543

Chen et al., 2015 Chen, C., Seff, A., Kornhauser, A., & Xiao, J. (2015). Deepdriving: Learning affordance for direct perception in autonomous driving. In Proceedings of the IEEE International Conference on Computer Vision (pp. 2722-2730).

Clarke et al., 2016 Clarke, P., O'Connor, R.V., Leavy, B.: A complexity theory viewpoint on the software development process and situational context. In: Proceedings of the International Conference on Software and Systems Process (ICSSP), Co-Located with the International Conference on Software Engineering (ICSE), pp. 86–90 (2016). https://doi.org/10.1145/2904354.2904369

Corbett-Davies and Goel, 2018 Corbett-Davies, S., & Goel, S. (2018). The measure and mismeasure of fairness: A critical review of fair machine learning. arXiv preprint arXiv:1808.00023.

Dalpiaz and Niu, 2020 Dalpiaz, F. & Niu, N. (2020) "Requirements Engineering in the Days of Artificial Intelligence," in IEEE Software, vol. 37, no. 4, pp. 7-10, July-Aug. 2020, doi: 10.1109/MS.2020.2986047.

Dam et al., 2018 Dam, H. K., Tran, T., & Ghose, A. (2018, May). Explainable software analytics. In Proceedings of the 40th International Conference on Software Engineering: New Ideas and Emerging Results (pp. 53-56).

DARPA, 2020 Defense Advanced Research Projects Agency, Assured Autonomy, https://www.darpa.mil/program/assured-autonomy. Last accessed on 12 April 2020.

Devlin et al., 2018 Devlin, J., Chang, M. W., Lee, K., & Toutanova, K. (2018). Bert: Pre-training of deep bidirectional transformers for language understanding. arXiv preprint arXiv:1810.04805.

Dixon et al., 2020 Dixon, M. F., Halperin, I., & Bilokon, P. (2020). Machine Learning in Finance. Springer Verlag Berlin Heidelberg.

Domingos, 2015 Domingos, P. (2015). The master algorithm: How the quest for the ultimate learning machine will remake our world. Basic Books.

Druffel and Little, 1990 Druffel, L., & Little, R. (1990). Software engineering for AI based software products. Data & Knowledge Engineering, 5(2), 93-103.

Dua and Graff, 2019 Dua, D. and Graff, C. (2019). UCI Machine Learning Repository [http://archive.ics.uci.edu/ml]. Irvine, CA: University of California, School of Information and Computer Science.

Durelli et al., 2019 Durelli, V. H., Durelli, R. S., Borges, S. S., Endo, A. T., Eler, M. M., Dias, D. R., & Guimaraes, M. P. (2019). Machine learning applied to software testing: A systematic mapping study. IEEE Transactions on Reliability, 68(3), 1189-1212.

Ereth, 2018 Ereth, J. (2018, August). DataOps-Towards a Definition. In LWDA (pp. 104-112).

European Commission, 2019 High-Level Expert Group on Artificial Intelligence, 'Ethics Guidelines for Trustworthy AI', Technical report, European Commission, (2019).

Fei-Fei, 2010 Fei-Fei, L. (2010, March). ImageNet: crowdsourcing, benchmarking & other cool things. In CMU VASC Seminar (Vol. 16, pp. 18-25).

Felderer et al., 2019 Felderer, M., Russo, B., & Auer, F. (2019). On Testing Data-Intensive Software Systems. In Security and Quality in Cyber-Physical Systems Engineering (pp. 129-148). Springer, Cham.

Fernández et al., 2017 Fernández, D. M., Wagner, S., Kalinowski, M., Felderer, M., Mafra, P., Vetrò, A., ... & Männistö, T. (2017). Naming the pain in requirements engineering. Empirical software engineering, 22(5), 2298-2338.

Ferreira et al., 2019 Ferreira, F., Silva, L. L., & Valente, M. T. (2019). Software Engineering Meets Deep Learning: A Literature Review. arXiv preprint arXiv:1909.11436.

Fink et al., 2020 Fink, O., Wang, Q., Svensén, M., Dersin, P., Lee, W. J., & Ducoffe, M. (2020). Potential, challenges and future directions for deep learning in prognostics and health management applications. Engineering Applications of Artificial Intelligence, 92, 103678.

Flaounas, 2017 Flaounas, I. (2017). Beyond the technical challenges for deploying Machine Learning solutions in a software company. arXiv preprint arXiv:1708.02363.

Floridi, 2019 Floridi, L. (2019). Establishing the rules for building trustworthy AI. Nature Machine Intelligence, 1(6), 261-262.

Fowler, 2002 Fowler, M. (2002). Patterns of enterprise application architecture. Addison-Wesley Longman Publishing Co., Inc..

Fung et al., 2007 Fung, B. C., Wang, K., & Philip, S. Y. (2007). Anonymizing classification data for privacy preservation. IEEE transactions on knowledge and data engineering, 19(5), 711-725.

Gade et al., 2019 Gade, K., Geyik, S. C., Kenthapadi, K., Mithal, V., & Taly, A. (2019, July). Explainable AI in industry. In Proceedings of the 25th ACM SIGKDD International Conference on Knowledge Discovery & Data Mining (pp. 3203-3204).

Galhotra et al., 2017 Galhotra, S., Brun, Y., & Meliou, A. (2017, August). Fairness testing: testing software for discrimination. In Proceedings of the 2017 11th Joint Meeting on Foundations of Software Engineering (pp. 498-510).

Gamma et al., 1995 Gamma, E., Helm, R., Johnson, R., Vlissides, J., & Patterns, D. (1995). Elements of Reusable Object-Oriented Software. Design Patterns. Massachusetts: Addison-Wesley Publishing Company.

Garousi and Mäntylä, 2016 Garousi, V., & Mäntylä, M. V. (2016). When and what to automate in software testing? A multi-vocal literature review. Information and Software Technology, 76, 92-117.

Garousi et al., 2016 Garousi, V., Felderer, M., & Mäntylä, M. V. (2016, June). The need for multivocal literature reviews in software engineering: complementing systematic literature reviews with grey literature. In Proceedings of the 20th international conference on evaluation and assessment in software engineering (pp. 1-6).

Garousi et al., 2018 Garousi, V., Felderer, M., Karapıçak, Ç. M., & Yılmaz, U. (2018). Testing embedded software: A survey of the literature. Information and Software Technology, 104, 14-45.

Garousi et al., 2019 Garousi, V., Giray, G., Tüzün, E., Catal, C., & Felderer, M. (2019). Aligning software engineering education with industrial needs: A meta-analysis. Journal of Systems and Software, 156, 65-83.

Gartner, 2020 Gartner, (October 19-22, 2020). Gartner Identifies the Top Strategic Technology Trends for 2021. At Gartner IT Symposium/Xpo 2020 Americas. https://www.gartner.com/en/newsroom/press-releases/2020-10-19-gartner-identifies-the-top-strategic-technology-trends-for-2021. Last accessed on 10 November 2020.

Gebru et al., 2018 Gebru, T., Morgenstern, J., Vecchione, B., Vaughan, J. W., Wallach, H., Daumeé III, H., & Crawford, K. (2018). Datasheets for datasets. arXiv preprint arXiv:1803.09010.

Gharibi et al., 2019 Gharibi, G., Walunj, V., Alanazi, R., Rella, S., & Lee, Y. (2019, June). Automated Management of Deep Learning Experiments. In Proceedings of the 3rd International Workshop on Data Management for End-to-End Machine Learning (pp. 1-4).

Giray and Tüzün, 2018 Giray, G., & Tüzün, E. (2018). A systematic mapping study on the current status of total cost of ownership for information systems. Information Technologies Journal, 11(2), 131-145.

Godin et al., 2015 Godin, K., Stapleton, J., Kirkpatrick, S. I., Hanning, R. M., & Leatherdale, S. T. (2015). Applying systematic review search methods to the grey literature: a case study examining guidelines for school-based breakfast programs in Canada. Systematic reviews, 4(1), 138.

Golovin et al., 2017 Golovin, D., Solnik, B., Moitra, S., Kochanski, G., Karro, J., & Sculley, D. (2017, August). Google vizier: A service for black-box optimization. In Proceedings of the 23rd ACM SIGKDD international conference on knowledge discovery and data mining (pp. 1487-1495).

Goodfellow et al., 2016 Goodfellow, I., Bengio, Y., & Courville, A. (2016). Deep learning. The MIT Press.

Google, 2020 "Problem Framing" course at developers.google.com. https://developers.google.com/machine-learning/problem-framing/cases. Last accessed on 30 October 2020.

Hartsell et al., 2019 Hartsell, C., Mahadevan, N., Ramakrishna, S., Dubey, A., Bapty, T., Johnson, T., Koutsoukos, X., Sztipanovits, J. & Karsai, G. (2019, October). CPS Design with Learning-Enabled Components: A Case Study. In Proceedings of the 30th International Workshop on Rapid System Prototyping (RSP'19) (pp. 57-63).

Hazelwood et al., 2018 Hazelwood, K., Bird, S., Brooks, D., Chintala, S., Diril, U., Dzhulgakov, D., Fawzy, M., Jia, B., Jia, Y., Kalro, A. & Law, J. (2018, February). Applied machine learning at facebook: A datacenter infrastructure perspective. In 2018 IEEE International Symposium on High Performance Computer Architecture (HPCA) (pp. 620-629). IEEE.

He and Garcia, 2009 He, H., & Garcia, E. A. (2009). Learning from imbalanced data. IEEE Transactions on knowledge and data engineering, 21(9), 1263-1284.

Heaton et al., 2017 Heaton, J. B., Polson, N. G., & Witte, J. H. (2017). Deep learning for finance: deep portfolios. Applied Stochastic Models in Business and Industry, 33(1), 3-12.

Heck, 2019 Heck, P. (2019). Software engineering for machine learning applications. Fontys Blogt. https://fontysblogt.nl/software-engineering-for-machine-learning-applications/. Last accessed on 29 October 2020.

Hesenius et al., 2019 Hesenius, M., Schwenzfeier, N., Meyer, O., Koop, W., & Gruhn, V. (2019, May). Towards a software engineering process for developing data-driven applications. In 2019 IEEE/ACM 7th International Workshop on Realizing Artificial Intelligence Synergies in Software Engineering (RAISE) (pp. 35-41). IEEE.

Howard and Gugger, 2020 Howard, J., & Gugger, S. (2020). Fastai: A layered API for deep learning. Information, 11(2), 108.

Huang, 2017 Huang, S. J. (2017). The design of a software engineering life cycle process for big data projects. IT Professional.

Hulten, 2018 Hulten, G. (2018). Building Intelligent Systems. Apress, Berkeley, CA. DOI: https://doi.org/10.1007/978-1-4842-3432-7

Hummer et al., 2019 Hummer, W., Muthusamy, V., Rausch, T., Dube, P., El Maghraoui, K., Murthi, A., & Oum, P. (2019, June). Modelops: Cloud-based lifecycle management for reliable and trusted ai. In 2019 IEEE International Conference on Cloud Engineering (IC2E) (pp. 113-120). IEEE.

IIBA, 2015 IIBA. (2015). A Guide to the Business Analysis Body of Knowledge, v3. International Institute of Business Analysis.

Jia and Harman, 2010 Jia, Y., & Harman, M. (2010). An analysis and survey of the development of mutation testing. IEEE transactions on software engineering, 37(5), 649-678.

Jiang et al., 2017 Jiang, F., Jiang, Y., Zhi, H., Dong, Y., Li, H., Ma, S., ... & Wang, Y. (2017). Artificial intelligence in healthcare: past, present and future. Stroke and vascular neurology, 2(4), 230-243.

Jiarpakdee et al., 2020 Jiarpakdee, J., Tantithamthavorn, C., Dam, H. K., & Grundy, J. (2020). An empirical study of model-agnostic techniques for defect prediction models. IEEE Transactions on Software Engineering.

Jordan and Mitchell, 2015 Jordan, M. I., & Mitchell, T. M. (2015). Machine learning: Trends, perspectives, and prospects. Science, 349(6245), 255-260.

Kamishima et al., 2011 Kamishima, T., Akaho, S., & Sakuma, J. (2011, December). Fairness-aware learning through regularization approach. In 2011 IEEE 11th International Conference on Data Mining Workshops (pp. 643-650). IEEE.

Kanewala and Bieman, 2014 Kanewala, U., & Bieman, J. M. (2014). Testing scientific software: A systematic literature review. Information and software technology, 56(10), 1219-1232.

Karpathy, 2017 Karpathy, A. (2017) Software 2.0. Available at https://medium.com/@karpathy/software-2-0-a64152b37c35. Last accessed on 23 April 2021.

Kästner and Kang, 2020 Kästner, C., & Kang, E. (2020). Teaching Software Engineering for AI-Enabled Systems. In Proceedings of the ACM/IEEE 42nd International Conference on Software Engineering: Software Engineering Education and Training (ICSE-SEET '20) (pp. 45-48). ACM. DOI: https://doi.org/10.1145/3377814.3381714

Kelleher, 2019 Kelleher, J. D. (2019). Deep learning. The MIT Press.

Kitchenham and Charters, 2007 Kitchenham, B., & Charters, S. (2007). Guidelines for performing systematic literature reviews in software engineering.

Kitchenham et al., 2015 Kitchenham, B. A., Budgen, D., & Brereton, P. (2015). Evidence-based software engineering and systematic reviews (Vol. 4). CRC press.

Klünder et al., 2019 Klünder, J., Hebig, R., Tell, P., Kuhrmann, M., Nakatumba-Nabende, J., Heldal, R., Krusche, S., Fazal-Baqaie, M., Felderer, M., Bocco, M.F.G. & Küpper, S. (2019, May). Catching up with method and process practice: An industry-informed baseline for researchers. In 2019 IEEE/ACM 41st International Conference on Software Engineering: Software Engineering in Practice (ICSE-SEIP) (pp. 255-264). IEEE.

Kohavi, 1996 Kohavi, R. (1996, August). Scaling up the accuracy of naive-bayes classifiers: A decision-tree hybrid. In Proceedings of the Second International Conference on Knowledge Discovery and Data Mining (Vol. 96, pp. 202-207).

Korel, 1990 Korel, B. (1990). Automated software test data generation. IEEE Transactions on software engineering, 16(8), 870-879.

Kotonya and Sommerville, 1998 Kotonya, G., & Sommerville, I. (1998). Requirements engineering: processes and techniques. Wiley Publishing.

Kovitz, 1998 Kovitz, B. L. (1998). Practical software requirements: a manual of content and style. Manning Publications Co..

Kriens and Verbelen, 2019 Kriens, P., & Verbelen, T. (2019). Software Engineering Practices for Machine Learning. arXiv preprint arXiv:1906.10366.

Krizhevsky and Hinton, 2009 Krizhevsky, A., & Hinton, G. (2009). Learning multiple layers of features from tiny images.

Kumeno, 2019 Kumeno, F. (2019). Software engineering challenges for machine learning applications: A literature review. Intelligent Decision Technologies, 13(4), 463-476.

Kuwajima et al., 2020 Kuwajima, H., Yasuoka, H., & Nakae, T. (2020). Engineering problems in machine learning systems. Machine Learning, 1-24.

Larman, 2005 Larman, C. (2005). Applying UML and patterns: an introduction to object oriented analysis and design and iterative development, 3rd Edition. Pearson Education.

LeCun et al., 1998 LeCun, Y., Bottou, L., Bengio, Y., & Haffner, P. (1998). Gradient-based learning applied to document recognition. Proceedings of the IEEE, 86(11), 2278-2324.

Li and Li, 2009 Li, T., & Li, N. (2009, June). On the tradeoff between privacy and utility in data publishing. In Proceedings of the 15th ACM SIGKDD international conference on Knowledge discovery and data mining (pp. 517-526).

Lin and Kolcz, 2012 Lin, J., & Kolcz, A. (2012, May). Large-scale machine learning at twitter. In Proceedings of the 2012 ACM SIGMOD International Conference on Management of Data (pp. 793-804).

Liu et al., 2019 Liu, Y., Ma, L., & Zhao, J. (2019, November). Secure Deep Learning Engineering: A Road Towards Quality Assurance of Intelligent Systems. In International Conference on Formal Engineering Methods (pp. 3-15). Springer, Cham.

Lorica and Loukides, 2018 Lorica, B., & Loukides, M. (2018). What machine learning means for software development. https://www.oreilly.com/radar/what-machine-learning-means-for-software-development/. Last accessed on 29 October 2020.

Loshin, 2011 Loshin, D. (2011). The practitioner's guide to data quality improvement. Morgan Kaufmann.

Louridas and Ebert, 2016 Louridas, P. & Ebert, C. (2016). Machine Learning. IEEE Software, 33(5), 110-115.

Luo, Q., Hariri, F., Eloussi, L., & Marinov, D. (2014, November). An empirical analysis of flaky tests. In Proceedings of the 22nd ACM SIGSOFT International Symposium on Foundations of Software Engineering (pp. 643-653).

Lwakatare et al., 2020 Lwakatare, L. E., Raj, A., Crnkovic, I., Bosch, J., & Olsson, H. H. (2020). Large-scale machine learning systems in real-world industrial settings: A review of challenges and solutions. Information and Software Technology, 127, 106368.

Maimon and Rokach, 2010 Maimon, O., & Rokach, L. (Eds.). (2010). Data mining and knowledge discovery handbook, 2nd Edition. Springer.

Malhotra, 2015 Malhotra, R. (2015). A systematic review of machine learning techniques for software fault prediction. Applied Soft Computing, 27, 504-518.

Martin, 2002 Martin, R. C. (2002). Agile software development: principles, patterns, and practices. Prentice Hall.

Masuda et al., 2018 Masuda, S., Ono, K., Yasue, T., & Hosokawa, N. (2018, April). A Survey of Software Quality for Machine Learning Applications. In 2018 IEEE International Conference on Software Testing, Verification and Validation Workshops (ICSTW) (pp. 279-284). IEEE.

McMahan et al., 2013 McMahan, H. B., Holt, G., Sculley, D., Young, M., Ebner, D., Grady, J., Nie, L., Phillips, T., Davydov, E., Golovin, D. & Chikkerur, S. (2013, August). Ad click prediction: a view from the trenches. In Proceedings of the 19th ACM SIGKDD international conference on Knowledge discovery and data mining (pp. 1222-1230).

Meacham et al., 2019 Meacham, S., Isaac, G., Nauck, D., & Virginas, B. (2019, July). Towards Explainable AI: Design and Development for Explanation of Machine Learning Predictions for a Patient Readmittance Medical Application. In Intelligent Computing-Proceedings of the Computing Conference (pp. 939-955). Springer, Cham.

Meade et al., 2019 Meade, E., O'Keeffe, E., Lyons, N., Lynch, D., Yilmaz, M., Gulec, U., O'Connor, R.V. & Clarke, P. M. (2019, September). The Changing Role of the Software Engineer. In European Conference on Software Process Improvement (pp. 682-694). Springer, Cham.

Menzies, 2019 Menzies, T. (2019). The Five Laws of SE for AI. IEEE Software, 37(1), 81-85.

Miao et al., 2017 Miao, H., Li, A., Davis, L. S., & Deshpande, A. (2017, April). Modelhub: Deep learning lifecycle management. In 2017 IEEE 33rd International Conference on Data Engineering (ICDE) (pp. 1393-1394). IEEE.

Microsoft, 2009 Microsoft. (2009). Microsoft application architecture guide. Microsoft Press.

Miles and Huberman, 1994 Miles, M. B., & Huberman, A. M. (1994). Qualitative data analysis: An expanded sourcebook. Sage Publications.

Mitchell, 2011 Mitchell, P. (2011). A systematic review of the efficacy and safety outcomes of anti-VEGF agents used for treating neovascular age-related macular degeneration: comparison of ranibizumab and bevacizumab. Current medical research and opinion, 27(7), 1465-1475.

Mjeda and Botterweck, 2019 Mjeda, A., & Botterweck, G. (2019). Scalable Software Testing and Verification for Industrial-Scale Systems: The Challenges. Electronic Communications of the EASST, 77.

Mohanani et al., 2018 Mohanani, R., Salman, I., Turhan, B., Rodríguez, P., & Ralph, P. (2018). Cognitive biases in software engineering: a systematic mapping study. IEEE Transactions on Software Engineering.

Molléri et al., 2019 Molléri, J. S., Petersen, K., & Mendes, E. (2019). Cerse-catalog for empirical research in software engineering: A systematic mapping study. Information and Software Technology, 105, 117-149.

Moore, 1990 Moore, A. W. (1990). Efficient memory-based learning for robot control. PhD thesis, University of Cambridge.

Morgenthaler et al., 2012 Morgenthaler, J. D., Gridnev, M., Sauciuc, R., & Bhansali, S. (2012, June). Searching for build debt: Experiences managing technical debt at Google. In 2012 Third International Workshop on Managing Technical Debt (MTD) (pp. 1-6). IEEE.

Mosley and Posey, 2002 Mosley, D. J., & Posey, B. A. (2002). Just enough software test automation. Prentice Hall Professional.

Motta et al., 2018 Motta, R. C., de Oliveira, K. M., & Travassos, G. H. (2018, September). On challenges in engineering IoT software systems. In Proceedings of the XXXII Brazilian Symposium on Software Engineering (pp. 42-51).

Muntés-Mulero and Nin, 2009 Muntés-Mulero, V., & Nin, J. (2009, November). Privacy and anonymization for very large datasets. In Proceedings of the 18th ACM conference on Information and knowledge management (pp. 2117-2118).

Nascimento et al., 2020 Nascimento, E., Nguyen-Duc, A., Sundbø, I., & Conte, T. (2020). Software engineering for artificial intelligence and machine learning software: A systematic literature review. arXiv preprint arXiv:2011.03751.

Nassif et al., 2016 Nassif, A. B., Azzeh, M., Capretz, L. F., & Ho, D. (2016). Neural network models for software development effort estimation: a comparative study. Neural Computing and Applications, 27(8), 2369-2381.

Nemecek and Bemley, 1993 Nemecek, S., & Bemley, J. (1993, March). A Model for Estimating the Cost of AI Software Development: What to do if there are no Lines of Code?. In [1993] Proceedings IEEE International Conference on Developing and Managing Intelligent System Projects (pp. 2-9). IEEE.

Netzer et al., 2011 Netzer, Y., Wang, T., Coates, A., Bissacco, A., Wu, B., & Ng, A. Y. (2011). Reading digits in natural images with unsupervised feature learning.

Ng, 2018 Ng, A. (2018). Machine learning yearning. URL: https://www.deeplearning.ai/machine-learning-yearning/.

Northrop et al., 2019 Northrop, L., Ozkaya, I., Fairbanks, G. and Keeling, M. (2019). Designing the Software Systems of the Future. SIGSOFT Softw. Eng. Notes 43, 4 (January 2019), 28–30.

Nushi et al., 2018 Nushi, B., Kamar, E., & Horvitz, E. (2018, June). Towards accountable ai: Hybrid human-machine analyses for characterizing system failure. In Sixth AAAI Conference on Human Computation and Crowdsourcing.

Pan and Yang, 2009 Pan, S. J., & Yang, Q. (2009). A survey on transfer learning. IEEE Transactions on knowledge and data engineering, 22(10), 1345-1359.

Panigrahi et al., 2019 Panigrahi, R., Kuanar, S. K., Kumar, L., Padhy, N., & Satapathy, S. C. (2019). Software reusability metrics prediction and cost estimation by using machine learning algorithms. International Journal of Knowledge-based and Intelligent Engineering Systems, 23(4), 317-328.

Partridge, 1988 Partridge, D. (1988). Artificial intelligence and software engineering: a survey of possibilities. Information and Software Technology, 30(3), 146-152.

Pedreshi et al., 2008 Pedreshi, D., Ruggieri, S., & Turini, F. (2008, August). Discrimination-aware data mining. In Proceedings of the 14th ACM SIGKDD international conference on Knowledge discovery and data mining (pp. 560-568).

Petersen et al., 2008 Petersen, K., Feldt, R., Mujtaba, S., & Mattsson, M. (2008, June). Systematic mapping studies in software engineering. In 12th International Conference on Evaluation and Assessment in Software Engineering (EASE) 12 (pp. 1-10).

Petersen et al., 2015 Petersen, K., Vakkalanka, S., & Kuzniarz, L. (2015). Guidelines for conducting systematic mapping studies in software engineering: An update. Information and Software Technology, 64, 1-18.

Polyzotis et al., 2018 Polyzotis, N., Roy, S., Whang, S. E., & Zinkevich, M. (2018). Data lifecycle challenges in production machine learning: a survey. ACM SIGMOD Record, 47(2), 17-28.

Pospieszny et al., 2018 Pospieszny, P., Czarnacka-Chrobot, B., & Kobylinski, A. (2018). An effective approach for software project effort and duration estimation with machine learning algorithms. Journal of Systems and Software, 137, 184-196.

Protzel, P. W., Palumbo, D. L., & Arras, M. K. (1993). Performance and fault-tolerance of neural networks for optimization. IEEE transactions on Neural Networks, 4(4), 600-614.

Pruss, 2017 Pruss, L. (2017). Infrastructure 3.0: Building blocks for the AI revolution. Venture Beat. https://venturebeat.com/2017/11/28/infrastructure-3-0-building-blocks-for-the-ai-revolution/ Last accessed on 25 April 2020.

QA4AI, 2020 Consortium of Quality Assurance for Artificial-intelligence-based products and services. (2020). Guidelines for the Quality Assurance of AI Systems, 2020.02 edition. Available at http://www.qa4ai.jp/QA4AI.Guideline.202002.en.pdf. Last accessed on 2 May 2021.

Rahman et al., 2019 Rahman, M. S., Rivera, E., Khomh, F., Guéhéneuc, Y. G., & Lehnert, B. (2019). Machine learning software engineering in practice: An industrial case study. arXiv preprint arXiv:1906.07154.

Ralph, P. (2015). The Sensemaking-coevolution-implementation theory of software design. Science of Computer Programming, 101, 21-41.

Ramler and Wolfmaier, 2006 Ramler, R., & Wolfmaier, K. (2006, May). Economic perspectives in test automation: balancing automated and manual testing with opportunity cost. In Proceedings of the 2006 international workshop on Automation of software test (pp. 85-91).

Rao, 2019 Rao, D. J. (2019). Machine Learning Development Lifecycle. In Keras to Kubernetes®, D. J. Rao (Ed.). doi:10.1002/9781119564843.ch9

Ravichandran and Rothenberger, 2003 Ravichandran, T., & Rothenberger, M. A. (2003). Software reuse strategies and component markets. Communications of the ACM, 46(8), 109-114.

Rech and Althoff, 2004 Rech, J., & Althoff, K. D. (2004). Artificial Intelligence and Software Engineering: Status and Future Trends. In Special Issue on Artificial Intelligence and Software Engineering, KI.

Ribeiro et al., 2016 Ribeiro, M. T., Singh, S., & Guestrin, C. (2016, August). " Why should i trust you?" Explaining the predictions of any classifier. In Proceedings of the 22nd ACM SIGKDD international conference on knowledge discovery and data mining (pp. 1135-1144).

Riccio et al., 2020 Riccio, V., Jahangirova, G., Stocco, A., Humbatova, N., Weiss, M., & Tonella, P. (2020). Testing machine learning based systems: a systematic mapping. Empirical Software Engineering, 25(6), 5193-5254.

Rudraraju, N.V., & Boyanapally, V. (2019). Data Quality Model for Machine learning. MSc thesis, Faculty of Computing, Blekinge Institute of Technology.

Russakovsky et al., 2015 Russakovsky, O., Deng, J., Su, H., Krause, J., Satheesh, S., Ma, S., Huang, Z., Karpathy, A., Khosla, A., Bernstein, M., Berg, A.C. & Fei-Fei, L. (2015). Imagenet large scale visual recognition challenge. International journal of computer vision, 115(3), 211-252.

Russel and Norvig, 2013 Russel, S., & Norvig, P. (2013). Artificial intelligence: a modern approach. Pearson Education Limited.

Sapp, 2017 Sapp, C. E. (2017). Preparing and architecting for machine learning. Gartner Technical Professional Advice, 1-37.

Sato et al., 2019 Sato, D., Wider, A. & Windheuser, C. (2019). Continuous Delivery for Machine Learning. Available at https://martinfowler.com/articles/cd4ml.html. Last accessed 22 April 2020.

Sato, 2014 Sato, D. (2014). Canary Release. https://martinfowler.com/bliki/CanaryRelease.html. Last accessed on 29 October 2020.

Schelter et al., 2018 Schelter, S., Biessmann, F., Januschowski, T., Salinas, D., Seufert, S., Szarvas, G., Vartak, M., Madden, S., Miao, H. & Deshpande, A. (2018). On Challenges in Machine Learning Model Management. IEEE Data Eng. Bull., 41(4), 5-15.

Schwerdtner et al., 2020 Schwerdtner, P., Greßner, F., Kapoor, N., Assion, F., Sass, R., Günther, W., Hüger, F. & Schlicht, P. (2020). Risk Assessment for Machine Learning Models. Machine Learning for Autonomous Driving Workshop at the 34th Conference on Neural Information Processing Systems (NeurIPS 2020), Vancouver, Canada.

Sculley et al., 2014 Sculley, D., Holt, G., Golovin, D., Davydov, E., Phillips, T., Ebner, D., Chaudhary, V. & Young, M. (2014). Machine learning: The high interest credit card of technical debt.

Sculley et al., 2015 Sculley, D., Holt, G., Golovin, D., Davydov, E., Phillips, T., Ebner, D., Chaudhary, V., Young, M., Crespo, J.F. & Dennison, D. (2015). Hidden technical debt in machine learning systems. In Advances in neural information processing systems (pp. 2503-2511).

Segura et al., 2016 Segura, S., Fraser, G., Sanchez, A. B., & Ruiz-Cortés, A. (2016). A survey on metamorphic testing. IEEE Transactions on software engineering, 42(9), 805-824.

Shams, 2018 Shams, R. (2018). Developing machine learning products better and faster at startups. IEEE Engineering Management Review, 46(3), 36-39.

Sherin and Iqbal, 2019 Sherin, S., & Iqbal, M. Z. (2019). A Systematic Mapping Study on Testing of Machine Learning Programs. arXiv preprint arXiv:1907.09427.

Siebert et al., 2020 Siebert, J., Joeckel, L., Heidrich, J., Nakamichi, K., Ohashi, K., Namba, I., Yamamoto, R. & Aoyama, M. (2020, September). Towards Guidelines for Assessing Qualities of Machine Learning Systems. In International Conference on the Quality of Information and Communications Technology (pp. 17-31). Springer, Cham.

Smith and Topin, 2016 Smith, L. N., & Topin, N. (2016). Deep convolutional neural network design patterns. arXiv preprint arXiv:1611.00847.

Smith et al., 2020 Smith, M. J., Sala, C., Kanter, J. M., & Veeramachaneni, K. (2020, June). The machine learning bazaar: Harnessing the ML ecosystem for effective system development. In Proceedings of the 2020 ACM SIGMOD International Conference on Management of Data (pp. 785-800).

Sutton et al., 2018 Sutton, C., Hobson, T., Geddes, J., & Caruana, R. (2018, July). Data diff: Interpretable, executable summaries of changes in distributions for data wrangling. In Proceedings of the 24th ACM SIGKDD International Conference on Knowledge Discovery & Data Mining (pp. 2279-2288).

Szegedy et al., 2014 Szegedy, C., Zaremba, W., Sutskever, I., Bruna, J., Erhan, D., Goodfellow, I.J. & Fergus, R. (2014). Intriguing properties of neural networks. In 2nd International Conference on Learning Representations (ICLR).

Szegedy et al., 2016 Szegedy, C., Vanhoucke, V., Ioffe, S., Shlens, J., & Wojna, Z. (2016). Rethinking the inception architecture for computer vision. In Proceedings of the IEEE conference on computer vision and pattern recognition (pp. 2818-2826).

Tantithamthavorn et al., 2018 Tantithamthavorn, C., & Hassan, A. E. (2018, May). An experience report on defect modelling in practice: Pitfalls and challenges. In Proceedings of the 40th International conference on software engineering: Software engineering in practice (pp. 286-295).

Tantithamthavorn et al., 2020 Tantithamthavorn, C., Jiarpakdee, J., & Grundy, J. (2020). Explainable AI for Software Engineering. arXiv preprint arXiv:2012.01614.

Tarhan and Giray, 2017 Tarhan, A., & Giray, G. (2017, June). On the use of ontologies in software process assessment: a systematic literature review. In Proceedings of the 21st International Conference on Evaluation and Assessment in Software Engineering (pp. 2-11).

Tata et al., 2017 Tata, S., Popescul, A., Najork, M., Colagrosso, M., Gibbons, J., Green, A., Mah, A., Smith, M., Garg, D., Meyer, C. & Kan, R. (2017, August). Quick access: Building a smart experience for google drive. In Proceedings of the 23rd ACM SIGKDD International Conference on Knowledge Discovery and Data Mining (pp. 1643-1651).

Tripakis, 2018 Tripakis, S. (2018, May). Data-driven and model-based design. In 2018 IEEE Industrial Cyber-Physical Systems (ICPS) (pp. 103-108). IEEE.

Tsymbal, 2004 Tsymbal, A. (2004). The problem of concept drift: definitions and related work. Computer Science Department, Trinity College Dublin, 106(2), 58.

Tuncali et al., 2019 Tuncali, C. E., Fainekos, G., Prokhorov, D., Ito, H., & Kapinski, J. (2019). Requirements-driven test generation for autonomous vehicles with machine learning components. IEEE Transactions on Intelligent Vehicles, 5(2), 265-280.

Vartak et al., 2016 Vartak, M., Subramanyam, H., Lee, W. E., Viswanathan, S., Husnoo, S., Madden, S., & Zaharia, M. (2016, June). ModelDB: a system for machine learning model management. In Proceedings of the Workshop on Human-In-the-Loop Data Analytics (pp. 1-3).

Vartak, 2018 Vartak, M. (2018). Infrastructure for model management and model diagnosis (Doctoral dissertation, Massachusetts Institute of Technology).

Wang and Abraham, 2015 Wang, H., & Abraham, Z. (2015, July). Concept drift detection for streaming data. In 2015 International Joint Conference on Neural Networks (IJCNN) (pp. 1-9). IEEE.

Wang et al., 2018 Wang, A., Singh, A., Michael, J., Hill, F., Levy, O., & Bowman, S. R. (2018). GLUE: A multi-task benchmark and analysis platform for natural language understanding. arXiv preprint arXiv:1804.07461.

Washizaki et al., 2019a Washizaki, H., Uchida, H., Khomh, F. & Gueheneuc, Y. (2019). Machine Learning Architecture and Design Patterns. Available at http://www.washi.cs.waseda.ac.jp/wp-content/uploads/2019/12/IEEE_Software_19__ML_Patterns.pdf. Last accessed on 6 November 2020.

Washizaki et al., 2019b Washizaki, H., Uchida, H., Khomh, F., & Guéhéneuc, Y. G. (2019, December). Studying software engineering patterns for designing machine learning systems. In 2019 10th International Workshop on Empirical Software Engineering in Practice (IWESEP) (pp. 49-495). IEEE.

Watanabe et al., 2019 Watanabe, Y., Washizaki, H., Sakamoto, K., Saito, D., Honda, K., Tsuda, N., Fukazawa, Y. & Yoshioka, N. (2019). Preliminary Systematic Literature Review of Machine Learning System Development Process. arXiv preprint arXiv:1910.05528.

Wei et al., 2017 Wei, J., He, J., Chen, K., Zhou, Y., & Tang, Z. (2017). Collaborative filtering and deep learning based recommendation system for cold start items. Expert Systems with Applications, 69, 29-39.

Weiss et al., 2016 Weiss, K., Khoshgoftaar, T. M., & Wang, D. (2016). A survey of transfer learning. Journal of Big data, 3(1), 9.

Weyuker, 1982 Weyuker, E. J. (1982). On testing non-testable programs. The Computer Journal, 25(4), 465-470.

Wiegers and Beatty, 2013 Wiegers, K., & Beatty, J. (2013). Software requirements, 3rd Edition. Microsoft Press.

Wiegers, 1996 Wiegers, K. E. (1996). Creating a software engineering culture. New York: Dorset House Publishing.

Wiegers, 2005 Wiegers, K. (2005). More about software requirements: thorny issues and practical advice. Microsoft Press.

Wohlin, 2014 Wohlin, C. (2014, May). Guidelines for snowballing in systematic literature studies and a replication in software engineering. In Proceedings of the 18th international conference on evaluation and assessment in software engineering (pp. 1-10).

Workera, 2020 Workera, AI Career Pathways: Put Yourself on the Right Track, https://workera.ai/candidates/report/, Last accessed on 12 April 2020.

World Economic Forum, 2020 World Economic Forum. (2020, October). The future of jobs report 2020. Geneva: World Economic Forum. Available at https://www.weforum.org/reports/the-future-of-jobs-report-2020. Last accessed on 7 November 2020.

Wu et al., 2021 Wu, R., Guo, C., Hannun, A., & van der Maaten, L. (2021). Fixes That Fail: Self-Defeating Improvements in Machine-Learning Systems. arXiv preprint arXiv:2103.11766.

Wuest et al., 2016 Wuest, T., Weimer, D., Irgens, C., & Thoben, K. D. (2016). Machine learning in manufacturing: advantages, challenges, and applications. Production & Manufacturing Research, 4(1), 23-45.

Xiao et al., 2017 Xiao, H., Rasul, K., & Vollgraf, R. (2017). Fashion-mnist: a novel image dataset for benchmarking machine learning algorithms. arXiv preprint arXiv:1708.07747.

Xie, 2018 Xie, T. (2018, September). Intelligent software engineering: Synergy between ai and software engineering. In International Symposium on Dependable Software Engineering: Theories, Tools, and Applications (pp. 3-7). Springer, Cham.

Young et al., 2018 Young, T., Hazarika, D., Poria, S., & Cambria, E. (2018). Recent trends in deep learning based natural language processing. ieee Computational intelligenCe magazine, 13(3), 55-75.

Zaharia et al., 2018 Zaharia, M., Chen, A., Davidson, A., Ghodsi, A., Hong, S. A., Konwinski, A., Murching, S., Nykodym, T., Ogilvie, P., Parkhe, M. & Xie, F. (2018). Accelerating the Machine Learning Lifecycle with MLflow. IEEE Data Eng. Bull., 41(4), 39-45.

Zhang and Tsai, 2003 Zhang, D., & Tsai, J. J. (2003). Machine learning and software engineering. Software Quality Journal, 11(2), 87-119.

Zhang et al., 2018c Zhang, J., Zhang, L., Harman, M., Hao, D., Jia, Y., & Zhang, L. (2018). Predictive mutation testing. IEEE Transactions on Software Engineering, 45(9), 898-918.

Zhang et al., 2020c Zhang, J. M., Harman, M., Ma, L., & Liu, Y. (2020). Machine learning testing: Survey, landscapes and horizons. IEEE Transactions on Software Engineering.

Zhou et al., 2015 Zhou, Z. Q., Xiang, S., & Chen, T. Y. (2015). Metamorphic testing for software quality assessment: A study of search engines. IEEE Transactions on Software Engineering, 42(3), 264-284.

Zhou et al., 2018 Zhou, Z. H. (2018). Machine learning challenges and impact: an interview with Thomas Dietterich. National Science Review, 5(1), 54-58.

Zinkevich, 2017 Zinkevich, M. (2017). Rules of machine learning: Best practices for ML engineering. Google.[Online]. Available: https://developers. google. com/machine-learning/guides/rules-of-ml/.[Accessed Oct. 18, 2018].

Žliobaitė, 2010 Žliobaitė, I. (2010). Learning under concept drift: an overview. arXiv preprint arXiv:1010.4784.

Zoph et al., 2018 Zoph, B., Vasudevan, V., Shlens, J., & Le, Q. V. (2018). Learning transferable architectures for scalable image recognition. In Proceedings of the IEEE conference on computer vision and pattern recognition (pp. 8697-8710).